\documentclass[a4paper,11pt]{article}
\pdfoutput=1 
             
\usepackage{jheppub} 

\usepackage[bottom]{footmisc}
\usepackage{amssymb}
\usepackage{amsmath}
\usepackage{amsthm}
\usepackage[usenames,dvipsnames]{xcolor}
\usepackage{epsfig}
\usepackage{dcolumn}
\usepackage{tikz}
\usetikzlibrary{shapes.geometric, arrows}
\usetikzlibrary{calc}
\usepackage{upgreek}
\usepackage{setspace}
\usepackage{enumerate}
\usepackage[inline]{enumitem}

\usepackage{array,multirow,bigdelim,arydshln}
\usepackage{appendix}
\usepackage{xparse}
\usepackage{stmaryrd}
\usepackage[T1]{fontenc} 
\usepackage{mathtools}
\usepackage{physics} 
\usepackage{adjustbox}
\usepackage{multirow}
\usepackage{graphicx} 
\usepackage{float} 
\graphicspath{{./images/}}
\usepackage[nottoc]{tocbibind}
\usepackage{hyperref}
\usepackage[utf8]{inputenc}
\hypersetup{
	colorlinks,
	urlcolor=Maroon,
	linkcolor=Maroon,
	citecolor=Maroon
	}

\NewDocumentCommand{\binomial}{omm}
 {%
  \genfrac(){0pt}{}{#2}{#3}%
  \IfValueT{#1}{_{\!#1}}%
 }
\NewDocumentCommand{\eulerian}{omm}
 {%
  \genfrac<>{0pt}{}{#2}{#3}%
  \IfValueT{#1}{_{\!#1}}%
 }

\usepackage{latexsym}
\usepackage{tikz}

\theoremstyle{plain}

\theoremstyle{definition}

\def\bea#1\eea{\begin{eqnarray}#1\end{eqnarray}}
\def\be#1\ee{\begin{equation}#1\end{equation}}
\def\ba#1\ea{\begin{align}#1\end{align}}
\def\la{\label}
\def\nl{\nonumber\\} 
\def\non{\nonumber}

\def\yz#1\yz {{\color{blue} [[YZ: #1]] }}

\usepackage{amsmath}
\usepackage{multicol}
\usepackage{bbm}
\usepackage{enumerate}

\usepackage{amsthm}
\usepackage{mathrsfs}
\usepackage{upgreek}
\usepackage{amssymb}
\usepackage{bm}
\usepackage{setspace}
\usepackage{array,multirow,arydshln}
\usepackage{bigdelim}
\usepackage{scalerel}
\usepackage{diagbox}
\usepackage{caption}
\usepackage{tabularx}

\usepackage{tikz}

\usetikzlibrary{shapes.geometric,arrows,arrows.meta,decorations.pathmorphing,decorations.markings,patterns}

\usetikzlibrary{decorations.pathreplacing,calligraphy}

\def\<{\langle}
\def\>{\rangle}
\def\a{\alpha}

\def\e{\epsilon}

\font\tenshuffle=shuffle10 \font\sevenshuffle=shuffle7 \font\fiveshuffle=shuffle7 at 5pt
\def\shuffle{{%
		\def\Dshuffle{\mathbin{\hbox{\tenshuffle\char'001}}}%
		\def\Sshuffle{\mathbin{\hbox{\sevenshuffle\char'001}}}%
		\def\SSshuffle{\mathbin{\hbox{\fiveshuffle\char'001}}}%
		\mathchoice{\Dshuffle}{\Dshuffle}{\Sshuffle}{\SSshuffle}}}    

\newcommand*{\halfway}{0.5*\pgfdecoratedpathlength+4.2pt}

\def\beq{\begin{equation}}
\def\eeq{\end{equation}}

\let\Im\relax

\DeclareMathOperator{\Im}{Im}

\usepackage{enumitem}

\usepackage{enumerate}

\usepackage[percent]{overpic}
\usepackage{multirow} 
\usepackage{slashed}

\usepackage{cleveref}

\newcommand{\ap}{{\alpha'}}

\usepackage{subfig}
\usepackage{float}

\def \CC {{\bm C}}
\def \MM {{\bm M}}
\def  \OO {{\bm O}}
\def \CCF {{\tilde {\bm C}}}
\def \MMF { {\tilde {\bm  M}}}
\def  \OOF {{\tilde {\bm O}}}

\def\beq{\begin{equation}}
\def\eeq{\end{equation}}

\let\Im\relax

\DeclareMathOperator{\Im}{Im}

\usepackage{enumitem}

\usepackage{enumerate}

\allowdisplaybreaks[0]

\DeclareFontFamily{U}{mathx}{\hyphenchar\font45}
\DeclareFontShape{U}{mathx}{m}{n}{
      <5> <6> <7> <8> <9> <10>
      <10.95> <12> <14.4> <17.28> <20.74> <24.88>
      mathx10
      }{}
\DeclareSymbolFont{mathx}{U}{mathx}{m}{n}
\DeclareFontSubstitution{U}{mathx}{m}{n}
\DeclareMathAccent{\widecheck}{0}{mathx}{"71}

\usepackage{mathrsfs}
\usepackage{orcidlink}

\title{Advanced tools for basis decompositions of genus-one string integrals}

\author{Yong Zhang 
}

\affiliation[a]{Perimeter Institute for Theoretical Physics, Waterloo, ON N2L 2Y5, Canada}

\affiliation[b]{Department of Physics and Astronomy, Uppsala University, Box 516, 75120 Uppsala, Sweden}

\affiliation[c]{
School of Physical Science and Technology, Ningbo University, Ningbo 315211, China}

\emailAdd{yzhang@perimeterinstitute.ca}

\date{\today}

\abstract{In string theories, one-loop scattering amplitudes are characterized by integrals over genus-one surfaces using the Kronecker-Eisenstein series. A recent methodology proposed a genus-one basis formed from products of these series of chain topologies. A prior work further deconstructed cyclic products of the Kronecker-Eisenstein series on this basis. Building on it, our study further employs advanced and comprehensive combinatorial techniques to decompose more general genus-one integrands including a product of an arbitrary number of cyclic products of Kronecker-Eisenstein series, supplemented by {\tt Mathematica} codes. Our insights enhance the understanding of multiparticle amplitudes across various string theories and illuminate loop-level parallels with string tree-level amplitudes.
}

\begin{document}

\maketitle{}
\addtocontents{toc}{\protect\setcounter{tocdepth}{2}}

\newpage

\setcounter{page}{1}

\setcounter{tocdepth}{2}

\numberwithin{equation}{section}

 \newpage

\section{Introduction}

In string theories, scattering amplitudes are drawn from moduli-space integrals on punctured worldsheets. The core of the associated integrands, represented by certain correlation functions of vertex operators, encapsulates the scattering data. Efforts to simplify these integrands, especially through decomposition into bases of functions of the worldsheet moduli, have unveiled intricate patterns in string amplitudes. For $n$-point tree-level amplitudes, the Parke-Taylor factors, dependent on $n$ punctures, play a significant role. Aomoto \cite{Aomoto87} showed that these factors, coupled with the Koba-Nielsen factor, fall into $(n{-}3)!$-dimensional bases and resonate with the framework of twisted (co)homologies, as outlined in various works, including \cite{Mizera:2017cqs, Mizera:2017rqa, Mizera:2019gea}. 

The deep interplay of these integration-by-parts relations transcends field theory, string theory, and mathematics \cite{Mafra:2022wml}. 
They clarify relationships among 
gauge-theory amplitudes \cite{Zfunctions, Stieberger:2014hba} which often have a simple uplift to all orders in the inverse string tension $\ap$ \cite{Schlotterer:2016cxa,Bjerrum-Bohr:2009ulz,Stieberger:2009hq}.
Field-theory structures in tree amplitudes and their associations with gravity and gauge theory amplitudes are also spotlighted \cite{Mafra:2011nv, Zfunctions, Carrasco:2016ldy, Azevedo:2018dgo, Kawai:1985xq}. Furthermore, manipulations of certain genus-zero correlators contribute to the understanding of braid matrices in the $\ap$-expansion of string amplitudes \cite{Broedel:2013aza, Mizera:2019gea, Kaderli:2019dny}.

Tree-level insights from Parke-Taylor bases have spurred investigations into analogous bases for loop-level correlators under integration by parts (IBP) and algebraic relations of the integrand, with a focus on one-loop string amplitudes. Specifically, correlators on genus-one surfaces like the torus are expressed using Jacobi theta functions, replacing the Koba-Nielsen factor with $|\theta_1(z_{i,j},\tau)|^{\ap k_i \cdot k_j}$ (with $z_{i,j} := z_i{-}z_j$ and $k_i$ the external momenta). This study centers on functions of the punctures $z_i$ and the modular parameter $\tau$ that supplement the one-loop Koba-Nielsen factor and can be viewed as the loop analogs of the Parke-Taylor factors. These functions are systematically examined under IBP and Fay relations, collectively termed F-IBP.

Genus-one correlators for various string amplitudes are expressed using coefficients $f^{(w)}(z_{i,j},\tau)$ from the Kronecker-Eisenstein series \cite{Dolan:2007eh, Broedel:2014vla, Gerken:2018jrq} of modular weight $w \in {\mathbb N}_0$. A recent proposal for F-IBP bases \cite{Mafra:2019ddf, Mafra:2019xms, Gerken:2019cxz} is formulated in terms of their generating series $\Omega(z,\eta, \tau)$ which, in contrast to the individual $f^{(w)}$, close under tau-derivatives.
These series are imperative as $\tau$-derivatives augment the modular weight. The proposed bases are constructed from chains of the Kronecker-Eisenstein series, defined as 
\beq 
\label{intro1}
{\pmb \Omega}_{12\cdots n}:=\Omega(z_{1,2},\eta_2{+}\eta_3{+}\ldots{+}\eta_n,\tau) \ldots \Omega(z_{n-1,n},\eta_n,\tau) \,,
\eeq 
with $n{-}1$ bookkeeping variables $\eta_i$. Under $\tau$--derivatives, Koba-Nielsen integrals over these chains satisfy KZB-type differential equations. Solutions to these equations shed light on the $\ap$-expansions in string integrals \cite{Enriquez:Emzv, Broedel:2014vla, Mafra:2019ddf, Mafra:2019xms,Gerken:2020yii}. This foundation has led to breakthroughs in the relation \cite{Brown:2017qwo, Brown:2017qwo2, Gerken:2020yii, Gerken:2020xfv,Dorigoni:2022npe} between modular graph forms \cite{DHoker:2015wxz, DHoker:2016mwo} and iterated Eisenstein integrals \cite{Brown:2014pnb, Broedel:2015hia}.
Nevertheless, it is still an unproven conjecture that permutations of the chains \eqref{intro1} form an F-IBP basis -- their established closure under $\partial_{\tau}$ is a necessary but not a sufficient condition.

Rather than presenting a rigorous mathematical proof, we offer compelling evidence for \eqref{intro1} forming an F-IBP basis by decomposing a range of Kronecker-Eisenstein series into the chain form, thereby bolstering the credibility of the conjectural basis. In a companion paper \cite{Rodriguez:2023qir}, Rodriguez, Schlotterer and the author have made notable strides in advancing the IBP methodology for one-loop string integrals of the Koba-Nielsen type. Specifically, we have transformed cyclic products of the Kronecker-Eisenstein series denoted as 
\beq
\label{intro2}
{\pmb C}_{(12\cdots m)}(\xi):= \Omega(z_{1,2},\eta_2{+}\eta_3{+}\ldots{+}\eta_m+\xi,\tau) \ldots \Omega(z_{m-1,m},\eta_m+\xi,\tau) \Omega(z_{m,1},\xi,\tau) 
\,,
\eeq 
and their coefficients $f^{(w)}(z_i{-}z_j,\tau)$ into conjectural bases of one-loop string integrals derived from Kronecker-Eisenstein products of chain topology \cite{Mafra:2019ddf,Mafra:2019xms, Gerken:2019cxz}. This effort has not only validated the chain bases referenced but also yielded explicit formulae for the basis decompositions of one or two cycles of Kronecker-Eisenstein series regardless of their respective lengths.

In this paper, we aim to broaden the recursive approach from the previous work to encompass any number of Kronecker-Eisenstein cycles,
pinpointing the combinatorial structure underlying their integration-by-parts reduction. More explicitly, 
 let us refer to all cycles as \(W_1, W_2, \ldots, W_r\) and the remaining puncture set as \(R\). We will examine an open-string integrand or chiral sector of a closed-string integrand described by 
\ba\label{cycleproduct}
K_n^{(r+)}=
\CC_{W_1} (\xi_1) \CC_{W_2} (\xi_2) \ldots \CC_{W_r}(\xi_r)\,,\,\,{\rm where}~ W_1\sqcup W_2\sqcup \ldots W_r\sqcup R=\{1,2,\ldots, n\} \,.
\ea
We will use the single-cycle formulae derived in \cite{Rodriguez:2023qir} recursively to decompose \eqref{cycleproduct} to chain basis at the cost of introducing some total Koba-Nielsen derivative terms.

We will also delve into more intricate configurations of the Kronecker-Eisenstein series and coefficients beyond just cyclic products, introducing representations through tadpoles, multibranchs, and even interconnected multiloop graphs. Additionally, we will be offering a {\tt Mathematica} rendition of our principal formulae.
The methodologies we adopt serve as practical tools to streamline genus-one correlators and simplify $\ap$-expansions of genus-one integrals, thereby aiding computations within specific string theories \cite{Mafra:2019ddf, Mafra:2019xms, Gerken:2020yii} and ultimately shedding light on the physical implications of the basis coefficients.
 Building upon tree-level computations \cite{Huang:2016tag, Schlotterer:2016cxa, He:2018pol, He:2019drm}, our explicit basis breakdowns might pave the way for a deeper understanding, possibly connecting expansion coefficients
 with a generalized notion of intersection numbers and making contact with the twisted-(co)homology setting of \cite{Bhardwaj:2023vvm}.

Originating from conventional string theories with infinite spectra, our findings are applicable to ambitwistor strings \cite{Mason:2013sva, Berkovits:2013xba} and chiral strings \cite{Hohm:2013jaa, Huang:2016bdd}. Integration-by-parts techniques for moduli-space integrands transition smoothly between these string theories, as highlighted in multiple studies \cite{Gomez:2013wza, He:2017spx, He:2018pol, He:2019drm, Kalyanapuram:2021xow}, and may even involve a $\ap \rightarrow \infty$ limit. These results could illuminate massive loop amplitudes in both conventional and chiral string theories, reminiscent of tree-level work in \cite{Guillen:2021mwp}. Within the chiral splitting framework \cite{DHoker:1988pdl, DHoker:1989cxq}, introducing loop momenta simplifies closed-string loop amplitudes. Yet, F-IBP reductions of chiral amplitudes present challenges beyond the standard doubly-periodic $f^{(w)}(z_{i,j},\tau)$-integrands. We will address the impact of certain derivatives in chiral amplitudes leading to boundary terms in the chiral-splitting context of $(n{-1})!$ genus-one bases.

The present work is organized as follows: We review the genus-one string integrand and the single-cycle formula in 
\cref{section:two}. After introducing compact notations including open cycles and fusions in \cref{sec333}, we demonstrate how to break a product of two Kronecker-Eisenstein cycles in the presence of additional punctures in \cref{sec444} and a product of three cycles with or without additional punctures in \cref{sec555}. Then, after introducing the notion of labeled forests to capture increasingly complicated terms free of cycles in the procedure of basis decomposition in \cref{sec666}, we propose the general formula to break the product of an arbitrary number of both doubly periodic cycles and meromorphic Kronecker-Eisenstein cycles in \cref{sec777}. Practical applications of these formulae are presented in a {\tt Mathematica} code and we explain how to use the code in \cref{seccode}. Section \ref{multisec} delves into more complex scenarios including a product of multibranchs and connected multiloop graphs. Finally, our conclusions and future perspectives are discussed in  \cref{secdis}. 
More detailed explanations of the concise notations used in sections \ref{section:two} and \ref{seccode} can be found in \cref{appendixno}.

\section{Review of string integrals and single-cycle formulae}
\label{section:two}

In this companion paper to \cite{Rodriguez:2023qir}, we provide a succinct overview of the genus-one string integrands and present the closed-form formulae for decomposing a cycle product of the Kronecker-Eisenstein series. While our goal is to ensure this paper is self-contained, we direct readers to \cite{Rodriguez:2023qir} and the associated references for a more in-depth exploration.

\subsection{Kronecker-Eisenstein series, its doubly-periodic completion, and their products}
\label{sec:2.2}

The computation of one-loop string amplitudes relies on moduli-space integrals across correlation functions for specific worldsheet fields containing external-state data. The Kronecker-Eisenstein series \cite{Kronecker} informs the entire dependence of these genus-one correlators on punctures $z \in \mathbb C$ and the modular parameter $\tau \in \mathbb C$ with $\Im \tau>0$,
\begin{equation}
F(z,\eta,\tau) := \frac{ \theta_1'(0,\tau) \theta_1(z{+}\eta,\tau) }{\theta_1(z,\tau) \theta_1(\eta,\tau)}
\label{1.2}\,.
\end{equation}
Here, the standard odd Jacobi theta function is defined with $q := \exp (2 \pi i \tau)$ as
\ba
\theta_{1}(z, \tau) := 2 q^{1 / 8} \sin (\pi z) \prod_{n=1}^{\infty}\left(1-q^{n}\right)\left(1-q^{n} e^{2 \pi i z}\right)\left(1-q^{n} e^{-2 \pi i z}\right)\,.
\ea
In the context of the non-holomorphic admixture detailed in the exponent from \cite{BrownLev},
\begin{equation}
\Omega(z,\eta,\tau) = \exp\Big( 
2\pi i \eta \, \frac{ \Im z}{\Im \tau} \Big)
F(z,\eta,\tau),
\label{1.1}
\end{equation}
we achieve a doubly-periodic refinement of the meromorphic Kronecker-Eisenstein series as seen in (\ref{1.2}). This refinement satisfies the conditions 
$ \Omega(z,\eta,\tau) =\Omega(z{+}1,\eta,\tau) =\Omega(z{+}\tau,\eta,\tau) $.

\subsubsection{Properties}

The Kronecker-Eisenstein series and its doubly periodic completion both satisfy the antisymmetry property expressed by 
\beq
F(-z,-\eta,\tau) = - F(z,\eta,\tau) 
\,, \quad
\Omega(-z,-\eta,\tau) = - \Omega(z,\eta,\tau)\,.
\label{antisyprop}
\eeq
Additionally, they satisfy the Fay identities shown below,
\beq \label{fayid}
F(z_1,\eta_1,\tau)F(z_2,\eta_2,\tau) =
F(z_1,\eta_1{+}\eta_2,\tau) F(z_2{-}z_1,\eta_2,\tau)
+F(z_2,\eta_1{+}\eta_2,\tau) F(z_1{-}z_2,\eta_1,\tau)\,.
\eeq
The above identities remain consistent when substituting $F(z,\eta,\tau)$ with $\Omega(z,\eta,\tau)$.

\subsubsection{Coefficients and concise notations}

With the Laurent expansions using the bookkeeping variables $\eta \in \mathbb C$, we can define Kronecker-Eisenstein coefficients $g^{(w)},f^{(w)}$ for $w \in \mathbb N_0$. Specifically,
\begin{equation}
F(z,\eta,\tau) =: \sum_{w=0}^{\infty} \eta^{w-1} g^{(w)}(z,\tau)\,,
\qquad
\Omega(z,\eta,\tau) = :\sum_{w=0}^{\infty} \eta^{w-1} f^{(w)}(z,\tau),
\label{1.1b}
\end{equation}
with the notable points that $g^{(0)}(z,\tau)=f^{(0)}(z,\tau)=1$, $g^{(1)}(z,\tau) = \partial_z \log \theta_1(z,\tau)$, and $ f^{(1)}(z,\tau) = g^{(1)}(z,\tau)+ 2\pi i \frac{ \Im z}{\Im \tau}$.

Given that the primary results of this work  focus on configuration-space integrals over multiple punctures, namely \(z_1, z_2, \ldots\), we introduce a concise notation for ease of reference. Using 
$
\partial_j := \frac{ \partial}{\partial z_j}
$,
we define 
\beq
g_{i j}^{(w)} := g^{(w)}(z_{i}{-}z_j,\tau)\,,
\qquad
f_{i j}^{(w)} := f^{(w)}(z_{i}{-}z_j,\tau)\,.
\eeq
Likewise, we present
\beq
F_{i j}(\eta) := F(z_{i}{-}z_j,\eta,\tau)\,,
\qquad
\Omega_{i j}(\eta) := \Omega(z_{i}{-}z_j,\eta,\tau).
\eeq

\subsubsection{Chain and cycle products}

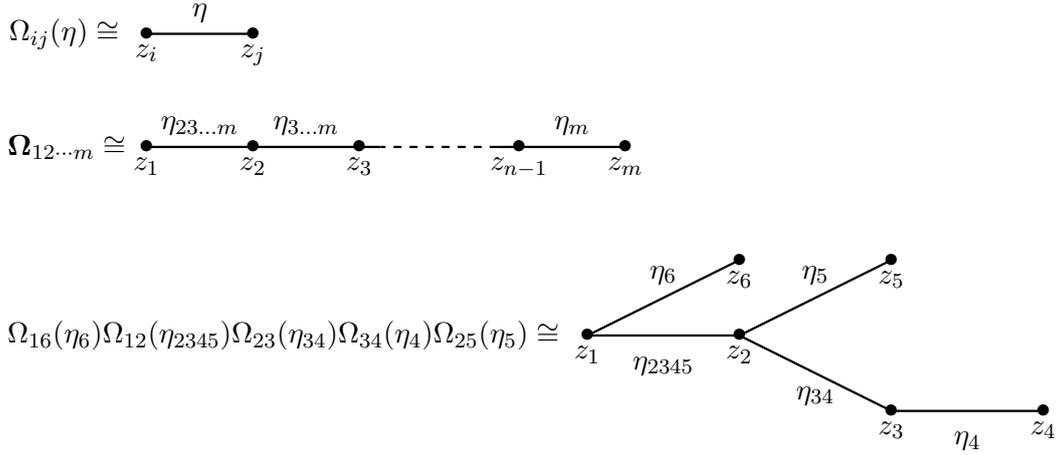
\begin{figure}[h]
  \centering
\begin{tikzpicture}
\begin{scope}[xshift=0cm, yshift=0cm,line width = 0.3mm, scale=0.7]
\draw(-1.55,0)node{$\Omega_{ij}(\eta) \cong$};
\draw (0,0) -- (2,0);  
\draw (0,0)node{$\bullet$}node[below]{$z_i$};
\draw (2,0)node{$\bullet$}node[below]{$z_j$};
\draw(1,0.4)node{$\eta$};
\end{scope}
\begin{scope}[xshift=-5.6cm, yshift=-1.5cm,line width = 0.3mm, scale=0.7]
%
\draw(6.5,0)node{${\bm \Omega}_{12\cdots m} \cong$};
\draw(8,0) -- (12.4,0);
\draw(17,0) -- (14.6,0);
\draw[dashed](14.6,0) -- (12.4,0);
\draw(8,0)node{$\bullet$} node[below]{$z_1$};
\draw(10,0)node{$\bullet$} node[below]{$z_2$};
\draw(12,0)node{$\bullet$} node[below]{$z_3$};
\draw(15,0)node{$\bullet$} node[below]{$z_{n-1}$};
\draw(17,0)node{$\bullet$} node[below]{$z_m$};
\draw(9,0.4)node{$\eta_{23\ldots m}$};
\draw(11,0.4)node{$\eta_{3\ldots m}$};
\draw(16,0.4)node{$\eta_{m}$};
\end{scope}
\begin{scope}[xshift=-2.2cm, yshift=-4cm, line width = 0.3mm, scale=1]
%
\draw(4,0)node{$ {\Omega}_{16}(\eta_6)
{\Omega}_{12}(\eta_{2345})
{\Omega}_{23}(\eta_{34})
{\Omega}_{34}(\eta_4)
{\Omega}_{25}(\eta_5)
 \cong$};
\draw(8,0) -- (10,0);
\draw(8,0) -- (10,1);
\draw(10,0) -- (12,-1);
\draw(10,0) -- (12,1);
\draw(12,-1) -- (14,-1);
\draw(8,0)node{$\bullet$} node[below]{$z_1$};
\draw(10,0)node{$\bullet$} node[below]{$z_2$};
\draw(10,1)node{$\bullet$} node[below]{$z_6$};
\draw(12,1)node{$\bullet$} node[below]{$z_{5}$};
\draw(12,-1)node{$\bullet$} node[below]{$z_3$};
\draw(14,-1)node{$\bullet$} node[below]{$z_4$};
\draw(9,0.8)node{$\eta_{6}$};
\draw(9,-0.4)node{$\eta_{2345}$};
\draw(11,0.8)node{$\eta_{5}$};
\draw(11,-0.8)node{$\eta_{34}$};
\draw(13,-1.4)node{$\eta_{4}$};
\end{scope}
\end {tikzpicture}
 \caption{Graphical representation of Kronecker-Eisenstein series $\Omega_{ij}(\eta) = {\Omega}({z}_{i}{-}{z}_j,{\eta},{\tau})$, their chain and tree  products }
  \label{fig:graph2}
\end{figure}

As schematically shown by the second graph in \cref{fig:graph2},
we define a specific chain product of the doubly-periodic Kronecker-Eisenstein series as 
\begin{equation}
\label{deflongomega2}
{\bm\Omega}_{\alpha (1) \a (2) \cdots \a (m) } := \delta\left( \sum_{i=1}^m \eta_{\a(i)} \right) \prod_{i=1}^{m-1}  
\Omega_{\alpha(i)\, \alpha({i+1})} ( \eta_{\alpha({i+1})\, \cdots\, \alpha({m}) }) \,,
\end{equation} 
where \( \eta_{ij\cdots k} = \eta_{i} + \eta_{j} + \ldots + \eta_{k} \).
The set \(\{\alpha (1), \a(2),\cdots, \a(m)\} \) with entirely unique elements can represent any subset of \(\{1,2,\cdots, n\}\) containing at least two elements. The delta constraints reveal that any \(\eta_{\alpha(i)}\), like \(\eta_{\alpha(1)}\), can be represented using all other terms. In particular, when $m=n$, $\a(i)=i$ and expressing all $\eta_1=-\eta_{23\cdots n}$, it reduces to \eqref{intro1}. Using the Fay identities \eqref{fayid} for pairs of Kronecker-Eisenstein series, we derive the chain identities
\begin{equation}
\label{fayprac}
{\bm\Omega}_{\alpha,i,\beta} = (-1)^{|\alpha|} {\bm\Omega}_{i,\alpha^{\rm T}} {\bm\Omega}_{i,\beta} = (-1)^{|\alpha|} \sum_{\rho\in \alpha^{\rm T}\shuffle \beta} {\bm\Omega}_{i,\rho} \,.\end{equation} 
This resembles the Kleiss-Kuijf relations of gauge-theory tree amplitudes referenced in \cite{Kleiss:1988ne}. Only \( (m{-}1)! \) of the \( m! \) permutations of \( {\bm\Omega}_{\alpha (1) \a(2)\cdots \a(m) } \) remain independent as discussed in \cite{Mafra:2019xms, Gerken:2019cxz, Broedel:2020tmd}. The shuffle \( \alpha\shuffle \beta \) for ordered sets \( \alpha,\beta \) in the summation of (\ref{fayprac}) captures all permutations of the combined set \( \alpha \beta \) that maintain the original order of \( \alpha \) and \( \beta \) elements.

Moreover, products of doubly-periodic Kronecker-Eisenstein series, which exhibit tree topologies similar to the one shown in the last graph of \cref{fig:graph2}, can be expanded into chains \( {\bm\Omega}_{\alpha (1) \alpha(2) \cdots \alpha(m) } \) with a fixed $\alpha(1)$. This expansion uses the chain identities from \eqref{fayprac} iteratively and may require a redefinition of the bookkeeping variables.

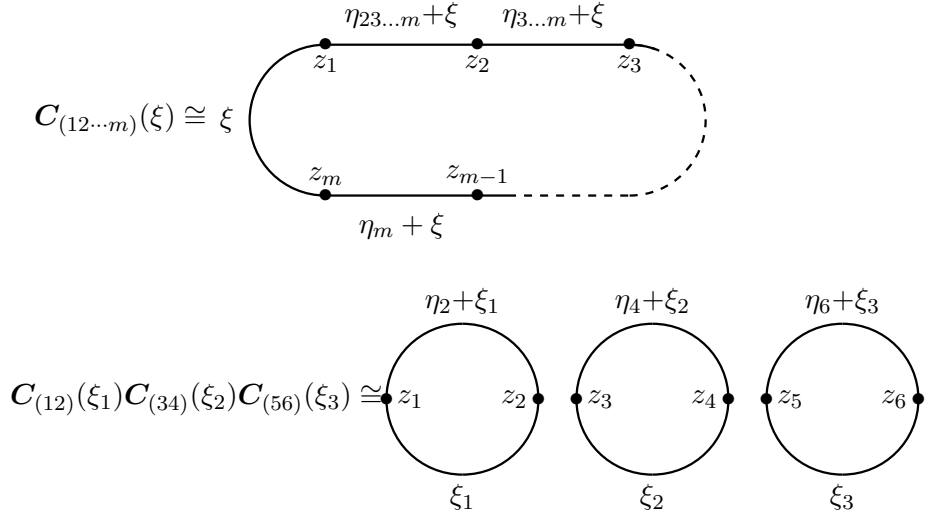
\begin{figure}[h]
  \centering
\begin{tikzpicture}
\begin{scope}[xshift=-6cm, yshift=-6.3cm, line width = 0.3mm, scale=1]
%
\draw(5.3,-1)node{${\CC}_{(12\cdots m)} (\xi)\cong$};
\draw(6.7,-1)node{$\xi$};
\draw(8,0) -- (12,0);
\draw(8,-2) -- (10.4,-2);
\draw (8,0) arc (90:270:1);
\draw (12,0) arc (90:70:1); 
\draw (12,-2)[dashed] arc (-90:90:1); 
\draw[dashed](10.4,-2) -- (12,-2);
\draw(8,0)node{$\bullet$} node[below]{$z_1$};
\draw(10,0)node{$\bullet$} node[below]{$z_2$};
\draw(12,0)node{$\bullet$} node[below]{$z_3$};
\draw(8,-2)node{$\bullet$} node[above]{$z_m$};
\draw(10,-2)node{$\bullet$} node[above]{$z_{m-1}$};
\draw(9,0.4)node{$\eta_{23\ldots m}{+}\xi$};
\draw(11,0.4)node{$\eta_{3\ldots m}{+}\xi$};
\draw(9,-2.4)node{$\eta_{m}+\xi$};
\end{scope}
\begin{scope}[xshift=-.2cm, yshift=-10cm, line width = 0.3mm, scale=1]
%
\draw(0.5,-1)node{${\CC}_{(12)} (\xi_1){\CC}_{(34)} (\xi_2){\CC}_{(56)} (\xi_3)\cong$};
\draw (4,0) arc (90:270:1);
\draw (4,0) arc (90:-90:1); 
\draw(4,.3)node{$\eta_2{+}\xi_1$};
\draw(4,-2.3)node{$\xi_1$};
\draw(3,-1)node{$\bullet$} node[right]{$z_1$};
\draw(5,-1)node{$\bullet$} node[left]{$z_2$};
\draw (6.5,0) arc (90:270:1);
\draw (6.5,0) arc (90:-90:1); 
\draw(6.5,.3)node{$\eta_4{+}\xi_2$};
\draw(6.5,-2.3)node{$\xi_2$};
\draw(5.5,-1)node{$\bullet$} node[right]{$z_3$};
\draw(7.5,-1)node{$\bullet$} node[left]{$z_4$};
\draw (9,0) arc (90:270:1);
\draw (9,0) arc (90:-90:1); 
\draw(9,.3)node{$\eta_6{+}\xi_3$};
\draw(9,-2.3)node{$\xi_3$};
\draw(8,-1)node{$\bullet$} node[right]{$z_5$};
\draw(10,-1)node{$\bullet$} node[left]{$z_6$};
\end{scope}
\end {tikzpicture}

  \caption{Graphical representation of a cyclic product of Kronecker-Eisenstein series ${\CC}_{(12\cdots m)} (\xi)$ and their product}
  \label{fig:graph3}
\end{figure}

The cyclic product of the doubly-periodic Kronecker-Eisenstein series, as depicted in the first graph of \cref{fig:graph3}, contrasts with a chain product as defined in \eqref{deflongomega2}. Specifically, we introduce a cyclic product as follows,
\begin{equation}\label{defc}
\CC_{(12\cdots m)}(\xi):=\delta \bigg( \sum_{i=1}^m \eta_{i} \bigg) \Omega_{12}(\eta_{23\cdots m}{+}\xi)\Omega_{23}(\eta_{3\cdots m}{+}\xi) 
\cdots 
\Omega_{m-1,m}(\eta_{ m}{+}\xi)
\Omega_{m,1}(\xi) \,,
\end{equation}
applicable for general multiplicities within the range $2 \leq m \leq n$. Notably, when we set $\eta_1 = -\eta_{23\cdots m}$, this cyclic product simplifies to the form described in \eqref{intro2}.

In this case, the cycle topology of $\CC_{(12\cdots m)}(\xi)$ precludes a straightforward algebraic expansion into chains as defined in \eqref{deflongomega2} using  Fay identities \eqref{fayid} or the identities in \eqref{fayprac} in practice. Instead, this expansion necessitates the application of IBP, which we will review in the next subsection. Our focus in this paper is on the analysis of products of doubly-periodic Kronecker-Eisenstein cycles, as represented in \eqref{cycleproduct} and illustrated by the last graph in \cref{fig:graph3}.

Building on our definitions \eqref{deflongomega2} and \eqref{defc} for the products of doubly-periodic Kronecker-Eisenstein series, we introduce the notation for chains and cycles of their meromorphic counterparts, $F_{ij}(\eta) = F(z_i - z_j, \eta, \tau)$, as follows:
\begin{align}
&
\label{deflongF}
{\bm F}_{\alpha (1) \a(2)\cdots \a(m) }:= \delta\bigg( \sum_{i=1}^m \eta_{\a(i)} \bigg) \prod_{i=1}^{m-1}  
F_{\alpha(i)\, \alpha({i+1})} ( \eta_{\alpha({i+1})\, \cdots\, \alpha({m} ) })    \,,
\\
\label{defcf}
&\CCF_{(12\cdots m)}(\xi):= \delta\bigg( \sum_{i=1}^m \eta_{i} \bigg)  F_{12}(\eta_{23\cdots m}{+}\xi)F_{23}(\eta_{3\cdots m}{+}\xi) 
\cdots 
F_{m-1,m}(\eta_{ m}{+}\xi)
F_{m 1}(\xi)\,.
\end{align}
Given the universality of the Fay identity (\ref{fayid}) for both $F_{ij}(\eta)$ and $\Omega_{ij}(\eta)$, the relations in (\ref{fayprac}) for doubly-periodic chains with ordered sets $\alpha,\beta$ naturally extend to the $F$-chain in (\ref{deflongF}), with a simple substitution of ${\bm F}$ for ${\bm \Omega}$. Furthermore, of the $m!$ permutations of ${\bm F}_{\alpha(1) \alpha(2) \cdots \alpha(m)}$, only $(m-1)!$ are algebraically independent. For example, one can choose the chains with $\alpha(1) = 1$ as the independent set. In the case of meromorphic Kronecker-Eisenstein cycles $\CCF_{(12\cdots m)}(\xi)$ and their products,
\ba\label{cycleproductF}
\tilde K_n^{(r+)}=
\tilde\CC_{W_1} (\xi_1) \tilde\CC_{W_2} (\xi_2) \ldots \tilde\CC_{W_r}(\xi_r)\,,{\rm where}~  W_1\sqcup W_2\sqcup \ldots W_r\sqcup R=\{1,2,\ldots, n\} \,.
\ea
decomposition requires the application of F-IBP.

\subsection{Sting integrals, IBP relations, and single-cycle formulae}

The \( n \)-point genus-one string amplitudes are derived from worldsheets characterized by cylinder and Moebius-strip topologies for open strings, and torus topologies for closed strings. Each of these configurations is described with a modular parameter \( \tau \) and features punctures located on the boundary. The amplitudes are mathematically represented as follows,
\begin{align}\label{openamp}
& \mathcal{A}_{n}= \int_{\rm op} d {\bm \mu}_n^{\rm op}
\, \mathcal{I}_{n}^{\text {op}}(z_i,\tau,k_i) \,
K_{n}^{\text {op}}
(f^{(w)}, \tau, k_i, \epsilon_i,\cdots)\,,
\\ 
& \label{closedamp}
\mathcal{M}_{n}= \int_{\rm cl} d {\bm \mu}_n^{\rm cl}\,
\mathcal{I}_{n}^{\rm cl}(z_i,\tau,k_i) \,
K_{n}^{\text {cl}}
(f^{(w)},{\bar f}^{(w)}, \tau,k_i, \epsilon_i, {\bar \e}_i,\cdots)\,.
\end{align}

\subsubsection{Integration domains, measures and Koba-Nielsen factors}

The specific details regarding the integration domains and the measures $d {\bm \mu}_n^{\rm \bullet}$, along with the Koba-Nielsen factors $\mathcal{I}_{n}^{\rm \bullet}(z_i,\tau,k_i)$ for both open and closed strings, are comprehensively outlined in \cref{appendixno}. Despite the distinctions between open and closed strings, their IBP relations can be expressed in a standardized manner. Primarily, the integration of total derivatives acting on doubly periodic functions yields zero, as shown below
\ba \label{total}
\int_\bullet d {\rm \mu}_n^\bullet \partial_i \left( \mathcal{I}^{\bullet}_{n} \varphi \right) =0\,, \qquad \forall\, \,{\text{doubly periodic}}~ \varphi(z_j, \tau)\,,
\ea 
where \(\partial_i\) represents a real analysis derivative for open strings (\(\bullet \to {\rm op}\)) and transitions to a holomorphic derivative for closed strings (\(\bullet \to {\rm cl}\)), with \(\bar\partial_{ i}\) denoting the conjugate derivative.

Furthermore, the derivatives of the Koba-Nielsen factors for both open and closed strings can be encapsulated in a unified expression:
\ba\label{baseafter}
\partial_i \mathcal{I}^{\bullet}_{n}=-\bigg(\sum_{j \neq i}^{n} x_{i,j} \bigg) \mathcal{I}^{\bullet}_{n}\,,\qquad x_{i,j}:=s_{ij} f^{(1)}_{ij}\, .
\ea

We define the dimensionless Mandelstam invariants used throughout our analysis as follows:
\begin{equation}
s_{i j}=- k_i\cdot k_j\, , \quad s_{i_1 i_2\ldots i_r}=- \sum_{1\leq p<q \leq r} k_{i_p}\cdot k_{i_q} \, ,
\quad \alpha' = \left\{ \begin{array}{cl} 1/2 &\ {\rm open} \ {\rm strings}\, , \\
2 &\ {\rm closed} \ {\rm strings} \ .
\end{array} \right.
\end{equation} 

Combining \eqref{total} and \eqref{baseafter}, we derive an IBP relation:
\ba 
\label{IBPre}
\int_\bullet d {\rm \mu}_n^\bullet  \mathcal{I}^{\bullet}_{n} \underbrace{ \left(\partial_i - \sum_{j \neq i}^{n} x_{i,j} \right) }_{:= \nabla_i} \varphi  =0
\,, \qquad \forall\, {\text{doubly periodic}}~ \varphi(z_j, \tau)\,,
\ea 
where we defined operator \( \nabla_i \), termed the Koba-Nielsen derivative. 
In the context of our analysis, this IBP relation, now compactly denoted as \( \nabla_i\varphi \overset{\rm IBP}{=} 0 \), emphasizes the transformational properties of the integrands. Note that the operators \( \nabla_{i} \) and \( \nabla_{j} \) exhibit commutativity.

\subsubsection{Integrands and basis}

The integrands  $K_{n}^{\bullet}$  vary across different theories. Typically, these are polynomials involving doubly-periodic Kronecker-Eisenstein coefficients, $f^{(w)}_{ij}$, as defined in \eqref{1.1b}, and, for closed strings, their complex conjugates ${\bar f}^{(w)}_{ij}$ as well. The polynomial coefficients incorporate the modular parameter $\tau$ and various physical parameters such as polarizations $\epsilon_i$ and momenta$k_i$, yet they remain independent of the puncture locations $z_i$.

Conjectures, as proposed in the studies by Mafra and Schlotterer \cite{Mafra:2019ddf, Mafra:2019xms}, suggest a structured formulation for the string theory integrands. Specifically, for any $n$-point open-string one-loop integrand related to $K_n^{\rm op}$, as seen in (\ref{openamp}), it can be aligned with an $(n{-}1)!$-basis of generating functions. An example of such a basis is ${\bm \Omega}_{1,\alpha(2),\ldots, \alpha(n)}$, where $\alpha$ varies over the symmetric group $S_{n-1}$. For the closed-string counterpart, the one-loop integrand $K_n^{\rm cl}$ is conjectured to be representable as a linear combination of products like ${\bm \Omega}_{1\alpha(2)\alpha(3)\ldots \alpha(n)} \bar{\bm \Omega}_{1\beta(2)\beta(3)\ldots \beta(n)}$, with $\alpha, \beta \in S_{n-1}$, where $\bar{\bm \Omega}$ denotes the complex conjugate of ${\bm \Omega}$.

\subsubsection{Single-cycle formulae}

In the case of products of $\Omega$ without cycles, such as those represented by the last graph in \cref{fig:graph2}, these can be algebraically expanded into basis elements using  Fay identities \eqref{fayid} or \eqref{fayprac} in practice. In the previous companion paper \cite{Rodriguez:2023qir}, we found a general formula to decompose an arbitrary $\Omega$-cycle into basis elements:
\begin{align}\label{rhorhog}
( 1{+}s_{12\cdots m}) \CC_{(12\cdots m)}(\xi)
=\,\,& \MM_{12\cdots m} (\xi)-
  \sum_{\substack{
b   =1 \\ b\neq a}
}^{m} \sum_{\rho \in  A \shuffle  B^{\rm T} \atop{(a, A, b, B)=\mathbb I_m}} 
\! \! (-1)^{|B|} \Bigg(
 \sum_{i=m+1}^{n}  x_{b,i} 
 +\nabla_b   
\Bigg){\bm \Omega}_{a,\rho,b} 
 \,.
 \end{align}
In this equation, \(a\) represents any element from \( \{1,2,\ldots, m\}\). On the right-hand side,  \(A\) and \(B\) are determined by the relationship \((a, A, b, B )=(1,2, \cdots ,m)\), subjected to cyclic transformations. The chains \( {\bm \Omega}_{a,\rho,b} \) are then aligned with the foundational basis \( {\bm \Omega}_{1\ldots} \), following the guidance of \eqref{fayprac}. The term \(\MM_{12\cdots m}(\xi)\) pertains to the proposed \((m{-}1)!\) basis for \(m\)-point chains, delineated as \({\bm \Omega}_{1,\alpha(2),\a(3),\ldots,\a(m)}\) where $\alpha \in S_{m-1}$. The complete form of \(\MM_{12\cdots m}(\xi)\) is detailed in \eqref{rhorhoMM} in the appendix.

Further, our previous findings demonstrated the recursive application of single-cycle formulae to dissect products of two $\Omega$-cycles. This paper aims to broaden this methodology to encompass products involving an arbitrary number of cycles as indicated in \eqref{cycleproduct}.

\subsubsection{Chiral splitting \label{sec:revCS}}

Chiral splitting, as detailed in the works \cite{DHoker:1988pdl, DHoker:1989cxq}, serves as a foundational method for deriving open and closed string amplitudes from the same chiral function, denoted as ${\cal K}_n(\ell)$. This function encapsulates the kinematic data of the system. The amplitudes for open and closed strings are computed through the following integrals,
\ba\label{openampF}
&\mathcal{A}_{n}= \frac{1}  {(2 \pi i)^{D}} \int_{\rm op}
d {\bm \mu}^{\rm op}_{n} 
\int_{\mathbb R^D} d^{D} \ell\left|\mathcal{J}_{n}(\ell)\right|\mathcal{K}_{n}(\ell,g^{(w)}, \tau,k_i, \epsilon_i, \cdots) \,,
\\
&\label{closedampF}
\mathcal{M}_{n}= \frac{1}  {(2 \pi i)^{D}} \int_{\rm cl}
d {\bm \mu}^{\rm cl}_{n}  
\int_{\mathbb R^D} d^{D} \ell\left|\mathcal{J}_{n}(\ell)\right|^{2} 
\mathcal{K}_{n}(\ell) 
\tilde{\mathcal{K}}_{n}(-\ell, {\bar g}^{(w)}, \tau,k_i, {\bar \e}_i,\cdots)
\,,
\ea
where $D$ represents the spacetime dimension. Here, the chiral Koba-Nielsen factor, $\mathcal{J}_{n}(\ell)$  reflects the loop momentum $\ell$ and shows a meromorphic dependency on both the puncture locations $z_i$ and the modular parameter $\tau$. The specific form of $\mathcal{J}_{n}(\ell)$ is elaborated in \eqref{chiKN} in the appendix. Furthermore, the $z$-dependence within $\mathcal{K}_{n}(\ell)$ is dictated by the $g^{(w)}_{ij}$ functions, which are derivable from the meromorphic Kronecker-Eisenstein series as outlined in \eqref{1.1b}.

The derivatives of the chiral Koba-Nielsen factor $\mathcal{J}_{n}(\ell)$ with respect to the worldsheet positions $z_i$ now incorporate the loop momentum,
\ba\label{basebefore}
\partial_i \mathcal{J}_{n}(\ell)=\bigg( \ell \!\cdot\! k_{i}  -\sum_{j\neq i}^{n}  {\tilde  x}_{i,j}  \bigg) 
\mathcal{J}_{n}(\ell)\,,\qquad {\rm with~}  {\tilde  x}_{i,j} :=s_{i j} g_{i j}^{(1)} ~ {\rm for~} j\neq i \,.
\ea
Similar to \eqref{IBPre}, we define the operators ${\tilde\nabla}_i$  to encapsulate the Koba-Nielsen derivatives,
\ba
\label{defnablaJ}
{\tilde\nabla}_i {\tilde \varphi}:= \partial_i {\tilde \varphi} + \bigg ( \ell \!\cdot\! k_{i}-\sum_{j \neq i}^{n} {\tilde x}_{i j} \bigg) {\tilde \varphi}=\frac{1}{\mathcal{J}_{n}}\partial_i (\tilde \varphi\,  \mathcal{J}_{n} )\,,
\ea 
applicable to any meromorphic function $\tilde \varphi = \tilde \varphi (z_i,\tau)$ relevant to the chiral integrands $\mathcal{K}_{n}(\ell)$. Note that the operators $\tilde\nabla_{i}$ and $\tilde\nabla_{j}$ commute.

Following the logic in \eqref{rhorhog}, substitution rules can be applied as
\begin{equation}
\label{subsrule}
{\bm \Omega}_{\alpha(1) \alpha(2)\ldots } \to {\bm F}_{\alpha(1) \alpha(2)\ldots }, \quad x_{ij} \to \tilde x_{ij}, \quad  \MM_{12\cdots m} (\xi)  \to  \MMF_{12\cdots m} (\xi), \quad \nabla_b \to {\tilde\nabla}_b - \ell \cdot k_b,
\end{equation}
to deconstruct the meromorphic cycles \eqref{defcf} as shown below,
\begin{align}\label{rhorhogf}
( 1{+}s_{12\cdots m}) \CCF_{(12\cdots m)}(\xi)&
=\,\, \MMF_{12\cdots m} (\xi)
\\
&
\nonumber
-
  \sum_{\substack{
b   =1 \\ b\neq a}
}^{m} \sum_{\substack{\rho \in  A \shuffle  B^{\rm T}
\\
(a, A, b, B )=
 \mathbb I_m
 }} 
 (-1)^{|B|} \bigg({-}\ell \!\cdot\! k_b+ 
 \sum_{i=m+1}^{n}  {\tilde x}_{b,i} 
 +{\tilde\nabla}_b   
\bigg){\bm F}_{a,\rho,b} 
 \,,
 \end{align}
where $ \MMF_{12\cdots m} (\xi)$ is a linear combination of $m$-point basis, ${\bm F}_{1,\a(2),\cdots,\a(m)}$ with $\a \in S_{m-1}$. See \eqref{MMFeq}.

We aim to recursively apply \eqref{rhorhogf} to decompose products of meromorphic cycles \eqref{cycleproductF} into a chain basis \({\bm F}_{1,\alpha(2),\a(3),\ldots,\a(n)}\) with $\alpha \in S_{n-1}$, introducing some total Koba-Nielsen derivative terms in the process. These derivative terms, unlike in previous doubly-periodic cases \eqref{total}, cannot be neglected but can be simplified.

 We will explore that many techniques applicable in doubly-periodic scenarios are equally viable for chiral splitting cases. Sections \ref{sec333} through \ref{sec666} will primarily focus on the doubly-periodic cases, and towards the end in \cref{sec777}, we will detail the approach for handling an arbitrary number of meromorphic cycles. Additionally, in \cref{multisec}, we will discuss the handling of integrands beyond the scope of meromorphic cycle products.

\section{Open cycles and fusions \label{sec333}}

Given that we will address an arbitrary number of series of double periodic Kronecker-Eisenstein  cycles, simplifying the representation of a single-cycle formula is beneficial. Initially, let us introduce the notion of open cycle \(\OO_{a,i}^W\) defined as
\ba
\OO_{a,i}^W:= - \sum_{\substack{
b   \in W \\ b\neq a}
} \sum_{\substack{\rho \in  A \shuffle  B^{\rm T}
\\ 
 (a, A, b, B )=W }} 
 (-1)^{|B|} 
{\bm \Omega}_{a,\rho,b}    x_{b,i}  \,, \quad {\rm with}~ a\in W, {\rm and}~ i\notin W\,.
\label{defoblock}
\ea 
Here are some examples,
\ba
\OO_{1,i}^{(12)}=&\,\,- \Omega_{12}(\eta_2) x_{2,i}\,,
\qquad\qquad\qquad \OO_{2,i}^{(12)}= \Omega_{12}(\eta_2) x_{1,i}\,,
\nl
\OO_{1,i}^{(123)}= &\,\,- {\bm \Omega}_{123}x_{3,i}+ {\bm \Omega}_{132}x_{2,i} \,,
\qquad 
\OO_{2,i}^{(123)}= -{\bm \Omega}_{231}x_{1,i}+ {\bm \Omega}_{213}x_{3,i} \,,
\nl
\OO_{1,i}^{(1234)}=&\,\, -{\bm \Omega}_{1234}x_{4,i}+( {\bm \Omega}_{1243} + {\bm \Omega}_{1423})x_{3,i}-{\bm \Omega}_{1432}x_{2,i}  \,.
\ea

Note that the terms \(x_{b,i}\) and \(\nabla_b\) in \eqref{rhorhog} exhibit similar characteristics. To formalize this, we introduce an auxiliary puncture \(0\) and set the operator \(x_{b,0}=\nabla_b\). With this introduction, we can extend the definition \eqref{defoblock} to
\ba
\OO_{a,0}^W:= - \sum_{\substack{
b   \in W \\ b\neq a}
} \sum_{\rho \in  A \shuffle  B^{\rm T}} 
 (-1)^{|B|}
 x_{b,0}
{\bm \Omega}_{a,\rho,b}      \,, \qquad {\rm with}~ a\in W. 
\label{defoblock}
\ea 

With the above definitions, the single-cycle formula \eqref{rhorhog} can be compactly represented as 
 \begin{align}\label{stibp}
( 1{+}s_{W}) \CC_{W} (\xi)
=\,\,& \MM_{W} (\xi) 
+
 \sum_{\substack{ i\notin W,\,  i \in \widehat{[ n]}}}  \OO_{a,i}^W\,,  \qquad \forall ~ a\in W.
 \end{align}
where we have defined \(\widehat{[ n]}\) as \(\{0\}\cup [n]  = \{0,1,2,\cdots,n\}\). For streamlined referencing, we will also use \(\widehat R= \{0\}\cup R\).

With a little abuse of notation, we introduce the abbreviation
\be
{\bm \Omega}_{a, A\shuffle B^T,b}:=\sum_{\sigma\in  A\shuffle B^T}{\bm \Omega}_{a, \sigma,b}\,.
\ee
Besides, we use an oriented  dotted line 
\tikz[{dir/.style={decoration={markings, mark=at position \halfway with {\arrow{Latex}}},postaction={decorate}}}]{
\draw[thick,dir, dotted] (1.5,0) node [above] {$b$}-- (2.5,0)  node [above] {$j$};}
and  a wavy line 
\begin{tikzpicture}[baseline={([yshift=-1.5ex]current bounding box.center)},every node/.style={font=\footnotesize,},vertex/.style={inner sep=0,minimum size=3pt,circle,fill},wavy/.style={decorate,decoration={coil,aspect=0, segment length=2.2mm, amplitude=0.5mm}},dir/.style={decoration={markings, mark=at position \halfway with {\arrow{Latex}}},postaction={decorate}}]
\node at (-0,0) [label={above:{${a}$}},vertex] {};
\node at (1.5,0) [label={above:{${b}$}},vertex] {};
\draw[thick,wavy] (0,0) -- (1.5,0);
\end{tikzpicture}
to represent $  -x_{b,j} (-1)^{|B|+1} {\bm \Omega}_{a, A\shuffle B^T,b} $ such that 
\begin{align}
\OO^W_{a,j}=\sum_{b\in W/\{a\}}\begin{tikzpicture}[baseline={([yshift=-.5ex]current bounding box.center)},every node/.style={font=\footnotesize,},vertex/.style={inner sep=0,minimum size=3pt,circle,fill},wavy/.style={decorate,decoration={coil,aspect=0, segment length=2.2mm, amplitude=0.5mm}},dir/.style={decoration={markings, mark=at position \halfway with {\arrow{Latex}}},postaction={decorate}}]
\node at (-0,0) [label={above:{${a}$}},vertex] {};
\node at (1.5,0) [label={above:{${b}$}},vertex] {};
\draw[thick,wavy] (0,0) -- (1.5,0) node[pos=0.5,below=0pt]{$\vphantom{A\shuffle B^T}$};
\draw[thick,dir, dotted] (1.5,0) -- (2.5,0) node[label={above:{$j$}},vertex] {};
\end{tikzpicture}\,.
\end{align}
Then the single-cycle formula \eqref{stibp} can be schematically shown by
\begin{align}
(1+s_W)
\begin{tikzpicture}[baseline={([yshift=0.3 ex]current bounding box.center)},every node/.style={font=\footnotesize,},vertex/.style={inner sep=0,minimum size=3pt,circle,fill},wavy/.style={decorate,decoration={coil,aspect=0, segment length=2.2mm, amplitude=0.5mm}},dir/.style={decoration={markings, mark=at position \Halfway with {\arrow{Latex}}},postaction={decorate}}]
\draw [thick] (0.35,0) ++(-110:0.35) arc (-110:-470:0.35);
\node at (0.4,-.6) {$\CC_{W}$};
\end{tikzpicture}
=\begin{tikzpicture}[baseline={([yshift=0.3ex]current bounding box.center)},every node/.style={font=\footnotesize,},vertex/.style={inner sep=0,minimum size=3pt,circle,fill},wavy/.style={decorate,decoration={coil,aspect=0, segment length=2.2mm, amplitude=0.5mm}},dir/.style={decoration={markings, mark=at position \Halfway with {\arrow{Latex}}},postaction={decorate}}]
\draw [thick] (0.35,0) ++(-110:0.35) arc (-110:-430:0.35);
\node at (0.4,-.6) {$\MM_{W}$};
\end{tikzpicture}
+ 
\sum_{j\in \widehat {[n]}/ W } \sum_{b\in W/\{a\}}\begin{tikzpicture}[baseline={([yshift=-.5ex]current bounding box.center)},every node/.style={font=\footnotesize,},vertex/.style={inner sep=0,minimum size=3pt,circle,fill},wavy/.style={decorate,decoration={coil,aspect=0, segment length=2.2mm, amplitude=0.5mm}},dir/.style={decoration={markings, mark=at position \halfway with {\arrow{Latex}}},postaction={decorate}}]
\node at (-0,0) [label={above:{${a}$}},vertex] {};
\node at (1.5,0) [label={above:{${b}$}},vertex] {};
\draw[thick,wavy] (0,0) -- (1.5,0) node[pos=0.5,below=0pt]{$\vphantom{A\shuffle B^T}$};
\draw[thick,dir, dotted] (1.5,0) -- (2.5,0) node[label={above:{$j$}},vertex] {};
\end{tikzpicture}\,,
\quad  \qquad \forall ~ a\in W \,.
\end{align}
From now on, whenever we see $\OO^W_{a,j}$, we assume $a\in W$. Besides, we define ${\bm \Omega}_{a, A\shuffle B^T,a} =0$ such that we can simply write the  summation $\sum_{b\in W/\{a\}}$ in the above equation as $\sum_{b\in W}$.
\begin{figure}[]
    \centering

    \begin{align}
\non
\sum_{\substack{b_1,j_1 \in W_1\\ j_2,b_2\in W_2}}\begin{tikzpicture}[scale=1.25, baseline={([yshift=-0ex]current bounding box.center)},every node/.style={font=\footnotesize,},vertex/.style={inner sep=0,minimum size=3pt,circle,fill},wavy/.style={decorate,decoration={coil,aspect=0, segment length=2.2mm, amplitude=0.5mm}},dir/.style={decoration={markings, mark=at position \halfway with {\arrow{Latex}}},postaction={decorate}}]
\node at (-0,0) [label={below:{${p_1}$}},vertex] {};
\node at (1,0) [label={below:{${j_1}$}},vertex] {};
\node at (1.6,0) [vertex] {};
\node at (4,0) [vertex] {};
\draw[thick,wavy] (0,0) -- (1,0) -- (1.5,0) node [below] {$b_1$} (2.5,0) -- (4,0) node [below] {$b_2$} ;
\draw[thick,dir,dotted] (1.6,0) -- (2.5,0) node[label={below:{$j_2$}},vertex] {};
\draw[thick,dir,dotted] (4,0) .. controls (3.5,1) and (1.5,1) .. (1,0);
\end{tikzpicture}
= 
\sum_{\substack{a_1,b_1 \in W_1\\ a_2,b_2\in W_2}}\begin{tikzpicture}[scale=1.25,baseline={([yshift=-0ex]current bounding box.center)},every node/.style={font=\footnotesize,},vertex/.style={inner sep=0,minimum size=3pt,circle,fill},wavy/.style={decorate,decoration={coil,aspect=0, segment length=2.2mm, amplitude=0.5mm}},dir/.style={decoration={markings, mark=at position \halfway with {\arrow{Latex}}},postaction={decorate}}]
\node at (-0,0) [label={below:{${a_1}$}},vertex] {};
\node at (1.6,0) [vertex] {};
\node at (4,0) [vertex] {};
\draw[thick,wavy] (0,0) -- (1.5,0) node [below] {$b_1$} (2.5,0) -- (4,0) node [below] {$b_2$} ;
\draw[thick,dir,dotted] (1.6,0) -- (2.5,0) node[label={below:{$a_2$}},vertex] {};
\draw[thick,dir,dotted] (4,0) .. controls (3.5,1) and (.5,1) .. (0,0);
\end{tikzpicture}
\end{align}

    \caption{From tadpoles to isolated cycles}
    \label{fig:1}
\end{figure}

\subsection{Fusion of two cycles}

As we will see, open cycles can not only be used to simplify the single-cycle formula, but also have a very important property. 
Consider a particular combination of open cycles $\sum_{\substack{j_2\in W_2, j_1 \in W_1}} 
\OO^{W_1}_{p_1,j_2} \OO^{W_2}_{j_2,j_1}$, naively it would form lots of tadpoles as shown on the left of \cref{fig:1}. The key observation is that all of such tadpoles can reorganized to produce new isolated cycles without tails planting on them on the support of \eqref{fayprac} as shown on the left of \cref{fig:1} when we average the summation index,
\begin{align}\label{fusionCC}
\sum_{\substack{j_2\in W_2\\ j_1 \in W_1}} 
\OO^{W_1}_{p_1,j_2} \OO^{W_2}_{j_2,j_1}
=&
\!\!\!\!\!\!\!\!\!
\sum_{\substack{b_1\in W_1 \\ (p_1, A_1, b_1,B_1) = (W_1)}}
\!\!\!\!\!\!\!\!\!\!\!\!\!\!\!\!
(-1)^{|B_1|}{\bm \Omega}_{p_1, A_1\shuffle B_1^T,b_1}
\!\!\!\!\!\!\!\!\!\!\!\!\!\!\!\!
\sum_{\substack{j_2,b_2\in W_2 \\ (j_2, A_2, b_2,B_2) = (W_2)}} 
\!\!\!\!\!\!\!\!\!\!\!\!\!\!\!\!
(-1)^{|B_2|} x_{b_1,j_2}
{\bm \Omega}_{j_2, A_2\shuffle B_2^T,b_2}
\sum_{j_1\in W_1}
x_{b_2,j_1}
\non
\\
=&\,\,\,\,\,\,\,\,\,\,\,\,\,\,
\,\,\,\,\,
\frac{1}{2}
\!\!\!\!\!
\!\!\!\!\!
\!\!\!\!\!\!\!\!\!\!\!\!\!\!\!\!
\sum_{\substack{a_1,b_1\in W_1, (a_1, A_1, b_1,B_1) = (W_1) \\ a_2,b_2\in W_2, (a_2, A_2, b_2,B_2) = (W_2)}}
\!\!\!\!\!\!\!\!\!\!\!\!\!\!\!\!
\!\!\!\!\!\!\!\!\!\!\!
(-1)^{|B_1|+|B_2|}{\bm \Omega}_{a_1, A_1\shuffle B_1^T,b_1}x_{b_1,a_2}
{\bm \Omega}_{a_2, A_2\shuffle B_2^T,b_2}x_{b_2,a_1}
\nl
=: & \langle\, W_1,W_2\,\rangle
\,,
\end{align}
We already see the application of such simplification in the decomposition of double cycles in the companion paper \cite{Rodriguez:2023qir}. 
We call the last line of \eqref{fusionCC} as a fusion of two cycles and denote it as 
$\langle\, W_1,W_2\,\rangle= \langle\, W_2,W_1\,\rangle$.
Here are some concrete examples,
%
\begin{align}
\langle (12),(34)\rangle=\,\,& 
{\bm \Omega}_{12}x_{2,3}{\bm \Omega}_{34}x_{4,1}+{\bm \Omega}_{12}x_{2,4}{\bm \Omega}_{43}x_{3,1} \,,
\\
\langle (12),(345)\rangle=\,\,& 
{\bm \Omega}_{12}x_{2,3}{\bm \Omega}_{345}x_{5,1}+{\bm \Omega}_{12}x_{2,4}{\bm \Omega}_{453}x_{3,1}+{\bm \Omega}_{12}x_{2,5}{\bm \Omega}_{534}x_{4,1} 
\nl\,\,&
-{\bm \Omega}_{12}x_{2,5}{\bm \Omega}_{543}x_{3,1}-{\bm \Omega}_{12}x_{2,3}{\bm \Omega}_{354}x_{4,1}-{\bm \Omega}_{12}x_{2,4}{\bm \Omega}_{435}x_{5,1} 
\,,
\nl
\langle (123),(456)\rangle=\,\,&\Big[ \big[
{\bm \Omega}_{123} {\bm \Omega}_{456} \Big( x_{3,4}x_{6,1}- x_{3,6}x_{4,1}\big) +
{\rm cyc}(456) \big] + {\rm cyc}(123) \Big]\,,
\nl
\langle (12),(3456)\rangle=\,\,&{\bm \Omega}_{12} \Big[\Big(
{\bm \Omega}_{3456} x_{2,3}x_{6,1}
+{\bm \Omega}_{3654} x_{2,3}x_{4,1} 
- ( {\bm \Omega}_{3465}+{\bm \Omega}_{3645}) x_{2,3}x_{5,1} \Big)+{\rm cyc}(3456)\Big]
\,.
\non
\end{align}
In general, there are $2^{|W_1|+|W_2|-5}\,|W_1|\,|W_2| $ terms in $\<W_1,W_2\>$ as one can check in the above examples.

\subsection{Fusion of multiple cycles}
The generalization of \eqref{fusionCC} is straightforward 
\begin{align}\la{cquef}
&\sum_{j_2\in W_2}\! \OO^{W_1}_{a_1,j_2}\sum_{j_3\in W_3}\! \OO^{W_2}_{j_2,j_3}\cdots 
\sum_{j_r\in W_r}\! \OO^{W_{r-1}}_{j_{r-1},j_r}
\sum_{j_1\in W_1}\! \OO^{W_r}_{j_{r},j_1}
\nl&
+
\sum_{j_r\in W_r}\! \OO^{W_1}_{a_1,j_r}\sum_{j_{r-1}\in W_{r-1}}\! \OO^{W_r}_{j_r,j_{r-1}}\cdots 
\sum_{j_2\in W_2}\! \OO^{W_3}_{j_{3},j_2}
\sum_{j_1\in W_1}\! \OO^{W_2}_{j_{2},j_1}
=2\<W_1,W_2,\ldots,W_r\>\,,
\end{align}
with the generalization to the fusion of $r$ cycles defined by (with $r\geq 2$ and $a_{r+1}:=a_1$):
\begin{align}\label{eq:fusionr}
\langle\, W_1,W_2,\ldots,W_r\rangle :\!\!&=\frac{(-1)^r}{2}\Bigg[\prod_{i=1}^r\sum_{a_i,b_i\in W_i}\Omega_{a_i, A_i\shuffle B_i^T,b_i}x_{b_i,a_{i+1}} \Bigg]\nonumber\\
&=\frac{(-1)^r}{2}\Bigg[\prod_{i=1}^{r}\sum_{a_i,b_i\in W_i}\Bigg]\begin{tikzpicture}[baseline={([yshift=-.5ex]current bounding box.center)},every node/.style={font=\footnotesize,},vertex/.style={inner sep=0,minimum size=3pt,circle,fill},wavy/.style={decorate,decoration={coil,aspect=0, segment length=2.2mm, amplitude=0.5mm}},dir/.style={decoration={markings, mark=at position \halfway with {\arrow{Latex}}},postaction={decorate}}]
\pgfmathsetmacro{\spa}{1};
\pgfmathsetmacro{\spb}{0.75};
\node (a1) at (0,0) [label={above:{$a_1$}},vertex] {};
\node (b1) at (\spa,0) [label={above:{$b_1$}},vertex] {};
\node (a2) at (\spa+\spb,0) [label={above:{$a_2$}},vertex] {};
\node (b2) at (2*\spa+\spb,0) [label={above:{$b_2$}},vertex] {};
\node (br) at (0,-\spb) [label={below:{$b_r$}},vertex] {};
\node (ar) at (\spa,-\spb) [label={below:{$a_r$}},vertex] {};
\node (a3) at (2*\spa+\spb,-\spb) [vertex] {};
\node (b3) at (\spa+\spb,-\spb) [vertex] {};
\draw[thick,wavy] (a1.center) -- (b1.center);
\draw[thick,wavy] (a2.center) -- (b2.center);
\draw[thick,wavy] (ar.center) -- (br.center);
\draw[thick,dir,dotted] (b1.center) -- (a2.center);
\draw[thick,dir,dotted] (br.center) -- (a1.center);
\draw[thick,dir,dotted] (b2.center) -- (a3.center) node [pos=0.36,right=0pt]{$\vdots$};
\draw[thick,wavy] (a3.center) -- (b3.center) node [pos=0.5,below=0pt]{$\cdots$};
\draw[thick,dir,dotted] (b3.center) -- (ar.center);
\end{tikzpicture}\,.
\end{align}
Each wavy line represents a summation over $A_i\shuffle B_i^T$ with $i$ inferred by the endpoints. The definition~\eqref{eq:fusionr} is clearly cyclic, 
\ba
\langle W_1,W_2,\ldots,W_r\rangle=\langle W_2,W_3,\ldots,W_r,W_1\rangle\,,
\ea
and the factor $\frac{1}{2}$ cancels the double counting due to the reflection symmetry 
\ba 
\langle W_1,W_2,\ldots,W_r\rangle=\langle W_r,\ldots,W_2,W_1\rangle\,.
\ea
For instance,
\ba
\sum_{j_2\in W_2}\! \OO^{W_1}_{a_1,j_2}\sum_{j_3\in W_3}\! \OO^{W_2}_{j_2,j_3} 
\sum_{j_1\in W_1}\! \OO^{W_3}_{j_{3},j_1}
+&
\sum_{j_3\in W_3}\! \OO^{W_1}_{a_1,j_3}\sum_{j_{2}\in W_{2}}\! \OO^{W_3}_{j_3,j_{2}} \sum_{j_1\in W_1}\! \OO^{W_2}_{j_{2},j_1}
\nl 
&=\,2\<W_1,W_2, W_3\>
=\,2\<W_1,W_3, W_2\>\,,
\ea
which, in the simplest non-trivial case of two-element cycles $W_1,W_2,W_3$ specializes to \ba
&\<(12),(34), (56)\>
\\
=&\,\, \OO^{(12)}_{1,3}\OO^{(34)}_{3,5}\OO^{(56)}_{5,1}+\OO^{(12)}_{1,4}\OO^{(34)}_{4,5}\OO^{(56)}_{5,1}+\OO^{(12)}_{1,3}\OO^{(34)}_{3,6}\OO^{(56)}_{6,1}+\OO^{(12)}_{1,4}\OO^{(34)}_{4,6}\OO^{(56)}_{6,1}
\nl
=&\,\, -\Omega_{12}(\eta_2) \Omega_{34}(\eta_4) \Omega_{56}(\eta_6) \Big(x_{2,3}x_{4,5} x_{6,1} -x_{2,3}x_{4,6} x_{5,1} -x_{2,4}x_{3,5} x_{6,1} +x_{2,4}x_{3,6} x_{5,1}  \Big)\,.
\non
\ea
In general, there are $2^{-2r-1+\sum_{i=1}^r |W_i|} \,\prod_{i=1}^r |W_i| $ terms in $\<W_1,W_2,\cdots, W_r\>$ as one can check in the above example.

As we will see, such fusions $\<W_1,W_2,\cdots, W_r\>$ will appear when we deal with the product of several cycles. We will discuss the decomposition of two, three, and an arbitrary number of cycles gradually in the following sections.

\begin{figure}[t]
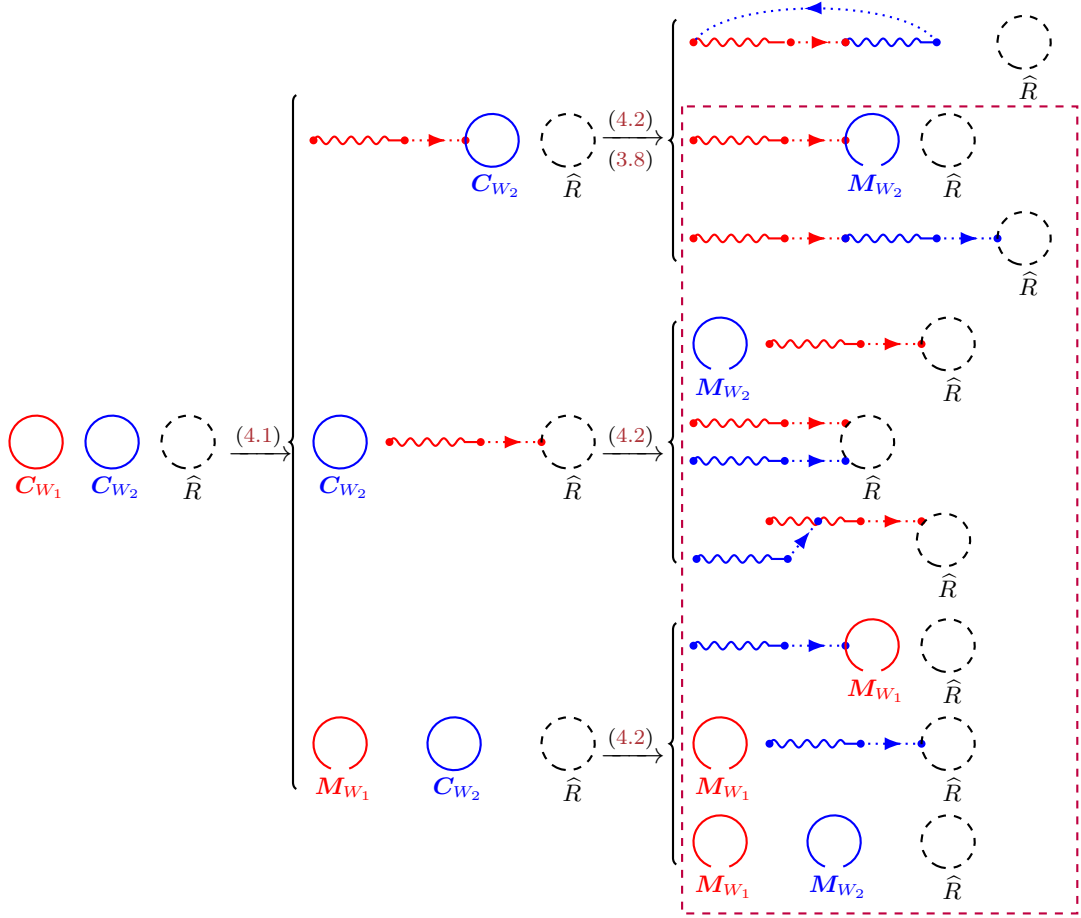

    \centering



  \caption{ The F-IBP reduction of the double cycle string integrand $\CC_{W_1}(\xi_1)\CC_{W_2}(\xi_2)$ in presence of additional punctures gathered in $\widehat R$. $\xi_i$ in $\CC_{W_i}(\xi_i)$ and $\MM_{W_i}(\xi_i)$ are suppressed in the graphs. All terms in the dashed rectangle are free of cycles. }
  \label{fig:2}
\end{figure}

\section{Double cycles in presence of additional punctures \label{sec444}}
In the companion paper \cite{Rodriguez:2023qir}, we have showcased how to break the product of two general double periodic Kronecker-Eisenstein cycles without additional punctures.
 Here we extend to demonstrate how to break the product of two cycles in the presence of additional punctures, i.e., $r=2$ in \eqref{cycleproduct} and $W_1\sqcup W_2 \sqcup \widehat R= \{0,1,2,\cdots,n\}$ using the single-cycle formula \eqref{rhorhog} with compact notations. 
As shown in \cref{fig:2}, we use a dashed circle to represent the punctures in $\widehat R$ to remind us there are punctures beyond $W_1,W_2$. Note that this includes the auxiliary puncture $0$ which encodes the information of Koba-Nielsen derivatives. 
 We break $ \CC_{W_1}(\xi_1)$ at first,
 \begin{align} \label{twocycle22}
 ( 1+s_{W_1}) ( 1+s_{W_2}) \CC_{W_1}(\xi_1)&\CC_{W_2} (\xi_2)
 \\ 
 \non
= \,\,& 
 ( 1+s_{W_1})  
 (\MM_{W_1}(\xi_1)
 +\sum_{j \in \widehat R}  \OO^{W_1}_{a_1,j}
 +
 \sum_{j_2 \in W_2}  \OO^{W_1}_{a_1,j_2} ) \CC_{W_2} (\xi_2) \,,
 \end{align}
 then break $ \CC_{W_2}(\xi_2)$ according to the attaching point if there is,
 \begin{align} 
 \label{twocycle2211}
 ( 1+s_{W_1}) ( 1+s_{W_2}) \CC_{W_1}(\xi_1)\CC_{W_2} (\xi_2)
=
& (\MM_{W_1}(\xi_1)
 + 
 \sum_{j \in \widehat R}  \OO^{W_1}_{a_1,j})
  (\MM_{W_2}(\xi_2)
+
  \sum_{k \notin W_2}  \OO^{W_2}_{a_2,k} )
  \non
   \\ \,\,& 
 +\sum_{j_2 \in W_2}  \OO^{W_1}_{a_1,j_2} 
 (  \MM_{W_2}(\xi_2)
 +
 \sum_{k \notin W_2}  \OO^{W_1}_{j_2, k} )\,.
 \end{align}
 Using \eqref{fusionCC}, we get
 \begin{align} \label{twocycle222}
 & ( 1+s_{W_1}) ( 1+s_{W_2}) \CC_{W_1}(\xi_1)\CC_{W_2} (\xi_2)
 \\
 = \,\,&
\MM_{W_1}(\xi_1)\MM_{W_2}(\xi_2)
+\MM_{W_1} (\xi_1)
\sum_{j_1 \in W_1} \OO^{W_2}_{a_2,j_1 } 
+
\MM_{W_2} (\xi_2)
\sum_{j_2 \in W_2} \OO^{W_1}_{a_1,j_2 } 
 +\<W_1,W_2 \>
 \nl
&
+
\MM_{W_1} (\xi_1)
\sum_{j_1 \in \widehat R} \OO^{W_2}_{a_2,j_1 } 
+
\MM_{W_2} (\xi_2)
\sum_{j_2 \in \widehat R} \OO^{W_1}_{a_1,j_2 } 
+\sum_{p \in \widehat R} \OO^{W_1}_{a_1, p } \sum_{q \notin W_2} \OO^{W_2}_{a_2, q }  
 \non \\ 
& \qquad\qquad\qquad\qquad\qquad\qquad\qquad\qquad\qquad\qquad\qquad\qquad + \sum_{j_2 \in W_2} \OO^{W_1}_{a_1, j_2 } \sum_{q \in \widehat R} \OO^{W_2}_{a_2, q } \,.
\non
 \end{align}
The skeleton of the final result is shown on the right of the \cref{fig:2}. Except for the top graph there which corresponds to $\<W_1,W_2 \>$, the other terms in the dashed rectangle are free of cycles and can be easily expanded onto the basis \eqref{intro1} using Fay identities \eqref{fayid} or \eqref{fayprac}. $\<W_1,W_2 \>$ is just a sum of isolated cycles and can be decomposed again using the single-cycle formulae \eqref{rhorhog}. Here $a_1, a_2$ could be any point in $W_1,W_2$ respectively. 
 Different choices of them lead to equivalent results in the support of F-IBP.
 
 When $R=\emptyset$, i.e., $\widehat R=\{ 0\}$, the last two lines in \eqref{twocycle222} become total derivative terms and it reproduces the results in the companion paper.

Here is an example at $n=5$, 
\begin{align}
&(1{+}s_{12})(1{+}s_{34})\CC_{(12)}(\xi_1)\CC_{(34)} (\xi_2)
= \MM_{12}(\xi_1)\MM_{34}(\xi_2) \label{ofofr} \\
&\quad -\MM_{12} (\xi_1) x_{4,12} \Omega_{34}(\eta_{4}) -  x_{2,34} \Omega_{12}(\eta_{2}) \MM_{34}(\xi_2)
  +\big( 
  x_{2,3}x_{4,1} -  x_{2,4}x_{3,1} 
\big) \Omega_{12}(\eta_{2})\Omega_{34}(\eta_{4})
\non
\nl
&\quad + {\nabla}_2 {\nabla}_4\big( \Omega_{12}(\eta_{2})  \Omega_{34}(\eta_{4}) \big) 
 - {\nabla}_4\big( \MM_{12} (\xi_1) \Omega_{34}(\eta_{4}) - \Omega_{12}(\eta_{2}) \Omega_{34}(\eta_{4}) { x}_{2,3}  \big) 
\nl&\quad 
-{\nabla}_2\big( \MM_{34} (\xi_2) \Omega_{12}(\eta_{2}) - \Omega_{12}(\eta_{2}) \Omega_{34}(\eta_{2}) { x}_{4,12}  \big)
- {\nabla}_3 \big( 
\Omega_{12}(\eta_{2})
\Omega_{34}(\eta_{4})  { x}_{2,4} \big)
\nl
~
\non
\\[-.5cm] 
& \quad -\MM_{12} (\xi_1) \Omega_{34}(\eta_{4}) x_{4,5}-\MM_{34} (\xi_2) \Omega_{12}(\eta_{2}) x_{2,5}+ \Omega_{12}(\eta_{2}) \Omega_{34}(\eta_{4}) x_{2,5} x_{4,5}
\nl& \quad
+
\Omega_{12}(\eta_{2}) \Omega_{34}(\eta_{4}) \big( x_{2,5} x_{4,12}+ x_{2,3} x_{4,5} -x_{2,4} x_{3,5} \big)
\,,
\non
\end{align} 
where $x_{i,j\cdots p}:=x_{i,j}+\cdots x_{i,p}$.

 \paragraph{Appearance of reference ordering}
 
 In the above derivation, we break $W_1$ at first and then $W_2$.  
 In this sense, we say we have chosen a reference ordering ${\frak R}=W_1 \prec W_2$
when we break the cycles.  In the final result \eqref{twocycle222}, we see the last two terms on the  right-hand side are not manifestly symmetric under the exchange of two cycles $W_1,W_2$ because of this reference ordering. As a self-consistent method,  different  choices of reference orderings of course must lead to equivalent results. Actually, one can check on the support of F-IBP, we have
\ba
\sum_{p \in R} \OO^{W_1}_{a_1, p }  \sum_{q \in W_1} \OO^{W_2}_{a_2, q }   + \sum_{j_2 \in W_2} \OO^{W_1}_{a_1, j_2 }  \sum_{q \in R} \OO^{W_2}_{a_2, q }  \overset{\rm IBP}= 
\sum_{p \in R} \OO^{W_2}_{a_2, p }  \sum_{q \in W_2} \OO^{W_1}_{a_1, q }   + \sum_{j_1 \in W_1} \OO^{W_2}_{a_2, j_1 }  \sum_{q \in R} \OO^{W_1}_{a_1, q }  \,.
\ea

Let us illustrate this idea by 
 an explicit example of double pairs with non-vanishing $R=\{5\}$, i.e., $n=5$. Under the reference ordering ${\frak R}=(12) \prec (34)$, we have \eqref{ofofr}.

If we break $\CC_{(34)}(\xi_2)$ at first and then  $\CC_{(12)}(\xi_1)$, i.e., using the reference ordering ${\frak R}=(34) \prec (12)$ instead, we  get almost the same result except that the last line of   \eqref{ofofr}  
is replaced by its relabelling $(12)\leftrightarrow (34)$, that is
\ba
\Omega_{12}(\eta_{2}) \Omega_{34}(\eta_{4})  \big( 
x_{4,5} x_{2,34}+ x_{4,1} x_{2,5} -x_{4,2} x_{1,5} 
\big)\,.
\ea
They are equivalent since their difference
$-\Omega_{12}(\eta_{2}) \Omega_{34}(\eta_{4}) x_{2,4} \sum_{i=1}^4 x_{i,5}$ vanishes on the support of F-IBP.

\begin{figure}[t]
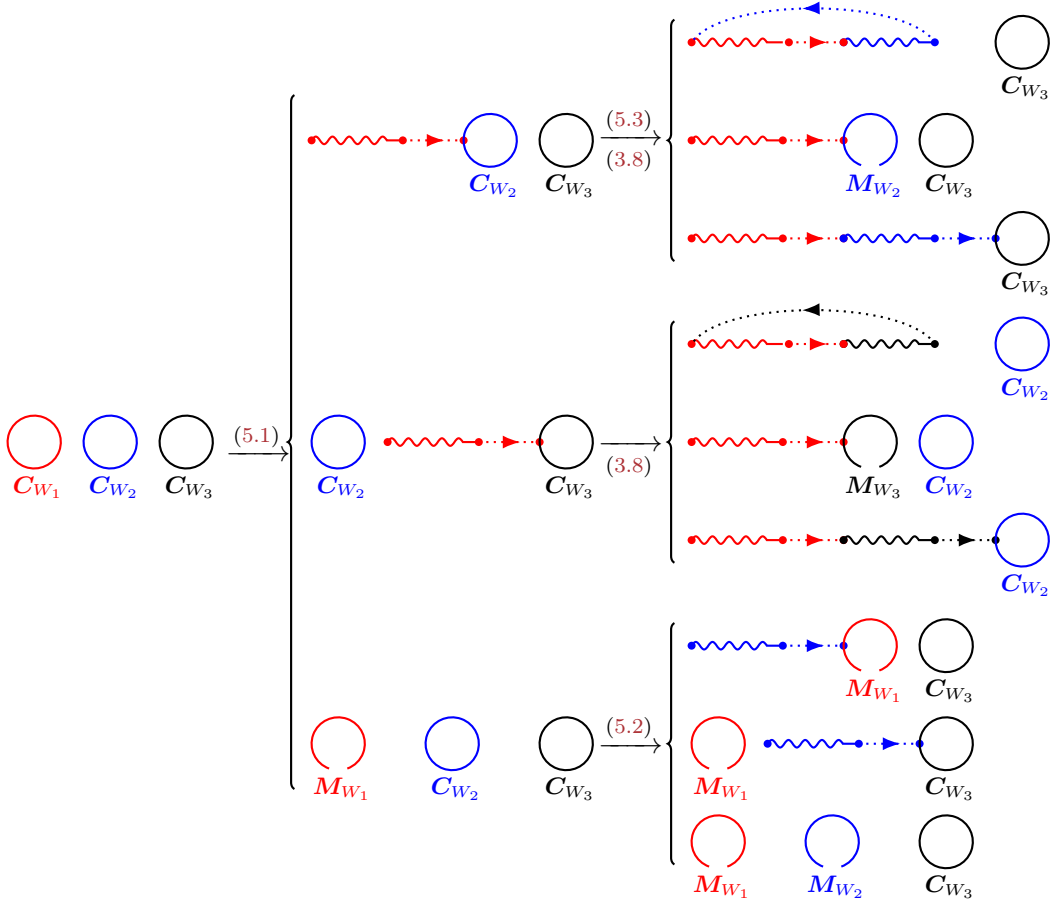

    \centering



  \caption{ The F-IBP reduction of triple cycle string integrand $\CC_{W_1}(\xi_1)\CC_{W_2}(\xi_2)\CC_{W_3}(\xi_3)$ with no additional punctures. Besides, we further ignore the total Koba-Nelson derivative terms here. $\xi_i$ in $\CC_{W_i}(\xi_i)$ and $\MM_{W_i}(\xi_i)$ are suppressed in the graphs. We break two cycles first with the last cycle left to be broken in \cref{fig:4}. }
  \label{fig:3}
\end{figure}

\begin{figure}
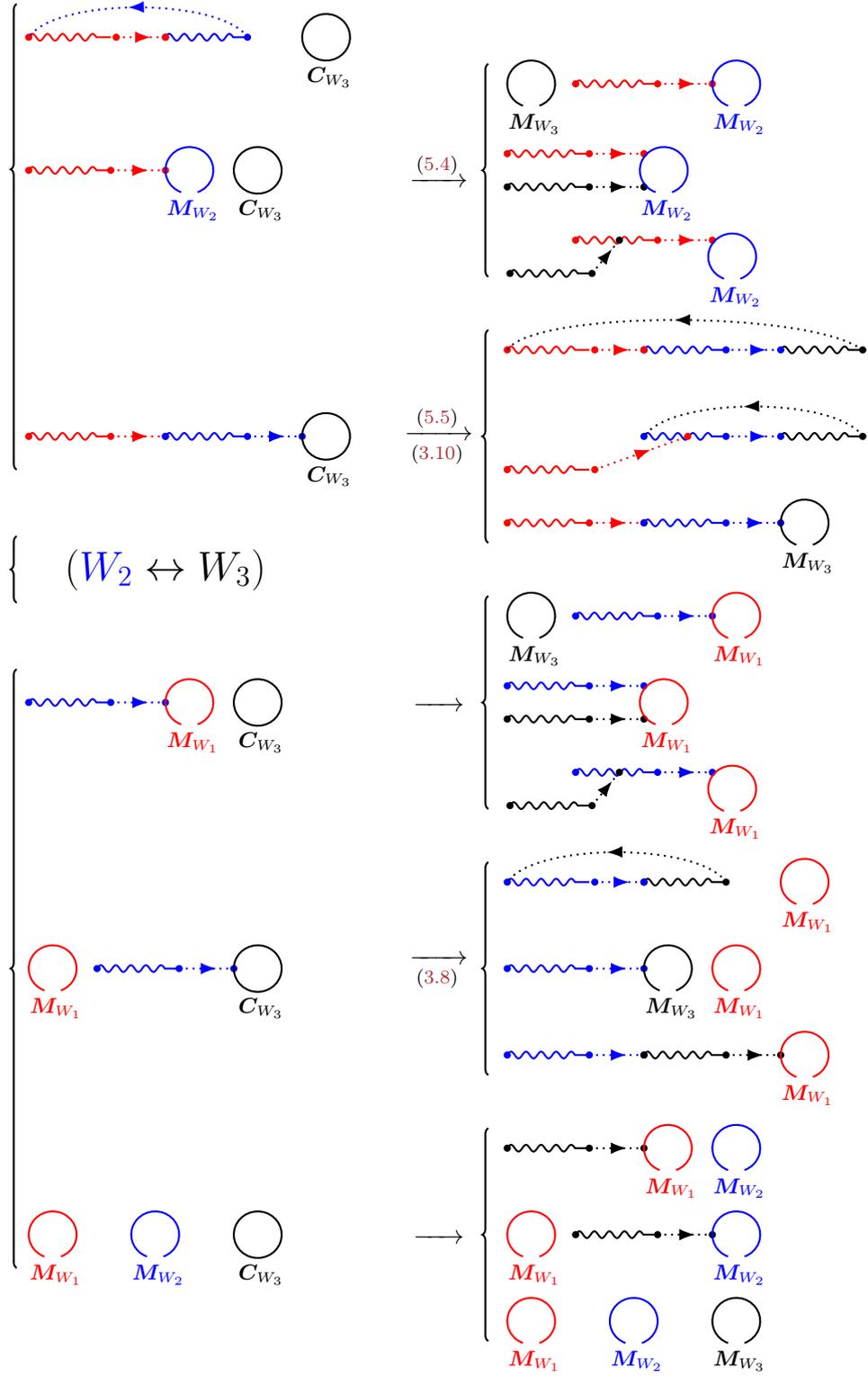

    \centering

%

    \caption{ The F-IBP reduction of  $\CC_{W_1}(\xi_1)\CC_{W_2}(\xi_2)\CC_{W_3}(\xi_3)$ with $R=\emptyset$ and the total Koba-Nelson derivative terms ignored. We continue to break the last  cycles left in \cref{fig:3}.  }
    \label{fig:4}
\end{figure}

\begin{figure} [t]
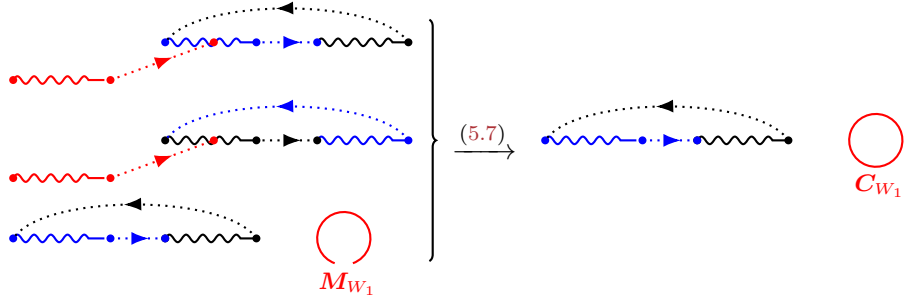

    \centering

%

    \caption{Recombine several terms as a sum of products of two cycles, $\CC_{W_1}(\xi_1)\<{W_2},{W_3}\>$}
    \label{fig:5}
\end{figure}

\section{Triple cycles \label{sec555}}

In our companion paper \cite{Rodriguez:2023qir}, we provided exemplifications for decomposing products of triple doubly periodic Kronecker-Eisenstein cycles at $n=6,7$. In this paper, we extend this to a derivation applicable to arbitrary sets of three cycles, streamlining the process in two main steps. Initially, we consider $R=\emptyset$ and temporarily set aside total derivative terms to foreground the central aspects of the demonstration, supplemented by illustrative sketch graphs to elucidate the process. Subsequently, we unfold the general result for an arbitrary $R$.

  \subsection{Ignoring total derivative terms and without additional punctures }
  In the analysis of the triple cycle product $\CC_{W_1} (\xi_1)\,\CC_{W_2} (\xi_2)\,\CC_{W_3}(\xi_3)$, depicted in \cref{fig:3}, we initiate the process by decomposing $W_1$ utilizing the relation detailed in \eqref{stibp}. This procedure yields,
\begin{align}\label{threeaf}
(1+s_{W_1})  \CC_{W_1}(\xi_1) \,\CC_{W_2}(\xi_2)&\,\CC_{W_3} (\xi_3)
\\
\non &\overset{\rm IBP}=  
\Bigg[ \MM_{W_1}(\xi_1)+  \sum\limits_{\substack{j_2\in W_2}} \OO^{W_1}_{a_1,{j_2}}  +   \sum\limits_{\substack{j_3\in W_3}} \OO^{W_1}_{a_1,j_3}   \Bigg]  \CC_{W_2}(\xi_2)\,\CC_{W_3} (\xi_3)\,,
\end{align}
where we have dropped a total Koba-Nielsen derivative term.

The right-hand side of \eqref{threeaf} entails three terms. The first term, when combined with $\CC_{W_2}(\xi_2)$ and $\CC_{W_3}(\xi_3)$, follows a similar breakdown as demonstrated in \eqref{twocycle222} since $\MM_{W_1}  (\xi_1)$ is free of punctures from $W_2$ or $W_3$, resulting in,
 \begin{align} \label{twocycle2}
  & ( 1+s_{W_2}) ( 1+s_{W_3})   \MM_{W_1}  (\xi_1)\CC_{W_2}(\xi_2)\CC_{W_3} (\xi_3)
   \\
 \overset{\rm IBP}=\,\,& 
\MM_{W_1}(\xi_1)\MM_{W_2}(\xi_2)\MM_{W_3} (\xi_3)
+
\MM_{W_1}(\xi_1)   \MM_{W_2} (\xi_2)
\sum_{k \notin W_3} \OO^{W_3}_{a_3,k} 
\nl 
&+
\MM_{W_1} (\xi_1) \MM_{W_3} (\xi_3)
\sum_{k \in W_2} \OO^{W_2}_{a_2, k} 
  +\MM_{W_1} (\xi_1)  \<W_2,W_3 \>
\nl
&+\MM_{W_1}  (\xi_1)\sum_{p \in W_1} \OO^{W_2}_{a_2, p }  \sum_{q \notin W_3} \OO^{W_3}_{a_3, q }   + \MM_{W_1}  (\xi_1)\sum_{j_3 \in W_3} \OO^{W_2}_{a_2, j_3 }  \sum_{q \in W_1} \OO^{W_3}_{a_3, q } \,.
\non
 \end{align}

 For the subsequent terms on the right side of \eqref{threeaf}, it is adequate to analyze the second term only, as the third term exhibits similar behavior through the interchange of $W_2$ and $W_3$. In the second term, we observe the formation of a tadpole structure as the chain $\OO^{W_1}$ connects with $W_2$. This necessitates the choice of $a_2 = j_2$ for every $\OO^W_{a_1,j_2}$, leading to
\begin{align}\label{three2}
 ({1+s_{W_2}}  )\CC_{W_3}(\xi_3)& \!\!\sum_{j_2\in W_2}\! \!\OO^{W_1}_{a_1,j_2}  \CC_{W_2}(\xi_2)
\\
\non
& \! \overset{\rm IBP}= \!  \CC_{W_3}(\xi_3) \Bigg[  \MM_{W_2}  (\xi_2)\sum_{\substack{j_2\in W_2}}\! \OO^{W_1}_{a_1,j_2}+ 
\<W_1, W_2\>+\sum_{\substack{j_2\in W_2\\ j_3 \in W_3}}\! \OO^{W_1}_{a_1,j_2} \OO^{W_2}_{j_2,j_3} 
\Bigg]\,,
\end{align}
where we have used \eqref{fusionCC} to derive the second term on the  right-hand side of   \eqref{three2}.
Following this, the cycle $\CC_{W_3}(\xi_3)$ is further decomposed as illustrated below
\ba
\label{eqw3last}
 \CC_{W_3}  (\xi_3)\MM_{W_2}  (\xi_2)\sum_{\substack{j_2\in W_2}}\! \OO^{W_1}_{a_1,j_2} 
\overset{\rm IBP}=\,\,
 \MM_{W_2} (\xi_2) \sum_{\substack{j_2\in W_2}}\! \OO^{W_1}_{a_1,j_2}  
( \MM_{W_3}(\xi_3) (\xi_3)+ \sum_{\substack{k \notin W_3}}\! \OO^{W_3}_{a_3, k}  
)\,.
\ea

Regarding the second term on the right-hand side of equation \eqref{three2}, the chain $\OO^{W_2}$ is connected to the cycle $W_3$. To break the cycle $W_3$, it becomes necessary to select $a_3 = j_3$ for each term $\OO^W_{a_2, j_3}$, in accordance with the relation specified in \eqref{stibp}
\ba
\label{eqw3last2}
  \CC_{W_3}(\xi_3) \sum_{\substack{j_2\in W_2\\ j_3 \in W_3}}\! \OO^{W_1}_{a_1,j_2} \OO^{W_2}_{j_2,j_3} 
\overset{\rm IBP}=\,\,
  \sum_{\substack{j_2\in W_2\\ j_3 \in W_3}}\! \OO^{W_1}_{a_1,j_2} \OO^{W_2}_{j_2,j_3} (   \MM_{W_3} (\xi_3)
+  \sum_{\substack{ k \notin W_3}}\! \OO^{W_3}_{j_3, k}  )\,.
\ea
Upon substituting the results derived from the equations mentioned above into equation \eqref{three2}, we obtain the final form for the second term on the right-hand side of equation \eqref{threeaf}. The last term in \eqref{threeaf} can be effectively addressed through a straightforward relabelling process.
Together with \eqref{twocycle2}, they lead to  (for $R=\emptyset $)
  \ba
  \label{threecycleRem}
(1+ &s_{W_1}) (1+s_{W_2}) (1+s_{W_3})  \CC_{W_1} (\xi_1)\CC_{W_2}(\xi_2)\CC_{W_3}(\xi_3) 
\\
\non
&\overset{\rm IBP}=      \MM_{W_1}(\xi_1)\MM_{W_2}(\xi_2)\MM_{W_3} (\xi_3)
 +\MM_{W_1}  (\xi_1) \MM_{W_2} (\xi_2)
\sum_{k \in W_{1,2}} \OO^{W_3}_{a_3,k} 
   \\
 &
 +
\MM_{W_1}  (\xi_1) \MM_{W_3} (\xi_3)
\sum_{k \in W_{1,3}} \OO^{W_2}_{a_2, k} 
 +
 \MM_{W_2}  (\xi_2)\MM_{W_3} (\xi_3)
\sum_{k \in W_{2,3}} \OO^{W_1}_{a_1, k} 
\nl
&+\Big[ \MM_{W_1} (\xi_1)\Big( \sum_{p \in W_1} \OO^{W_2}_{a_2, p }  \sum_{q \in W_ {1,2}} \OO^{W_3}_{a_3, q }   +  \sum_{j_3 \in W_3} \OO^{W_2}_{a_2, j_3 }  \sum_{q \in W_1} \OO^{W_3}_{a_3, q }  \Big)+ {\rm cyc}(W_1,W_2, W_3)\Big]
  \nl
  &+(1+s_{W_1}) \CC_{W_1}   \<W_2,W_3 \> +(1+s_{W_2}) \CC_{W_2}   \<W_1,W_3 \>+(1+s_{W_3}) \CC_{W_3}   \<W_1,W_2 \>
  \nl
  &+   2 \<W_1,W_2,W_3 \> 
  \,.
  \non
\ea
Here, $W_{1,2}= W_1 \cup W_2$.  The cyclic relabeling mentioned in the fourth line of course refers to the process of also relabeling the corresponding $\xi_i$, i.e., $(W_1,\xi_1)\to(W_2,\xi_2)\to(W_3,\xi_3)\to(W_1,\xi_1)$ or  $(W_1,\xi_1)\to(W_3,\xi_3)\to(W_2,\xi_2)\to(W_1,\xi_1)$. We have implemented \eqref{cquef} to get the fusions in the last two lines. Furthermore, we have reversed the relation~\eqref{stibp} to reconstruct several blocks as the original double periodic Kronecker-Eisenstein cycle $\CC_{W_1}(\xi_1)$ in the penultimate line,
\ba 
\label{instibp}
\sum_{p\in W_2}\OO_{a_1,p}^{W_1} \<{W_2},{W_3}\> + \sum_{p\in W_3}\OO_{a_1,p}^{W_1} \<{W_2},{W_3}\>  + \MM_{W_1}& (\xi_1)\<{W_2},{W_3}\> 
 \\
 \non
 &\overset{\rm IBP}= (1+s_{W_1})\CC_{W_1}(\xi_1)\<{W_2},{W_3}\> \,.
\ea 
This reconstruction is visually represented in \cref{fig:5}.

The initial two lines on the right-hand side of \eqref{threecycleRem} are devoid of cycles. Meanwhile, in the final two lines, the number of cycles of any type is diminished to either two or one, a situation we are already equipped to handle.

This is how we decompose $\CC_{(12)} (\xi_1)\,\CC_{(34)} (\xi_2)\,\CC_{(56)}(\xi_3)$ in the companion paper \cite{Rodriguez:2023qir} and we present the expression again here for $n=6$,
\begin{align} 
&(1{+}s_{12}) (1{+}s_{34}) (1{+}s_{56})  \CC_{(12)} (\xi_1)\,\CC_{(34)} (\xi_2)\,\CC_{(56)} (\xi_3) 
\overset{\rm IBP}= 
\MM_{12}(\xi_1) \MM_{34} (\xi_2)\MM_{56}(\xi_3)  \label{triple6} \\
&-\MM_{12}(\xi_1)\MM_{34}(\xi_2) {\bm \Omega}_{56}  x_{6,1234} 
-\MM_{12} (\xi_1) \MM_{56}(\xi_3){\bm \Omega}_{34} x_{4,1256}
- \MM_{34} (\xi_2)\MM_{56}(\xi_3){\bm \Omega}_{12} x_{2,3456}
\nl
&+ \MM_{12}(\xi_1){\bm \Omega}_{34} {\bm \Omega}_{56}  (x_{4,12}x_{6,1234}+x_{4,5}x_{6,12}-x_{4,6}x_{5,12})
\nonumber
\\
&+ \MM_{34}(\xi_2){\bm \Omega}_{12} {\bm \Omega}_{56} (x_{2,34}x_{6,1234}+x_{2,5}x_{6,34}-x_{2,6}x_{5,34})
\nonumber
\\
&+ \MM_{56}(\xi_3){\bm \Omega}_{12} {\bm \Omega}_{34}  (x_{2,56}x_{4,1256}+x_{2,3}x_{4,56}-x_{2,4}x_{3,56})
\nonumber
\\
&+(1{+}s_{12}) \CC_{(12)} (\xi_1) {\bm \Omega}_{34} {\bm \Omega}_{56} (x_{4,5}x_{6,3}-x_{4,6}x_{5,3})\nonumber
\\
&+  (1{+}s_{34}) \CC_{(34)}(\xi_2) {\bm \Omega}_{12} {\bm \Omega}_{56} (x_{2,5}x_{6,1}-x_{2,6}x_{5,1})
\nonumber
\\
&+  (1{+}s_{56}) \CC_{(56)} (\xi_3){\bm \Omega}_{12} {\bm \Omega}_{34} (x_{2,3}x_{4,1}-x_{2,4}x_{3,1})
\nonumber
\\
&
+{\bm \Omega}_{12} {\bm \Omega}_{34}  {\bm \Omega}_{56} \big(
x_{1,4} x_{2,6} x_{3,5}+x_{1,5} x_{2,4} x_{3,6} -x_{1,6} x_{2,4} x_{3,5}
-x_{1,4} x_{2,5} x_{3,6}
\nonumber
\\
&\quad\quad\quad\quad\quad\; \, +x_{1,6} x_{2,3} x_{4,5}-x_{1,3} x_{2,6} x_{4,5}-x_{1,5} x_{2,3} x_{4,6}+x_{1,3} x_{2,5} x_{4,6}\big)\,.
   \nonumber
\end{align}

Comparing \eqref{twocycle222}
and \eqref{threecycleRem}, one can find
\ba 
( 1+s_{W_1}) &( 1+s_{W_2})  \CC_{W_1}(\xi_1)\CC_{W_2} (\xi_2)
\\
\non
&\overset{\rm IBP}= 
(1+ s_{W_1}) (1+s_{W_2}) (1+s_{\hat R})  \CC_{W_1}(\xi_1)\CC_{W_2}(\xi_2)\CC_{\hat R}(\hat\xi) \big|_{ \MM_{\hat R}(\hat\xi)}  \,.
\ea 
This means, for the product of two cycles $W_1,W_2$ in the presence of additional cycles including the auxiliary puncture $0$, we can conceptualize $\widehat R$ to form a third auxiliary double periodic Kronecker-Eisenstein cycle, denoted as  $\CC_{W_3}(\hat \xi)=\CC_{\hat R}(\hat \xi)$. Subsequently, we apply the triple cycle formula \eqref{threecycleRem} for decomposing the integrands.
To finalize, extracting the coefficient of $\MM_{W_3}(\hat\xi)=\MM_{\widehat R}(\hat\xi)$ yields the result pertinent to the scenario of double cycles $W_1$ and $W_2$ in the presence of additional punctures. This methodology extends to the general scenarios involving more cycles, a topic we will delve into in \cref{sec777}.

 \subsection{Considering total derivative terms  and with additional punctures }

When considering a general ${\widehat R}$, attention must be given to the summation ranges while applying formula \eqref{stibp}, resulting in
\ba
\label{threecycle333}
&(1+s_{W_1}) (1+s_{W_2}) (1+s_{W_3})  \CC_{W_1} (\xi_1)\CC_{W_2} (\xi_2)\CC_{W_3}  (\xi_3)
\\
=\,\, &
   \big(    {\rm RHS~ of ~ \eqref{threecycleRem}} \big)
   +  \Big[   \MM_{W_1}    (\xi_1) \MM_{W_2}  (\xi_2)  \sum_{p \in {\widehat R}} \OO^{W_3}_{a_3, p } +{\rm cyc}(W_1,W_2,W_3) \Big] 
             \nl&
    +\Big[  \MM_{W_1}   (\xi_1) \Big( \sum_{p \in {\widehat R}} \OO^{W_2}_{a_2, p }  \sum_{q \notin W_3} \OO^{W_3}_{a_3, q }   +  \sum_{j_3 \in W_{1,3}} \OO^{W_2}_{a_2, j_3 }  \sum_{q \in {\widehat R}} \OO^{W_3}_{a_3, q }  \Big)+ {\rm cyc}(W_1,W_2, W_3)\Big]
              \nl&
    +   \sum_{p \in {\widehat R}} \OO^{W_1}_{a_1, p }   \Big( \sum_{p \in {\widehat R}\cup W_1} \OO^{W_2}_{a_2, p }  \sum_{q \notin W_3} \OO^{W_3}_{a_3, q }   +  \sum_{j_3 \in W_{3}} \OO^{W_2}_{a_2, j_3 }  \sum_{q \in {\widehat R}\cup W_1 } \OO^{W_3}_{a_3, q }  \Big)
    \nl
  &
  + \Big[\Big( \sum_{j_2 \in W_2} \OO^{W_1}_{a_1, j_2 }  \sum_{p \in {\widehat R}} \OO^{W_2}_{a_2, p }  \sum_{q \notin W_3} \OO^{W_3}_{a_3, q }    + \sum_{j_2 \in W_2} \OO^{W_1}_{a_1, j_2 }   \sum_{j_3 \in W_3} \OO^{W_2}_{a_2, j_3 }   \sum_{p \in {\widehat R}} \OO^{W_3}_{a_3, p } \Big) +(W_2\leftrightarrow W_3) \Big]
  \,.
  \non
\ea
Again, the cyclic relabeling mentioned in the second and third lines of course refers to the process of also relabeling the corresponding $\xi_i$.

Specifically, in the triple cycle case with $n=7$, additional terms based on \eqref{triple6} are expressed as
\begin{align} 
\label{triple67}
(1&+s_{12}) (1+s_{34}) (1+s_{56})  \CC_{(12)} (\xi_1)\,\CC_{(34)} (\xi_2)\,\CC_{(56)}  (\xi_3)
=
    \big(   {\rm RHS~ of ~ \eqref{triple6}} \big)
\\  &-
     x_{6,7}  \MM_{12} (\xi_1)\MM_{34}(\xi_2) {\bm \Omega}_{56}  - x_{4,7} \MM_{12}(\xi_1) \MM_{56}(\xi_3) {\bm \Omega}_{34}  - x_{2,7} \MM_{34}(\xi_2) \MM_{56}(\xi_3) {\bm \Omega}_{12}   
      \nl&
      \!+\!\Big[  \big(  x_{4,7}   x_{6,347} \!+\! x_{4,5} x_{6,7}-x_{4,6} x_{5,7}   \! +\!  x_{4,12} x_{6,7} \!+\! x_{6,12} x_{4,7}   \big) \MM_{12}(\xi_1)  {\bm \Omega}_{34} {\bm \Omega}_{56} 
 \!+\! {\rm cyc}(12,34,56) \Big] 
     \nl&
   -   (x_{2,7} x_{4,127} x_{6,12347} + x_{2,7} x_{4,5} x_{6, 127} + x_{2,7} x_{4,6} x_{5, 127}  
   \nl&
+  x_{2,3}x_{4,7} x_{6,12347}   +  x_{2,4}x_{3,7} x_{6,12347}   +  x_{2,5}x_{6,7} x_{4,12567}+  x_{2,6}x_{5,7} x_{4,12567}  
\nl&
 +x_{2,3}x_{4,5} x_{6,7} +x_{2,4}x_{3,5} x_{6,7}   +x_{2,3}x_{4,6} x_{5,7} +x_{2,4}x_{3,6} x_{5,7}  
 \nl&
 +x_{2,5}x_{6,3} x_{4,7} +x_{2,5}x_{6,4} x_{3,7}  +x_{2,6}x_{5,3} x_{4,7} +x_{2,6}x_{5,4} x_{3,7} 
) {\bm \Omega}_{12}{\bm \Omega}_{34} {\bm \Omega}_{56}
\nl&+{\text{(total Koba-Nielsen derivatives)}}\,,
      \non
\end{align}
where 
the total Koba-Nielsen derivatives in the last line can  be reinstated by
replacing $x_{i_1,i_2\cdots i_t 7}\rightarrow x_{i_1,i_2\cdots i_t 70}= x_{i_1,i_2\cdots i_t 7} + \nabla_{i_1}$. This is unambiguous for any number of factors $x_{i_1,i_2\cdots i_t 7}$ since any pair of $\nabla_i,\nabla_j$ commutes.

\paragraph{Reference ordering}

Throughout the derivation in this subsection, we consistently satisfy a specific ordering, $W_1\prec W_2 \prec W_3$, when deciding which original cycle to break in the absence of $f$-$\Omega$ tadpoles. As a guideline, we prioritize breaking cycles earlier in the ordering. For instance, we initiate the process by breaking $W_1$ in \eqref{threeaf}. Following this principle, $W_2$ is broken before $W_3$, evident from  \eqref{twocycle2}. Finally, $W_3$ is the last to be broken, only when it remains as the last isolated cycle, as demonstrated in  \eqref{eqw3last}. This approach defines a reference ordering, denoted as ${\frak R}=W_1\prec W_2 \prec W_3$, for breaking the product of triple cycles. It is crucial to note that different reference orderings can impact the parts of the final results devoid of new $f$-$\Omega$ cycles. Nevertheless, they all yield equivalent outcomes  on the support of F-IBP.

\section{Labeled  forest \label{sec666} }
The complexity of the results exponentially increases as more original cycles are broken in a product. Disregarding the terms that involve new $f$-$\Omega$ cycles, the remaining elements can be effectively structured within the labeled forest framework, which we will now demonstrate.

\subsection{Labeled  forest expansion}

In this subsection, we introduce the function $\mathcal{F}_{W_1,W_2,\cdots, W_t} ({\frak R})$, utilizing a labeled-forest expansion method. This approach systematically handles the increasing complexity resulting from breaking original cycles in a product. The function is defined as,
\begin{align}\label{treeExpansion}
\mathcal{F}_{W_1,W_2,\cdots, W_t}( {\frak R}) := \sum_{V \in\mathbf{V}_{W_1,W_2,\cdots, W_t}( {\frak R})}\,\mathcal{C}(V)\,,
\end{align}
where ${\frak R}$ denotes a reference ordering of the cycles $W_{t+1},W_{t+2},\cdots,W_r$. For instance, we can choose ${\frak R}=W_{t+1}\prec W_{t+2}\prec \cdots\prec W_r$. This reference ordering, crucially derived from the cycle-breaking order discussed in previous sections, influences the function ${\cal C}$, details of which will be provided in sections \ref{subsublabelforest} and \ref{subsublabelforestmap}. It is important to note that while different reference orderings yield algebraically varied results, they remain equivalent when subjected to F-IBP.

To facilitate the expansion in \eqref{treeExpansion}, our first task is to construct the associated labeled forests, followed by defining the mapping function $\mathcal{C}$ for each labeled forest. The summation encompasses $\mathbf{V}_{W_1,W_2,\cdots, W_t}({\frak R})$, referring to labeled forests with roots  $W_1,W_2,\cdots, W_t$, and nodes $W_{t+1},W_{t+2},\cdots, W_r$ , explained in detail in \cref{subsublabelforest}. The function $\mathcal{C}$, on the other hand, transforms the forest $V$ into a function of worldsheet and Mandelstam variables, further detailed in \cref{subsublabelforestmap}. The impact of the reference ordering ${\frak R}$ on the results will also be elucidated in these sections.

This approach of labeled forest serves as a genus-one extension to the labeled trees, initially introduced in \cite{Gao:2017dek} and extensively utilized in \cite{He:2019drm}. However, a key distinction exists between these two combinatorial tools. The tree-level cycles are  usually called  Parke-Taylor factors,   ${\rm PT}(12\cdots m):=$ $1/(z_{12}z_{23}\cdots z_{m1}) $ and for a product of tree-level cycles, ${\rm PT}(W_1) {\rm PT}(W_2) \cdots {\rm PT}(W_r)$ with $W_1\sqcup\cdots \sqcup W_r =\{1,2,\cdots,n\}$,
the ${\rm SL}(2,{\mathbb C})$ gauge fixing, setting $z_n \rightarrow \infty$, inherently breaks one tree-level cycle, transforming it into a tree-level open chain. Consequently, this open chain emerges as a root and all other punctures  will connect to it once all tree-level cycles are broken, necessitating only labeled trees for combinatorial purposes. In contrast, at the one-loop level, every cycle $\CC_{W_i} (\xi_i)$, when broken, yields a term $\MM_{W_i} (\xi_i)$, allowing each $\MM_{W_i} (\xi_i)$ to potentially serve as a root. This necessitates the use of labeled forests, essentially a product of genus-one labeled trees, to adequately address the problem.

\subsubsection{Forests\label{subsublabelforest}}
In this subsection, we delineate the process of constructing labeled forests $\mathbf{V}_{W_1,W_2,\cdots, W_t}( {\frak R})$ pertinent to our study. To begin, we enumerate all forests that are rooted at ${W_1,W_2,\cdots, W_t}$, extending nodes to $W_{t+1}, W_{t+2}, \ldots, W_r$. \footnote{The total number of such spanning forests is given by the formula,
\be
\sum_{\substack{0\leq u_1,u_2,\cdots,u_t \leq r-t\\ 
u_1+u_2+\cdots+u_t =r-t}}
\,\,
\prod_{i=1}^t {r-t-\sum_{j=1}^{i-1} u_j \choose u_i} (u_i+1)^{u_i-1}\,.
\ee 
This expression simplifies to $r^{r-2}$ when $t=1$. For $t=2$, it yields the sequence 1, 2, 8, 50, 432, 4802, 65536, $\cdots$ for $r-2=0,1,2,\cdots $, respectively.

} Consider, for example, $\mathbf{V}_{W_1,W_2}(W_{3}\prec W_{4})$, which is represented by the following 8 spanning forests,
\begin{equation}\label{treeExample}
	\begin{tikzpicture}[baseline={([yshift=-2ex]current bounding box.center)},every node/.style={font=\footnotesize},scale=1.1]
	\draw [thick] (0,0) -- (0,0.75) -- (0.75,0.75);
		\filldraw [thick] (0.75,0) circle (1pt) node[below=0pt]{$W_2$} ;
	\filldraw [thick] (0,0) circle (1pt) node[below=0pt]{$W_1$} (0,0.75) circle (1pt) node[above=0pt]{$W_4$};
	\filldraw [thick] (0.75,0.75) circle (1pt) node[above=0pt]{$W_3$};
	\begin{scope}[xshift=2.5cm]
	\draw [thick] (0,0) -- (0,0.75);
	\draw [thick] (0,0.75) -- (0.75,0.75);
			\filldraw [thick] (0.75,0) circle (1pt) node[below=0pt]{$W_2$} ;
	\filldraw [thick] (0,0) circle (1pt) node[below=0pt]{$W_1$} (0,0.75) circle (1pt) node[above=0pt]{$W_3$};
	\filldraw [thick] (0.75,0.75) circle (1pt) node[above=0pt]{$W_4$};
	\end{scope}
	\begin{scope}[xshift=5cm]
	\draw [thick] (0,0) -- (0.75,0.75);
	\draw [thick] (0,0) -- (0,0.75);
			\filldraw [thick] (0.75,0) circle (1pt) node[below=0pt]{$W_2$} ;
	\filldraw [thick] (0,0) circle (1pt) node[below=0pt]{$W_1$} (0,0.75) circle (1pt) node[above=0pt]{$W_3$};
	\filldraw [thick] (0.75,0.75) circle (1pt) node[above=0pt]{$W_4$};
		\end{scope}
			\begin{scope}[xshift=7.5cm]
	\draw [thick] (0.75,0)-- (0.75,0.75);
	\draw [thick] (0,0) -- (0,0.75);
			\filldraw [thick] (0.75,0) circle (1pt) node[below=0pt]{$W_2$} ;
	\filldraw [thick] (0,0) circle (1pt) node[below=0pt]{$W_1$} (0,0.75) circle (1pt) node[above=0pt]{$W_3$};
	\filldraw [thick] (0.75,0.75) circle (1pt) node[above=0pt]{$W_4$};
		\end{scope}
		
					\begin{scope}[xshift=10.5cm,yshift=.4cm]
\node at (0,0) {$\Bigg(W_1\leftrightarrow W_2\Bigg)$};
		\end{scope}		
	\end{tikzpicture}\,.
\end{equation}
Subsequently, with a reference ordering ${\frak R}$ established, we systematically disassemble each forest into an assortment of paths, simultaneously executing a precise blowup of the cycles in accordance with the stipulated procedures below:

\begin{enumerate}[label=(\arabic*)]

\item Initiate by sketching a trajectory originating from the initial element of ${\frak R}$, progressing directly towards the roots. Subsequently, forge another trajectory, this time commencing from the first untraversed element of ${\frak R}$, and extend it towards the roots. Inevitably, this path will culminate upon intersecting with a previously drawn path or at the roots themselves. Persevere in this methodology, iteratively repeating the process until every node has been systematically traversed, effectively decomposing each forest into a comprehensive set of paths. Notably, ensure that all the paths are meticulously oriented in the direction of the roots.
 
 \label{path}
\item Replace each root $W_i$ with $1\leq i\leq t$ by a chain representing $\MM_{W_i} (\xi_i)$:
\begin{align}
\begin{tikzpicture}[baseline={([yshift=-1.1ex]current bounding box.center)},every node/.style={font=\footnotesize,},vertex/.style={inner sep=0,minimum size=3pt,circle,fill},wavy/.style={decorate,decoration={coil,aspect=0, segment length=2.2mm, amplitude=0.5mm}},dir/.style={decoration={markings, mark=at position \Halfway with {\arrow{Latex}}},postaction={decorate}}]
\filldraw (-1,0) circle (1pt) node[below=0pt]{$W_{i}$};
\node at (-0.5,0) {$\rightarrow$};
\draw [thick] (0.35,0) ++(-110:0.35) arc (-110:-430:0.35);
\node at (0.4,-.6) {$\MM_{W_i}$};
\end{tikzpicture}\quad\,.
\end{align}
\item If a cycle $W_i$ with $t+1\leq i\leq r$ appears in the middle of a path, blow it up according to
\begin{align}\label{traceBlowUp}
\begin{tikzpicture}[baseline={([yshift=-1.ex]current bounding box.center)},every node/.style={font=\footnotesize,},vertex/.style={inner sep=0,minimum size=3pt,circle,fill},wavy/.style={decorate,decoration={coil,aspect=0, segment length=2.2mm, amplitude=0.5mm}}]
\filldraw (-1,0) circle (1pt) node[above=0pt]{$W_i$};
\node at (-0.5,0) {$\rightarrow$};
\node at (-0,0) [label={above:{${a_i}$}},vertex] {};
\node at (1.5,0) [label={above:{${b_i}$}},vertex] {};
\draw[thick,wavy] (0,0) -- (1.5,0);
\end{tikzpicture},
\end{align}
where by our convention $b_i$ is the end closer to the root. We will sum over all pairs of $a_i$ and $b_i$ in $W_i$.
\item If a cycle $W_i$ with $t+1\leq i\leq r$ appears at the start of a path, still blow it up as~\eqref{traceBlowUp}. However, only $b_i$ will be summed in $W_i$, while $a_i\in W_i$ is arbitrary but fixed. Across our construction, we keep the same choice of $a_i$ if this situation happens.
\item If a path ends on a trace $W_i$, then the endpoint can take any value in $W_i$. 
\end{enumerate}
Accordingly, the eight spanning forests in \eqref{treeExample} generate the following labeled forests:
\begin{align}\label{labeledTreesExample}
	& \mathbf{V}_{W_1,W_2}(W_{3}\prec W_{4}) :
	 \\
	 \nonumber
	 &\begin{tikzpicture}[baseline={([yshift=-1.1ex]current bounding box.center)},every node/.style={font=\scriptsize},wavy/.style={decorate,decoration={coil,aspect=0, segment length=2mm, amplitude=0.5mm}},dir/.style={decoration={markings, mark=at position \halfway with {\arrow{Latex[scale=0.9]}}},postaction={decorate}}]
	\draw [thick,blue,wavy] (0,0.75) -- (-0.75,0.75);
		\draw [thick,blue,wavy] (0.75,0.75) -- (1.5,0.75);
	\draw [thick,blue,dir,dotted] (0.75,0.75) -- (0,0.75) ;
	\draw [thick,blue,dir,dotted] (-0.75,0.75) -- (-0.75,0);
	\filldraw [thick] (-0.75,0) circle (1pt) node[below=0pt]{$j_1$} (0,0.75) circle (1pt) node[above=0pt]{$a_4$} (-0.75,0.75) circle (1pt) node[above=0pt]{$b_4$};
	\filldraw [thick] (0.75,0.75) circle (1pt) node[above=0pt]{$b_3$};
		\filldraw [thick] (1.5,0.75) circle (1pt) node[above=0pt]{$a_3$};
	\draw [thick] (-0.75,-0.35) ++(-110:0.35) arc (-110:-430:0.35);
	\node at (-0.75,-1) {$M_{W_1}$};
		\draw [thick] (0.75,-0.35) ++(-110:0.35) arc (-110:-430:0.35);
		\node at (0.75,-1) {$M_{W_2}$};
	\node at (0,-2) [align=center] {$b_3\in W_3$ \\$a_4,b_4\in W_3$ \\ $j_1\in W_1$};
	\begin{scope}[xshift=2.5cm]
	\draw [thick,blue,dir,dotted] (0,0.75) -- (0,0);
			\draw [thick,blue,wavy] (0,1.5) -- (0,0.75);
	\draw [thick,red,dir,dotted] (0.75,1.1) -- (0,1.1);
	\draw [thick,red,wavy] (0.75,1.1) -- (1.5,1.1);
	\filldraw [thick] (0,0) circle (1pt) node[below=0pt]{$j_1$} (0,0.75) circle (1pt) node[left =0pt]{$b_3$}(0,1.5) circle (1pt) node[left=0pt]{$a_3$}
	(0,1.1) circle (1pt) node[left=0pt]{$j_3$};
	\filldraw [thick] (0.75,1.1) circle (1pt) node[above=0pt]{$b_4$} (1.5,1.1) circle (1pt) node[above=0pt]{$a_4$};
	\draw [thick] (-0,-0.35) ++(-110:0.35) arc (-110:-430:0.35);
		\draw [thick] (1.5,-0.35) ++(-110:0.35) arc (-110:-430:0.35);
	\node at (0,-1) {$M_{W_1}$};	
		\node at (1.5,-1) {$M_{W_2}$};	
	\node at (.9,-2) [align=center] {$b_3\in W_3,\, j_1\in W_1$ \\ $b_4\in W_4,\, j_3\in W_3$};
	\end{scope}
	\begin{scope}[xshift=5.5cm]
		\draw [thick,blue,dir,dotted] (0,0.75) -- (0,0);
			\draw [thick,blue,wavy] (0,1.5) -- (0,0.75);
	\draw [thick,red,dir,dotted] (0.75,1.1) --(0.35,-.35) ;
	\draw [thick,red,wavy] (0.75,1.1) -- (1.5,1.1);
	\filldraw [thick] (0,0) circle (1pt) node[below=0pt]{$j_1$} (0,0.75) circle (1pt) node[left =0pt]{$b_3$}(0,1.5) circle (1pt) node[left=0pt]{$a_3$}
(0.35,-.35) circle (1pt) node[right=0pt]{$l_1$} ;
	\filldraw [thick] (0.75,1.1) circle (1pt) node[above=0pt]{$b_4$} (1.5,1.1) circle (1pt) node[above=0pt]{$a_4$};
	\draw [thick] (-0,-0.35) ++(-110:0.35) arc (-110:-430:0.35);
		\draw [thick] (1.5,-0.35) ++(-110:0.35) arc (-110:-430:0.35);
	\node at (0,-1) {$M_{W_1}$};	
		\node at (1.5,-1) {$M_{W_2}$};	
	\node at (.9,-2) [align=center] {$b_3\in W_3,\, j_1\in W_1$ \\ $b_4\in W_4,\, l_1\in W_1$};	
	\end{scope}
		\begin{scope}[xshift=8.5cm]
		\draw [thick,blue,dir,dotted] (0,0.75) -- (0,0);
			\draw [thick,blue,wavy] (0,1.5) -- (0,0.75);
			\draw [thick,blue,dir,dotted] (1,0.75) -- (1,0);
			\draw [thick,blue,wavy] (1,1.5) -- (1,0.75);
	\filldraw [thick] (0,0) circle (1pt) node[below=0pt]{$j_1$} (0,0.75) circle (1pt) node[left =0pt]{$b_3$}(0,1.5) circle (1pt) node[left=0pt]{$a_3$} ;
		\filldraw [thick] (1,0) circle (1pt) node[below=0pt]{$j_2$} (1,0.75) circle (1pt) node[right =0pt]{$b_4$}(1,1.5) circle (1pt) node[right=0pt]{$a_4$} ;
	\draw [thick] (-0,-0.35) ++(-110:0.35) arc (-110:-430:0.35);
		\draw [thick] (1,-0.35) ++(-110:0.35) arc (-110:-430:0.35);
	\node at (0,-1) {$M_{W_1}$};	
		\node at (1,-1) {$M_{W_2}$};	
	\node at (.9,-2) [align=center] {$b_3\in W_3,\, j_1\in W_1$ \\ $b_4\in W_4,\, j_2\in W_2$};	
	\end{scope}
	\begin{scope}[xshift=12cm,yshift=-.4cm]
\node at (0,0) {$\Bigg(W_1\leftrightarrow W_2\Bigg)$};
		\end{scope}
	\end{tikzpicture}\,\,,
\end{align}
in which we have used the reference order ${\frak R}=W_3\prec W_4$. All the paths are directed towards the roots, and different ones are illustrated by different colors.

\subsubsection{Map\label{subsublabelforestmap}}

 For each $V\in\mathbf{V}_{W_1,W_2,\cdots, W_t}({\frak R})$, the map $\mathcal{C}$ is defined as
\begin{align}\label{Zmap}
\begin{tikzpicture}[baseline={([yshift=0.5ex]current bounding box.center)},every node/.style={font=\footnotesize,},vertex/.style={inner sep=0,minimum size=3pt,circle,fill},wavy/.style={decorate,decoration={coil,aspect=0, segment length=2.2mm, amplitude=0.5mm}},dir/.style={decoration={markings, mark=at position \halfway with {\arrow{Latex[scale=0.9]}}},postaction={decorate}}]
\filldraw (3.5,0) circle (1pt) node[below=0]{$a_i$} (4.5,0) circle (1pt) node[below=0]{$b_i$}
(5.5,0) circle (1pt) node[below=0]{$j$}
;
\draw [thick,wavy] (3.5,0) -- (4.5,0);
\draw [thick,dir,dotted] (4.5,0) -- (5.5,0);
\end{tikzpicture}
\rightarrow\, (-1)^{|B_i|+1}
x_{b_i,j}
{\bm \Omega}_{a_i, A_i\shuffle B_i^T,b_i} \,,
\quad
&\begin{tikzpicture}[baseline={([yshift=0.5ex]current bounding box.center)},every node/.style={font=\footnotesize,},vertex/.style={inner sep=0,minimum size=3pt,circle,fill},wavy/.style={decorate,decoration={coil,aspect=0, segment length=2.2mm, amplitude=0.5mm}}]
	\draw [thick] (-0,-0.35) ++(-110:0.35) arc (-110:-430:0.35);
	\node at (0,-1) {$\MM_{W_i}$};
\end{tikzpicture}\rightarrow\, \MM_{W_i} (\xi_i) \,,
	\end{align}
and the map $\mathcal{C}(V)$ is given by the product of all these  factors.

According to  \eqref{Zmap}, the labeled forests in  \eqref{labeledTreesExample} are evaluated at  \cref{table} under the reference order ${\frak R}=W_3\prec W_4$.
	\begin{table}[!htb]
 \centering
	\begin{tabular}{|c|c|c|c|} \hline
		 & $\vphantom{\Big[}\sum_{V}$ &   $\mathcal{C}(V)/ (M_{W_1} M_{W_2})$ 
		 \\ \hline
		\begin{tikzpicture}[baseline={([yshift=-1.1ex]current bounding box.center)},every node/.style={font=\scriptsize},wavy/.style={decorate,decoration={coil,aspect=0, segment length=2mm, amplitude=0.5mm}},dir/.style={decoration={markings, mark=at position \halfway with {\arrow{Latex[scale=0.9]}}},postaction={decorate}}]
	\draw [thick,blue,wavy] (0,0.75) -- (-0.75,0.75);
		\draw [thick,blue,wavy] (0.75,0.75) -- (1.5,0.75);
	\draw [thick,blue,dir,dotted] (0.75,0.75) -- (0,0.75) ;
	\draw [thick,blue,dir,dotted] (-0.75,0.75) -- (-0.75,0);
	\filldraw [thick] (-0.75,0) circle (1pt) node[below=0pt]{$j_1$} (0,0.75) circle (1pt) node[above=0pt]{$a_4$} (-0.75,0.75) circle (1pt) node[above=0pt]{$b_4$};
	\filldraw [thick] (0.75,0.75) circle (1pt) node[above=0pt]{$b_3$};
		\filldraw [thick] (1.5,0.75) circle (1pt) node[above=0pt]{$a_3$};
	\draw [thick] (-0.75,-0.35) ++(-110:0.35) arc (-110:-430:0.35);
	\node at (-0.75,-1) {$M_{W_1}$};
		\draw [thick] (0.75,-0.35) ++(-110:0.35) arc (-110:-430:0.35);
		\node at (0.75,-1) {$M_{W_2}$};
		\end{tikzpicture} &

    $\displaystyle
         \sum_{b_3\in W_3}\sum_{a_4,b_4\in W_4}\sum_{j_1\in W_1}$ & $
             \begin{aligned}
         \displaystyle \vphantom{\sum_T}
		(-1)^{|B_3|+|B_4|}
  x_{b_3,a_4}
  {\bm \Omega}_{a_3, A_3\shuffle B_3^T, b_3} 
  \\
\times  
x_{b_4,j_1} 
{\bm \Omega}_{a_4, A_4\shuffle B_4^T, b_4}   
    \end{aligned} $

  \\ \hline
		\begin{tikzpicture}[baseline={([yshift=-0.ex]current bounding box.center)},every node/.style={font=\scriptsize},wavy/.style={decorate,decoration={coil,aspect=0, segment length=2mm, amplitude=0.5mm}},dir/.style={decoration={markings, mark=at position \halfway with {\arrow{Latex[scale=0.9]}}},postaction={decorate}},scale=0.8]
	\draw [thick,blue,dir,dotted] (0,0.75) -- (0,0);
			\draw [thick,blue,wavy] (0,1.5) -- (0,0.75);
	\draw [thick,red,dir,dotted] (0.75,1.1) -- (0,1.1);
	\draw [thick,red,wavy] (0.75,1.1) -- (1.5,1.1);
	\filldraw [thick] (0,0) circle (1pt) node[below=0pt]{$j_1$} (0,0.75) circle (1pt) node[left =0pt]{$b_3$}(0,1.5) circle (1pt) node[left=0pt]{$a_3$}
	(0,1.1) circle (1pt) node[left=0pt]{$j_3$};
	\filldraw [thick] (0.75,1.1) circle (1pt) node[above=0pt]{$b_4$} (1.5,1.1) circle (1pt) node[above=0pt]{$a_4$};
	\draw [thick] (-0,-0.35) ++(-110:0.35) arc (-110:-430:0.35);
		\draw [thick] (1.5,-0.35) ++(-110:0.35) arc (-110:-430:0.35);
	\node at (0,-1) {$M_{W_1}$};	
		\node at (1.5,-1) {$M_{W_2}$};	
		\end{tikzpicture} 
  & $\displaystyle\sum_{b_3\in W_3}\sum_{j_1\in W_1}\sum_{b_4\in W_4}\sum_{j_3\in W_3}$ &
  
  $\displaystyle
  \begin{aligned}
  \vphantom{\sum_T}
		(-1)^{|B_3|+|B_4|}
  x_{b_3,j_1}
  {\bm \Omega}_{a_3, A_3\shuffle B_3^T, b_3} 
  \\ \times 
x_{b_4,j_3}   {\bm \Omega}_{a_4, A_4\shuffle B_4^T, b_4} 
  \end{aligned}
		$ 
			 \\ \hline
		\begin{tikzpicture}[baseline={([yshift=-0.ex]current bounding box.center)},every node/.style={font=\scriptsize},wavy/.style={decorate,decoration={coil,aspect=0, segment length=2mm, amplitude=0.5mm}},dir/.style={decoration={markings, mark=at position \halfway with {\arrow{Latex[scale=0.9]}}},postaction={decorate}},scale=0.8] 
			\draw [thick,blue,dir,dotted] (0,0.75) -- (0,0);
			\draw [thick,blue,wavy] (0,1.5) -- (0,0.75);
	\draw [thick,red,dir,dotted] (0.75,1.1) --(0.35,-.35) ;
	\draw [thick,red,wavy] (0.75,1.1) -- (1.5,1.1);
	\filldraw [thick] (0,0) circle (1pt) node[below=0pt]{$j_1$} (0,0.75) circle (1pt) node[left =0pt]{$b_3$}(0,1.5) circle (1pt) node[left=0pt]{$a_3$}
(0.35,-.35) circle (1pt) node[right=0pt]{$l_1$} ;
	\filldraw [thick] (0.75,1.1) circle (1pt) node[above=0pt]{$b_4$} (1.5,1.1) circle (1pt) node[above=0pt]{$a_4$};
	\draw [thick] (-0,-0.35) ++(-110:0.35) arc (-110:-430:0.35);
		\draw [thick] (1.5,-0.35) ++(-110:0.35) arc (-110:-430:0.35);
	\node at (0,-1) {$M_{W_1}$};	
		\node at (1.5,-1) {$M_{W_2}$};	
		\end{tikzpicture} & $\displaystyle\sum_{b_3\in W_3}\sum_{b_4\in W_4}\sum_{j_1,l_1\in W_1}$ &  $\displaystyle
  \begin{aligned}
  \vphantom{\sum_T}
		(-1)^{|B_3|+|B_4|}
  x_{b_3,j_1}
  {\bm \Omega}_{a_3, A_3\shuffle B_3^T, b_3} 
  \\ 
  \times 
x_{b_4,l_1} 
{\bm \Omega}_{a_4, A_4\shuffle B_4^T, b_4} 
  \end{aligned}
		$ \\ \hline
		\begin{tikzpicture}[baseline={([yshift=-0.ex]current bounding box.center)},every node/.style={font=\scriptsize},wavy/.style={decorate,decoration={coil,aspect=0, segment length=2mm, amplitude=0.5mm}},dir/.style={decoration={markings, mark=at position \halfway with {\arrow{Latex[scale=0.9]}}},postaction={decorate}},scale=0.8] 
		\draw [thick,blue,dir,dotted] (0,0.75) -- (0,0);
			\draw [thick,blue,wavy] (0,1.5) -- (0,0.75);
			\draw [thick,blue,dir,dotted] (1,0.75) -- (1,0);
			\draw [thick,blue,wavy] (1,1.5) -- (1,0.75);
	\filldraw [thick] (0,0) circle (1pt) node[below=0pt]{$j_1$} (0,0.75) circle (1pt) node[left =0pt]{$b_3$}(0,1.5) circle (1pt) node[left=0pt]{$a_3$} ;
		\filldraw [thick] (1,0) circle (1pt) node[below=0pt]{$j_2$} (1,0.75) circle (1pt) node[right =0pt]{$b_4$}(1,1.5) circle (1pt) node[right=0pt]{$a_4$} ;
	\draw [thick] (-0,-0.35) ++(-110:0.35) arc (-110:-430:0.35);
		\draw [thick] (1,-0.35) ++(-110:0.35) arc (-110:-430:0.35);
	\node at (0,-1) {$M_{W_1}$};	
		\node at (1,-1) {$M_{W_2}$};	
		\end{tikzpicture} & $\displaystyle\sum_{b_3\in W_3}\sum_{j_1\in W_1}\sum_{b_4\in W_4}\sum_{j_2\in W_2}$ &  $
  \displaystyle
  \begin{aligned}
  \vphantom{\sum_T}
		(-1)^{|B_3|+|B_4|}
  x_{b_3,j_1}
  {\bm \Omega}_{a_3, A_3\shuffle B_3^T, b_3} 
  \\ 
  \times 
x_{b_4,j_2} 
{\bm \Omega}_{a_4, A_4\shuffle B_4^T, b_4} 
  \end{aligned}
		$ \\ \hline
\multicolumn{3}{|c|}{$
\displaystyle\vphantom{\Bigg(W_1\leftrightarrow W_2\Bigg)
}\Big(W_1\leftrightarrow W_2\Big)$}
			 \\ \hline
	\end{tabular}\,
	    \caption{Evaluation of labeled forest $\mathbf{V}_{W_1,W_2}(W_{3}\prec W_{4}) $}
	    \label{table}
	\end{table}
$\mathbf{V}_{W_1, W_2}(W_{3}\prec W_{4})$ is calculated by summing the eight rows directly. Notably, the $a_3\in W_3$ in both the second and third rows remains consistent and is not subject to summation. Opting for a different $a_3$ yields an equivalent $\mathbf{V}_{W_1,W_2}(W_{3}\prec W_{4})$ when F-IBP is applied, highlighting a redundancy in the generating functions of string integrands. Additionally, selecting ${\frak R}=W_4\prec W_3$ would lead to alterations in the first two categories of labeled forests presented in \cref{table},
\begin{subequations}
\begin{align}
	& 
	\begin{tikzpicture}[baseline={([yshift=-0.ex]current bounding box.center)},every node/.style={font=\scriptsize},wavy/.style={decorate,decoration={coil,aspect=0, segment length=2mm, amplitude=0.5mm}},dir/.style={decoration={markings, mark=at position \halfway with {\arrow{Latex[scale=0.9]}}},postaction={decorate}},scale=0.8]
	\draw [thick,blue,dir,dotted] (0,0.75) -- (0,0);
			\draw [thick,blue,wavy] (0,1.5) -- (0,0.75);
	\draw [thick,red,dir,dotted] (0.75,1.1) -- (0,1.1);
	\draw [thick,red,wavy] (0.75,1.1) -- (1.5,1.1);
	\filldraw [thick] (0,0) circle (1pt) node[below=0pt]{$j_1$} (0,0.75) circle (1pt) node[left =0pt]{$b_4$}(0,1.5) circle (1pt) node[left=0pt]{$a_4$}
	(0,1.1) circle (1pt) node[left=0pt]{$j_4$};
	\filldraw [thick] (0.75,1.1) circle (1pt) node[above=0pt]{$b_3$} (1.5,1.1) circle (1pt) node[above=0pt]{$a_3$};
	\draw [thick] (-0,-0.35) ++(-110:0.35) arc (-110:-430:0.35);
		\draw [thick] (1.5,-0.35) ++(-110:0.35) arc (-110:-430:0.35);
	\node at (0,-1) {$M_{W_1}$};	
		\node at (1.5,-1) {$M_{W_2}$};	
		\end{tikzpicture}
		 & 
		 \!\!\!\!\!\!\!\!\!\! \!\!\!
		 &\longrightarrow & & 
   \begin{aligned}
  \sum_{b_4\in W_4}\sum_{j_1\in W_1}\sum_{b_3\in W_3}\sum_{j_4\in W_4}
		(-1)^{|B_3|+|B_4|}x_{b_4,j_1}
  {\bm \Omega}_{a_4, A_4\shuffle B_4^T, b_4} 
  \\ \times
  x_{b_3,j_4}
  {\bm \Omega}_{a_3, A_3\shuffle B_3^T, b_3}      
   \end{aligned} 
				 \,,  \nonumber 
				 \\
	&
		\begin{tikzpicture}[baseline={([yshift=-1.1ex]current bounding box.center)},every node/.style={font=\scriptsize},wavy/.style={decorate,decoration={coil,aspect=0, segment length=2mm, amplitude=0.5mm}},dir/.style={decoration={markings, mark=at position \halfway with {\arrow{Latex[scale=0.9]}}},postaction={decorate}}]
	\draw [thick,blue,wavy] (0,0.75) -- (-0.75,0.75);
		\draw [thick,blue,wavy] (0.75,0.75) -- (1.5,0.75);
	\draw [thick,blue,dir,dotted] (0.75,0.75) -- (0,0.75) ;
	\draw [thick,blue,dir,dotted] (-0.75,0.75) -- (-0.75,0);
	\filldraw [thick] (-0.75,0) circle (1pt) node[below=0pt]{$j_1$} (0,0.75) circle (1pt) node[above=0pt]{$a_3$} (-0.75,0.75) circle (1pt) node[above=0pt]{$b_3$};
	\filldraw [thick] (0.75,0.75) circle (1pt) node[above=0pt]{$b_4$};
		\filldraw [thick] (1.5,0.75) circle (1pt) node[above=0pt]{$a_4$};
	\draw [thick] (-0.75,-0.35) ++(-110:0.35) arc (-110:-430:0.35);
	\node at (-0.75,-1) {$M_{W_1}$};
		\draw [thick] (0.75,-0.35) ++(-110:0.35) arc (-110:-430:0.35);
		\node at (0.75,-1) {$M_{W_2}$};
		\end{tikzpicture}
		&
		 \!\!\!\!\!\!\!\!\!\! \!\!\!\!
		  &\longrightarrow & &
    \begin{aligned}
  \sum_{b_4\in W_4}\sum_{a_3,b_3\in W_3}\sum_{j_1\in W_1} 
		(-1)^{|B_3|+|B_4|}x_{b_4,a_3} 
  {\bm \Omega}_{a_4, A_4\shuffle B_4^T, b_4} 
  \\ 
 \times x_{b_3,j_1}  {\bm \Omega}_{a_3, A_3\shuffle B_3^T, b_3}       
    \end{aligned}
    \,,
		\non
\end{align}
\end{subequations}
while the third and fourth categories of labeled forests presented in \cref{table}  remain the same. The outcome of $\mathbf{V}_{W_1,W_2}(W_{3}\prec W_{4})$ is equivalent to that of $\mathbf{V}_{W_1,W_2}(W_{4}\prec W_{3})$ under F-IBP, a property that persists across general cases,
\be
 \mathcal{F}_{W_1,W_2,\cdots, W_t}( {\frak R}) \overset{\rm IBP}= \mathcal{F}_{W_1,W_2,\cdots, W_t}( {\frak R}') \,,
 \ee
 with ${\frak R}$ and ${\frak R}'$ two different reference ordering of $W_{t+1},W_{t+2},\cdots, W_r$. Consequently, we define
 \begin{align}\label{treeExpansionnew}
\mathcal{F}_{W_1,W_2,\cdots, W_t}(W_{t+1},W_{t+2},\cdots,W_r):= \mathcal{F}_{W_1,W_2,\cdots, W_t}( {\frak R})\,,
\end{align}
where $ {\frak R}$ represents any reference ordering of $W_{t+1},W_{t+2},\cdots,W_r$.

\subsubsection{Examples}

Here, we provide additional examples of \eqref{treeExpansion} (or, more precisely, \eqref{treeExpansionnew}),
\ba
\mathcal{F}_{W_1,W_2,\cdots, W_r}():=\,\,&\prod_{i=1}^m \MM_{W_i} (\xi_i)\,,
\\
\mathcal{F}_{W_1,W_2,\cdots, W_{r-1}}(W_r):=\,\,&\sum_{p\in W_{1,2,\cdots,r-1}}O^{W_r}_{a_r, p} \prod_{i=1}^{r-1} \MM_{W_i} (\xi_i)\,,
\nl
\mathcal{F}_{W_1,W_2,\cdots, W_{r-2}}(W_{r-1},W_r):=\,\,& \left( \sum_{\substack{ p\in W_{1,2,\cdots,r-2}\\
q\in W_{1,2,\cdots,r-1}
}}
\!\!\!\! 
\!\!\!\! O^{W_{r-1} }_{a_{r-1}, p}  
 O^{W_r}_{a_r, q}  
 +
  \sum_{\substack{ p\in W_{r}\\
q\in W_{1,2,\cdots,r-2}
}}
\!\!\!\! 
\!\!\!\! O^{W_{r-1} }_{a_{r-1}, p}  
 O^{W_r}_{p, q}  
    \right) \prod_{i=1}^{r-2} \MM_{W_i} (\xi_i)\,.
\non
\ea

Concerning $\mathcal{F}_{W_1}(W_2,W_3,\cdots, W_r)$, the labeled forest actually simplifies to a labeled tree (specifically, a genus-one variant of those in \cite{Gao:2017dek, He:2019drm}). However, $\mathcal{F}_{\emptyset}(W_1, W_2, \cdots, W_r)$, which lacks any root, remains undefined. Furthermore, it is worth noting that the labeled forest expansion \eqref{treeExpansion} essentially consists of a combination of products of (genus-one version of) 
 labeled trees. For instance,
\ba
\mathcal{F}_{W_a,W_b}(W_c,W_d)=\,\,&\mathcal{F}_{W_a}(W_c,W_d) \mathcal{F}_{W_b}() +\mathcal{F}_{W_a}(W_c) \mathcal{F}_{W_b}(W_d)
\nl&
+\mathcal{F}_{W_a}(W_d) \mathcal{F}_{W_b}(W_c)  +\mathcal{F}_{W_a}() \mathcal{F}_{W_b}(W_c,W_d) \,.
\ea
More generally, this can be expressed as,
\ba
\mathcal{F}_{ {\bm W}_A, {\bm W}_B }( {\bm W}_H )=\,\,&\sum_{C\sqcup D= H}\mathcal{F}_{ {\bm W}_A  }( {\bm W}_C ) \mathcal{F}_{ {\bm W}_B }( {\bm W}_D )   \,.
\ea
Here, ${\bm W}_A$ denotes a set of cycles $\{ W_a | a\in A\}$ (e.g., ${\bm W}_{1,2}= \{W_1, W_2\}$), while ${ W}_A$ (previously defined in \eqref{threecycleRem}) refers to a single set of punctures (e.g., ${W}_{1,2}= W_1\cup W_2$). By recursively applying the above identities, any labeled forest can be reduced to a combination of labeled trees.

\subsection{Deformed labeled trees for total derivative terms and additional punctures}

When addressing total derivative terms or considering cases with $R \neq \emptyset$, there will be labeled trees planted on the punctures belonging to the set $\hat R$. To manage this, we can hypothetically treat the punctures in $\hat R$ as if they comprise an auxiliary double periodic Kronecker-Eisenstein cycle, denoted as $\CC_{W_{r+1}}(\hat \xi)=\CC_{\hat R}(\hat \xi)$. Utilizing the definition of $\mathcal{F}_{W_{r+1}}$, we introduce a new function, ${\mathcal{G}}_{\hat R}$, while ensuring that the term $\MM_{W_{r+1}} (\hat\xi)$ associated with the assumed cycle is removed in the final step,
\ba
\label{treeExpansionvariant}
{ \mathcal{G}}_{\hat R} (W_1,W_2,\cdots, W_r):=  \mathcal{F}_{W_{r+1}={\hat R}} (W_1,W_2,\cdots, W_r)/{\MM_{W_{r+1}} (\hat\xi)}\,.
\ea  
  For instance,
  \ba
 { \mathcal{G}}_{{\hat R}} (W_1) :=\,\,& \sum_{p\in {\hat R}}O^{W_1}_{a_1, p} \,,
  \nl
  { \mathcal{G}}_{{\hat R}}(W_{1},W_2):=\,\,& \sum_{\substack{ p\in {\hat R}\\
q\in W_{1} \cup {\hat R}
}}
\!\!\!\! O^{W_{1} }_{a_{1}, p}  
 O^{W_2}_{a_2, p}  
 +
 \sum_{\substack{ p\in W_{2}\\
q\in {\hat R}
}}
 O^{W_{1} }_{a_{1}, p}  
 O^{W_2}_{p, q} \,.
  \ea

\subsection{Rewriting formulae for two and three cycles}
  
 Using the labeled forest expansion, we can rewrite previous results for two cycles \eqref{twocycle222} and three cycles \eqref{threecycle333} as
 \ba\label{twocycle2222}
  ( 1+s_{W_1}) ( 1+s_{W_2}) \CC_{W_1}(\xi_1)\CC_{W_2} (\xi_2)
= & 
 \mathcal{F}_{W_1,W_2}()+\mathcal{F}_{W_1}(W_2)+\mathcal{F}_{W_2}(W_1)+ \big\< W_1,W_2\big\>
\non \\ 
 &
 + \mathcal{F}_{W_1}() { \mathcal{G}}_{\hat R}( W_2) + \mathcal{F}_{W_2}() { \mathcal{G}}_{\hat R}( W_1) +  { \mathcal{G}}_{\hat R}( W_1, W_2) 
\,,
\ea 
and 
\ba 
 &(1+s_{W_1}) (1+s_{W_2}) (1+s_{W_3}) \CC_{W_1} (\xi_1)\CC_{W_2}(\xi_2)\CC_{W_3} (\xi_3)
 \\
=\,\, & \mathcal{F}_{W_1,W_2,W_3}()
 +\mathcal{F}_{W_1,W_2}(W_3)+\mathcal{F}_{W_1,W_3}(W_2) 
   +\mathcal{F}_{W_2,W_3}(W_1) 
   \nl
   &  +\mathcal{F}_{W_1}(W_2,W_3) +\mathcal{F}_{W_2}(W_1,W_3) +\mathcal{F}_{W_3}(W_1,W_2)+2 \big\< W_1, W_2, W_3\big\>
     \nl
&+ (1+s_{W_1}) M_{W_1} \big\< W_2, W_3\big\> 
+ (1+s_{W_2}) M_{W_2} \big\< W_3, W_1 \big\>
+ (1+s_{W_3}) M_{W_3} \big\< W_1, W_2\big\>
\nl
&
 + \Big[\Big( \mathcal{F}_{W_1,W_2}() { \mathcal{G}}_{\hat R}( W_3) +  
 \mathcal{F}_{W_1}() { \mathcal{G}}_{\hat R}( W_2,W_3) +  
 \mathcal{F}_{W_1}(W_2) { \mathcal{G}}_{\hat R}( W_3) +  
 \mathcal{F}_{W_1}(W_3) { \mathcal{G}}_{\hat R}( W_2) \Big)
 \nl&
 +{\rm cyc}(W_1,W_2,W_3) \Big] 
+   { \mathcal{G}}_{\hat R}( W_1, W_2, W_3) 
 \,.
\nonumber
\ea

The clear and discernible pattern observed in these instances motivates us to propose a conjecture for the most general scenario in the following section.

 \section{An arbitrary number of cycles 
 \label{sec777}}

In this section, we put forth a comprehensive formula to address the product of any number of double periodic Kronecker-Eisenstein cycles \eqref{cycleproduct}, drawing on the fusion operations introduced in \cref{sec333} and the labeled forests conceptualized in \cref{sec666}, along with its meromorphic counterpart \eqref{cycleproductF} in parallel.

 \subsection{An arbitrary number of doubly periodic cycles}

The labeled forest adeptly characterizes terms devoid of any cycles after dismantling all original cycles. Next, we elucidate the structural pattern of remaining terms containing various cycles, including original cycles $W_i$ from the inverse operation of \eqref{stibp} like \eqref{instibp} and new $f$-$\Omega$ cycles emerging from fusions \eqref{eq:fusionr}. Intriguingly, their behavior mirrors that of standard cycle expansions in determinants, which we shall now explore.

Consider an $r\times r$ matrix $H$ with elements ${\bm h}_{i,j}$. Each permutation $\rho$ of $\{1,2,\ldots,r\}$ corresponds to a product of label-cycles,
\be\label{decom}
\rho \to (I)(J)\cdots (K) \,.
\ee
For example, the permutation $1324$ corresponds to $(1)(23)(4)$. Thus, the determinant of $H$ can be expressed as:
\begin{equation}
\label{matrixexpand}
\det H = \sum_{\rho\in S_r} H_{(I)}H_{(J)}\ldots H_{(K)}\,,
\end{equation}
where $H_{(1)}=h_{11}$, $H_{(12 \ldots r)} =h_{12}h_{23} \ldots h_{r1}$, etc.. For instance,
\be
\det\left(
\begin{array}{cc}
h_{11}&h_{12} \\
h_{21} &h_{22}
\end{array}
\right)=H_{(1)}H_{(2)}-  H_{(12)} = h_{11}h_{22} - h_{12}h_{21}\,.
\ee 

By extending the definition such that $h_{ii}=-\sum_{\substack{j=1\\ j\neq i}}^r h_{i,j}- h_{i,r+1}$, we obtain an alternate expansion of $\det H$ devoid of $H$-cycles, in alignment with the matrix tree theorem \cite{Feng:2012sy}. For  instance,
\be
\det\left(
\begin{array}{cc}
h_{11}&h_{12} \\
h_{21} &h_{22}
\end{array}
\right) = h_{13}h_{23} + h_{12}h_{23} + h_{21}h_{13} 
\,.
\ee 

These insights into cycle expansions and their connection to results free of $H$-cycles set the stage for discussing the product of $\Omega$-cycles. 

Now, we are ready to present our general ansatz, incorporating the labeled forest \eqref{treeExpansion} and its variant \eqref{treeExpansionvariant}, the cycle expansion of a matrix \eqref{matrixexpand}, and the fusion operation \eqref{eq:fusionr}.

\paragraph{Ansatz:}  
 
We introduce a general ansatz to dissect the product of  double periodic Kronecker-Eisenstein cycles \eqref{cycleproduct} as follows
 \ba
\label{ansa}
\prod_{i=1}^r ( 1+s_{W_i} )     \CC_{W_i }  = &\,\, 
-
   \!\!\!\!\!\!
   \sum_{\substack{\rho\in S_m \\ \rho\neq 12\cdots r }}
      \!\!\!\!
    \Psi_{(I)}\Psi_{(J)}\cdots \Psi_{(K)}
+
\!\!\!\!\!
   \sum_{\substack{{\bm W}\subset \{W_1,\cdots W_r\}\\
   {\bm W}\neq \emptyset
    }} 
   \!\!\!\!\!\!\!\!
   \mathcal{F}_{\bm W}(\overline{\bm W})  
   \\
   \non\,\,
   &+
         \!\!\!\!
       \sum_{ \substack{ {\bm W}_A \sqcup 
       {\bm W}_B \sqcup 
       {\bm W}_C 
       \\
       =\{W_1,\cdots W_r\}
       \\
              {\bm W}_A,{\bm W}_C \neq \emptyset
        }} 
                 \!\!\!\!
           \!\!\!\!
        \mathcal{F}_{{\bm W}_A}( {\bm W}_B) 
   { \mathcal{G}} _{\hat R}( {\bm W}_C) +  { \mathcal{G}} _{\hat R}( W_1,\cdots W_r )\,.
\ea
where ${\bm W}$ denotes a non-empty subset of $\{W_1,W_2,\ldots, W_r\}$, and ${\overline {\bm W}}$ its complement. The final two terms translate to a complete Koba-Nielsen derivative when $R$, as defined in \eqref{cycleproduct}, is empty. The initial summation spans all permutations of $\{1,2,\ldots,r\}$, excluding the identity permutation, with each permutation decomposed into label-cycles as in \eqref{decom}. 
We define the length-1 $\Psi$-cycle $\Psi_{(i)}$ as 
\be
\label{psiiii}
\Psi_{(i)}:= (1+s_{W_i}) \CC_{W_i}(\xi_i)\,,
\ee
and longer $\Psi$-cycles via fusion
\be
\Psi_{(i_1,i_2,\cdots,i_{|I|})}:=
-\big\< 
W_{i_1},\,W_{i_2},\,\cdots,W_{i_{|I|}}
\big\>   
\,.
   \ee

   The condition $\rho\neq 12\ldots m$ in \eqref{ansa} ensures that the product $  \Psi_{(I)}\Psi_{(J)}\ldots \Psi_{(K)}$ contains at most $m-1$ factors, implying a decrement in the total number of cycles, both original $\CC_{W_i}$ and new $f$-$\Omega$ cycles, on the right-hand side of \eqref{ansa}. This renders \eqref{ansa} an effective recursion formula, enabling the reduction of a cycle product to a basis within a finite number of steps.  

The exclusion of $\rho\neq 12\ldots r$ is intuitive, as the identity permutation represents the left-hand side of \eqref{ansa}, as indicated by \eqref{psiiii}. Note that the last three terms on the right-hand side are devoid of any cycles, drawing a parallel to the matrix tree theorem.

Two examples of \eqref{ansa}  are given by \eqref{twocycle2222}. Here we give another example of 4 cycles,
\ba
\label{fourcycleinge}
     & (1+s_{W_1}) (1+s_{W_2}) (1+s_{W_3})  (1+s_{W_4})  \CC_{W_1} (\xi_1)\CC_{W_2}(\xi_2)\CC_{W_3} (\xi_3)\CC_{W_4} (\xi_4)
        \\
    =\,\,& \mathcal{F}_{W_1,W_2,W_3,W_4}()
  +\Big[\mathcal{F}_{W_1,W_2}(W_3,W_4)+\Big(12\Big|13,14,23,24,34\Big)\Big]
  \nl
   &+\Big[\mathcal{F}_{W_1,W_2,W_3}(W_4)+\mathcal{F}_{W_1}(W_2,W_3,W_4)+{\rm cyc(W_1,W_2,W_3,W_4)}\Big]
         \nl
&
+ \Big[ (1+s_{W_3})(1+s_{W_4}) \CC_{W_3}(\xi_3)\CC_{W_4}(\xi_4)\big\< W_1,W_2\big\> +\Big(12\Big|13,14,23,24,34 \Big) \Big]
\nl
&+\Big[2 (1+s_{W_4})\CC_{M_4}(\xi_4) \big\< W_1,W_2,W_3\big\> +{\rm cyc}(W_1,W_2,W_3,W_4)\Big]
\nl
&
+\Big[\big\< W_1,W_2,W_3, W_4\big\> +{\rm perm}(W_2,W_3,W_4)\Big]- \Big[\big\< W_1,W_2\big\>\big\< W_3,W_4\big\> +{\rm cyc}(W_2,W_3,W_4)\Big] \,.
\nonumber 
\nl
& + 
     \sum_{ \substack{ {\bm W}_A \sqcup 
       {\bm W}_B \sqcup 
       {\bm W}_C 
       \\
       =\{W_1,W_2,W_3, W_4\}
       \\
              {\bm W}_A, {\bm W}_C \neq \emptyset
        }} 
                 \!\!\!\!
           \!\!\!\!
        \mathcal{F}_{{\bm W}_A}( {\bm W}_B) 
  \mathcal{G} _{\hat R}( {\bm W}_C) +   \mathcal{ G} _{\hat R}( W_1,W_2,W_3, W_4)
 \,.
 \non
\ea
Here the relabeling of cycles $W_i$ of course refers to the process of also relabeling the corresponding $\xi_i$.
We have verified it  for $ \CC_{(12)} (\xi_1) \,\CC_{(34)} (\xi_2)\,\CC_{(56)}  (\xi_3)\,\CC_{(789)} (\xi_4) $ with $n=10$.

\subsection{An arbitrary number of meromorphic cycles}

Our methodology to decompose a product of cycles in a doubly periodic scenario is readily applicable to meromorphic cases within the chiral splitting framework, as outlined in \cref{sec:revCS}, through a straightforward application of substitution \eqref{subsrule}.

Concretely, the expression for formula \eqref{rhorhogf} can be recast as  
\ba
\label{stibpf}
& ( 1+s_{W}) \CCF_{W}(\xi)
=\,\, \MMF_{W} (\xi)+
 \sum_{\substack{ i\notin W\\ 0 \leq i \leq n} } \OOF_{a,i}^W 
 \,,
\\ 
\non
&\quad 
{\rm with}~ \OOF_{a,i}^W:=- \sum_{\substack{
p  \in W \\ p\neq a}
} \sum_{
\substack{\rho \in A \shuffle B^{\rm T}\\
(a, A, p, B )=W
}} 
 (-1)^{|B|} 
 {\tilde x}_{p,i}
{\bm F}_{a,\rho, p}   \,, \qquad a\in W, i\notin W\,, 0 \leq i \leq n\,,
\ea 
where ${\tilde x}_{p,0}$ is introduced for notation compactness to include the total Koba-Nielsen derivative terms and the loop momentum terms,
\be
 {\tilde x}_{p,0}
{\bm F}_{a,\rho, p} := 
(- \ell\!\cdot \! k_p +\tilde \nabla_p) 
{\bm F}_{a,\rho, p}\,.
\ee 
 Example cases include,
\ba
&\OOF_{1,0}^{(12)}= F_{12}(\eta_2) \, \ell\!\cdot\! k_2\, - \tilde \nabla_2 F_{12}(\eta_2)\,,
\qquad \OOF_{1,i}^{(12)}=\,\,- F_{12}(\eta_2) {\tilde x}_{2,i} \quad {\rm for} ~i\geq3,
\\ & 
\OOF_{2,0}^{(12)}= - F_{12}(\eta_2) \, \ell\!\cdot\! k_1\, + \tilde \nabla_1 F_{12}(\eta_2)\,,
\qquad
\OOF_{2,i}^{(12)}= F_{12}(\eta_2) {\tilde x}_{1,i}
\quad {\rm for} ~i\geq3.
\ea

We 
proceed to the application of formula \eqref{stibpf} for a product of meromorphic cycles \eqref{cycleproductF}. First, we derive the result analogous to \eqref{twocycle222} for two cycles
 \begin{align} \label{twocycle222f}
 & ( 1+s_{W_1}) ( 1+s_{W_2}) \CCF_{W_1}(\xi_1)\CCF_{W_2} (\xi_2)
 \\
=\,\,&
\MMF_{W_1}(\xi_1)\MMF_{W_2}(\xi_2)
+\MMF_{W_1} (\xi_1)
\sum_{j_1 \in W_1} \OOF^{W_2}_{a_2,j_1 } +\MMF_{W_2} (\xi_2)
\sum_{j_2 \in W_2} \OOF^{W_1}_{a_1,j_2 } 
 +\<W_1,W_2 \>^{F}
 \nl
&
+\MMF_{W_1} (\xi_1)
\sum_{j_1 \in {\hat R} } \OO^{W_2}_{a_2,j_1 } +\MMF_{W_2} (\xi_2)
\sum_{j_2 \in {\hat R} } \OO^{W_1}_{a_1,j_2 } 
+\sum_{p \in {\hat R} } \OO^{W_1}_{a_1, p } \sum_{q \notin W_2} \OO^{W_2}_{a_2, q }  \nl
& \qquad\qquad\qquad\qquad\qquad\qquad\qquad\qquad\qquad\qquad\qquad\qquad
+ \sum_{j_2 \in W_2} \OO^{W_1}_{a_1, j_2 } \sum_{q \in {\hat R} } \OO^{W_2}_{j_2, q } \,,
\non
 \end{align}
 with $a_1\in W_1, a_2 \in W_2$. The fusion for meromorphic cycles is defined as, 
 \begin{align}
\label{eq:fusion2f}
\langle\, W_1,W_2\,\rangle^F:=\frac{1}{2}
\!\!\!\!\!
\sum_{\substack{a_1,b_1\in W_1,(a_1, A_1, b_1, B_1)=W_1 \\ a_2,b_2\in W_2,(a_2, A_2, b_2, B_2)=W_2}}
\!\!\!\!\!
(-1)^{|B_1|+|B_2|}{\bm F}_{a_1, A_1\shuffle B_1^T,b_1}x_{b_1,a_2}
{\bm F}_{a_2, A_2\shuffle B_2^T,b_2}x_{b_2,a_1}
\,.
\end{align}
 With a slight abuse of notation, we have 
 used the abbreviation
\ba
{\bm F}_{a, A\shuffle B^T,b}:=\sum_{\sigma\in A\shuffle B^T}{\bm F}_{a, \sigma,b}\,.
\ea
The generalization of \eqref{eq:fusion2f} for more meromorphic cycles is straightforward.

To extend this to the product of more meromorphic cycles, we introduce the labeled forest for $F$ functions analogous to \eqref{treeExpansion},
\begin{align}\label{treeExpansionF}
\tilde{ \mathcal{F}}_{W_1,W_2,\cdots, W_t}( {\frak R}):= \sum_{V\in\mathbf{V}_{W_1,W_2,\cdots, W_t}( {\frak R})}\,\mathcal{D}(V)\,.
\end{align}
The function $\mathbf{V}_{W_1,W_2,\cdots, W_t}( {\frak R})$ is the same as the one defined in  \cref{subsublabelforest} and the map ${\cal D}$ differs a little from ${\cal C}$ defined via \eqref{Zmap}. 
For each $V\in\mathbf{V}_{W_1,W_2,\cdots, W_t}({\frak R})$, the map $\mathcal{D}$ is defined as
 For each $V\in\mathbf{V}_{W_1,W_2,\cdots, W_t}({\frak R})$, the map $\mathcal{C}$ is defined as
\begin{align}\label{ZmapF}
\begin{tikzpicture}[baseline={([yshift=0.5ex]current bounding box.center)},every node/.style={font=\footnotesize,},vertex/.style={inner sep=0,minimum size=3pt,circle,fill},wavy/.style={decorate,decoration={coil,aspect=0, segment length=2.2mm, amplitude=0.5mm}},dir/.style={decoration={markings, mark=at position \halfway with {\arrow{Latex[scale=0.9]}}},postaction={decorate}}]
\filldraw (3.5,0) circle (1pt) node[below=0]{$a_i$} (4.5,0) circle (1pt) node[below=0]{$b_i$}
(5.5,0) circle (1pt) node[below=0]{$j$}
;
\draw [thick,wavy] (3.5,0) -- (4.5,0);
\draw [thick,dir,dotted] (4.5,0) -- (5.5,0);
\end{tikzpicture}
\rightarrow\, (-1)^{|B_i|+1}
{\tilde x}_{b_i,j}
{\bm F}_{a_i, A_i\shuffle B_i^T,b_i} \,,
\quad
&\begin{tikzpicture}[baseline={([yshift=0.5ex]current bounding box.center)},every node/.style={font=\footnotesize,},vertex/.style={inner sep=0,minimum size=3pt,circle,fill},wavy/.style={decorate,decoration={coil,aspect=0, segment length=2.2mm, amplitude=0.5mm}}]
	\draw [thick] (-0,-0.35) ++(-110:0.35) arc (-110:-430:0.35);
	\node at (0,-1) {$\MMF_{W_i}$};
\end{tikzpicture}\rightarrow\, \MMF_{W_i}(\xi_i) \,,
	\end{align}
and the map $\mathcal{D}(V)$ is given by the product of all these factors.

Different choices of reference orderings lead to the same ${\cal G}$ on the support of F-IBP.  Hence we define 
 \begin{align}\label{treeExpansionFF}
\tilde{ \mathcal{F}}_{W_1,W_2,\cdots, W_t}(W_{t+1},W_{t+2},\cdots,W_r):=\tilde{ \mathcal{F}}_{W_1,W_2,\cdots, W_t}( {\frak R})
\end{align}
analogous to \eqref{treeExpansionnew}, 
where $ {\frak R}$ could be any reference ordering of $W_{t+1},W_{t+2},\cdots,W_r$.

$\tilde{ \mathcal{F}}_{W_1,W_2,\cdots, W_t}(W_{t+1},W_{t+2},\cdots,W_r)$ is free of $\hat R$ and hence free of loop momentum.  All loop momentum dependence can be elegantly encapsulated in
\ba
\tilde{ \mathcal{G}}_{\hat R} (W_1,W_2,\cdots, W_r):=   \tilde{ \mathcal{F}}_{W_{r+1}= \hat R} (W_1,W_2,\cdots, W_r) /{\MMF_{W_{r+1}}(\hat\xi)} \,,
\ea  
analogous to \eqref{treeExpansionvariant}.

  \paragraph{Ansatz:}   
We propose a comprehensive ansatz to decompose a product of meromorphic  Kronecker-Eisenstein cycles \eqref{cycleproductF}, expressed as
   \ba\label{ansaF}
\prod_{i=1}^r ( 1+s_{W_i} )     \CCF_{W_i }   = & \,\,
-
   \!\!\!\!\!\!
   \sum_{\substack{\rho\in S_r \\ \rho\neq 12\cdots r }}
      \!\!\!\!
   \tilde \Psi_{(I)}\tilde \Psi_{(J)}\cdots \tilde \Psi_{(K)}
    +\!\!\!\!\!
   \sum_{\substack{{\bm W}\subset \{W_1,\cdots W_r\}\\
   {\bm W}\neq \emptyset
    }} 
   \!\!\!\!\!\!\!\!
 \tilde{ \mathcal{F}}_{\bm W}(\overline{\bm W})
 \\
 \non &   +
         \!\!\!\!
       \sum_{ \substack{ {\bm W}_A \sqcup 
       {\bm W}_B \sqcup 
       {\bm W}_C 
       \\
       =\{W_1,\cdots W_r\}
       \\
              {\bm W}_A, {\bm W}_C \neq \emptyset
        }} 
                 \!\!\!\!
           \!\!\!\!
     \tilde{ \mathcal{F}}_{{\bm W}_A}( {\bm W}_B) 
     \tilde{ \mathcal{G}} _{\hat R}( {\bm W}_C) + \tilde{ \mathcal{G}} _{\hat R}(W_1,\cdots W_r )
     \,,
\ea
where the length-1 $\tilde \Psi$-cycle $\tilde \Psi_{(i)}$ is defined as 
\be
\tilde \Psi_{(i)}:= (1+s_{W_i}) \CCF_{W_i}(\xi_i)\,,
\ee
and longer one is defined through a fusion
\be
\tilde \Psi_{(i_1,i_2,\cdots,i_{|I|})}:=
-\big\< 
W_{i_1},\,W_{i_2},\,\cdots,W_{i_{|I|}}
\big\>   ^F
\,.
   \ee
These definitions facilitate a structured and efficient decomposition of the product of meromorphic cycles.

\section{{\tt Mathematica} code and more examples \label{seccode}}

\begin{figure}
\centering
\includegraphics[scale=.37]{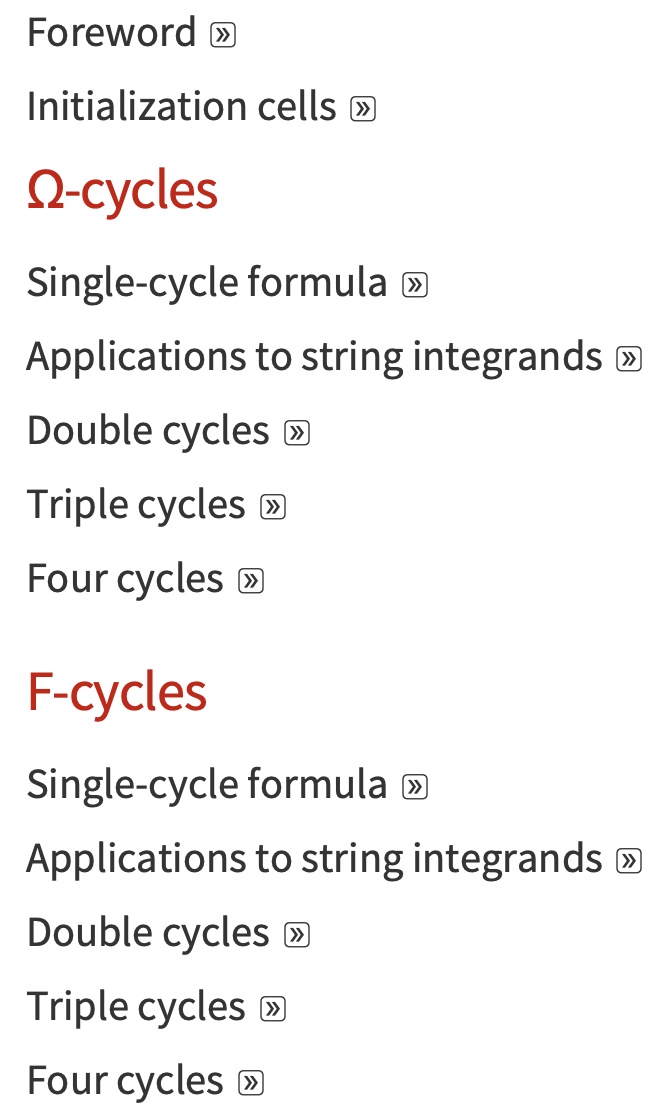}
\caption{
\label{contents} The organizational structure of the notebook. Following the execution of initialization cells, users can navigate to specific sections or subsections tailored to their specific objectives.
}
\end{figure}

We have successfully implemented formulae designed for breaking single Kronecker-Eisenstein cycles, addressing the intricacies of both doubly-periodic $\Omega$-cycles \eqref{rhorhog} and meromorphic $F$-cycles \eqref{rhorhogf}. Moreover, our implementation extends to efficiently handle products of two, three, or four cycles. To further illustrate the practical utility of these formulae, we have assembled a collection of numerous examples and demonstrated their applications to string integrands by generating identities for Kronecker-Eisenstein series coefficients. This comprehensive set of computational resources is conveniently packaged in an accompanying {\tt Mathematica} notebook titled {\texttt{breakingcycles.nb}}.
While some examples derived from our formulae \eqref{ansa} and \eqref{ansaF} are too intricate to be included directly in the paper due to their complexity, they are readily available for exploration in the notebook. The organizational structure of the notebook is detailed in \cref{contents}, where the outcomes for $\Omega$-cycles and $F$-cycles are thoughtfully presented in two parallel sections.

The codes in the notebook were made as transparent and intuitive as possible.
In doing so, we have named variables to closely mirror their representations in the paper. 
For instance, entering f[1][2,3] will yield the output $f_{2,3}^{(1)}$ (or Subsuperscript[f,"2,3", "(1)"] in full form in {\tt Mathematica}). 
Notably, the system recognizes $f_{2,3}^{(1)}$ and f[1][2,3] as synonymous, a fact verifiable using the command {\texttt{FullForm}} in {\tt Mathematica} as shown below,
\ba 
\non
{\color {blue} {\texttt {In[1]:=}} }& \,\texttt{f[1][2,3]}
\\
\non
{\color {blue} {\texttt {Out[1]=}}}& \,f_{2,3}^{(1)}
\\
\non
{\color {blue} {\texttt {In[2]:=}} }& \,\,f_{2,3}^{(1)}// {\texttt {FullForm}}  
\\
\non
{\color {blue} {\texttt {Out[2]=}}}& \,
\texttt{f[1][2,3]}
\ea 

In our approach, we interpret the equations in the paper as distinct replacement rules. 
Consequently, certain functions have been defined with the suffix ``{\texttt {-repl}}''. 
As an example, to dissect a Kronecker-Eisenstein series $\Omega$-cycle, denoted as $\CC_{(1,2)}(\xi_1)$, the command ${\texttt{Cboldrepl}}$ is employed according to \eqref{rhorhog}, \footnote{In {\tt {Mathematica}}, one can either type ${\texttt{Cboldrepl}}_{1,2}\texttt{[}\xi_1\texttt{]}$ using shortcuts for subscripts or directly type its full form, ${\texttt{Subscript[Cbold, 1, 2][Subscript[}}\xi {\texttt{, 1]]}}$. } \footnote{The command 
${\texttt{Cboldrepla}}_{i,j,\cdots,k}\texttt{[}\xi\texttt{]}$ is designed to automatically set $a=i$ when applying \eqref{rhorhog}. 
For situations requiring different choices of $a$, one should opt for using {\texttt{Cboldrepla}} directly, with an example being ${\texttt{Cboldrepla}_{3,4}\texttt{[}\xi_2\texttt{][4]}}$ to specify an alternative value for $a$.}
\ba 
\non
{\color {blue} {\texttt {In[3]:=}} }& \,{\texttt{Cbold}}_{1,2}\texttt{[}\xi_1\texttt{]}
\\
\non
{\color {blue} {\texttt {Out[3]=}}}& \,
{\CC}_{1,2}[\xi_1]
\\
\non
{\color {blue} {\texttt {In[4]:=}} }& \,{\texttt{Cboldrepl}}_{1,2}\texttt{[}\xi_1\texttt{]}
\\
\non
{\color {blue} {\texttt {Out[4]=}}}& \,
\frac{\MM_{1,2}[\xi
  _1]}{s_{1,2}+1}-\frac{\sum
  _i^{\text{holdComplement}[\text{punctureset},\{1,2\}]} {\bm \Omega} _{1,2}
  x_{2,i}+\nabla_2[ {\bm \Omega}
  _{1,2}]}{s_{1,2}+1}
\ea 
where ``punctureset'' represents the set of all punctures. Once assigned a value, for example, ${\text{punctureset}} =\{1,2,3,4,5\}$, one can replace {\texttt{holdComplement}} with {\texttt{Complement}} to obtain an explicit result,
\ba 
\non
{\color {blue} {\texttt {In[5]:=}} }& \,{\texttt{punctureset = Range[5]; }}
\\
\non
{\color {blue} {\texttt {In[6]:=}} }& \,{\texttt{Cboldrepl}}_{1,2}\texttt{[}\xi_1\texttt{]/.holdComplement->Complement}
\\
\non
{\color {blue} {\texttt {Out[6]=}}}& \,
\frac{\MM_{1,2}[ \xi
  _1]}{s_{1,2}+1}-\frac{ {\bm \Omega} _{1,2}
  x_{2,3}+{\bm \Omega} _{1,2}
  x_{2,4}+{\bm \Omega} _{1,2}
  x_{2,5}+\nabla_2[ {\bm \Omega}
  _{1,2}]}{s_{1,2}+1}
\ea 
Similarly, one can utilize {\texttt {Mblodrepl}}
to get the explicit expression of $\MM_{1,2}[\xi_1]$ according to \eqref{rhorhoMM},
\ba 
\non
{\color {blue} {\texttt {In[7]:=}} }& \,
{\texttt{Mbold}}_{1,2}\texttt{[}\xi_1\texttt{]}
\\
\non
{\color {blue} {\texttt {Out[7]=}}}& \,
{\MM}_{1,2}[\xi_1]
\\
\non
{\color {blue} {\texttt {In[8]:=}} }& \,
{\texttt{Mboldrepl}}_{1,2}\texttt{[}\xi_1\texttt{]}
\\
\non
{\color {blue} {\texttt {Out[8]=}}}& \,
{\bm \Omega} _{1,2} \left((s_{1,2}+1)
  v_1[\eta _2,\xi]-{\hat g}^{(1)}[\eta _2]\right)+s_{1,2}
  \partial_{\eta _2}[{\bm \Omega}
  _{1,2}]
\ea 

As elucidated in the companion paper \cite{Rodriguez:2023qir}, obtaining identities for Kronecker-Eisenstein coefficients $g^{(w)}, f^{(w)}$ defined by \eqref{1.1b} from the identities of their generating functions involves extracting the coefficients of bookkeeping variables $\eta_i$ and $\xi_j$ in a specific order, as illustrated in \eqref{vopen}. To facilitate this process, we have introduced the command {\texttt {OrderedCoefficient}}. For instance, to extract $\MM_{1,2}(\xi_1)\big|\big|_{\eta_2^0,\xi_1^0} := \big( \MM_{1,2}(\xi_1)\big|_{\eta_2^0} \big) \big|_{\xi_1^0}$, the command is executed as follows, \footnote{The command in fact consists of three parameters: \texttt{OrderedCoefficient[generatingFunction\_,list\_, truncate\_:6]}. Here, the third parameter, \texttt{truncate}, has a default value of 6. This parameter is instrumental in controlling the expansion in equation \eqref{1.1b}. Typically, the default setting of \texttt{truncate=6} suffices for examining cases where $n\leq6$. However, it is possible to specify a larger value for \texttt{truncate} to accommodate more complex analyses or studies requiring a broader expansion.} 
\ba 
\non
{\color {blue} {\texttt {In[9]:=}} }&  \,{\texttt{OrderedCoefficient[}}  {\texttt {Mbold}}_{1,2} , {\texttt  {\{}} \eta_2, \xi_1{\texttt {\}] }}   
\\
\non
{\color {blue} {\texttt {Out[9]=}}}& \,
2 s_{1,2} f_{1,2}^{(2)}+{\hat {\rm G}}_2
\ea 
where ${\hat {\rm G}}_2$ is defined in \eqref{G2hat}. 
On the other hand, $\CC_{(1,2)} [\xi_1] \big|\big|_{\eta_2^0,\xi_1^0} = V_2(1,2) =  2 f^{(2)}_{1,2} +  f^{(1)}_{1,2}f^{(1)}_{2,1} $.    Hence the code 
 successfully reproduces the elementary observation $V_2(1,2) \cong 2 s_{1,2} f_{1,2}^{(2)}+{\hat {\rm G}}_2$ within the F-IBP support for the case $n=2$.

The primary focus of this paper is to break a product of cycles. For this purpose, we have developed the command {\texttt{BreakingOmegaCycles}} to break a product of up to four Kronecker-Eisenstein series $\Omega$-cycles. For instance, to break $\CC_{(1,2)}(\xi_1)\CC_{(3,4)}(\xi_2)$ with $n=5$, the code is executed as follows,
\ba 
\non
{\color {blue} {\texttt {In[10]:=}} }&  \,{\texttt{BreakingOmegaCycles[\{1, 2\}, \{3, 4\}, Range[5]]  }}   
\\
\non
{\color {blue} {\texttt {Out[10]=}}}& \,
\frac{x_{1,3} x_{2,4} \bm\Omega _{1,2} \bm\Omega _{3,4}\!-\!x_{1,4} x_{2,3}
  \bm \Omega _{1,2} \bm\Omega _{3,4}}{(s_{1,2}+1)
   (s_{3,4}+1)}\!-\!\frac{x_{1,4} x_{2,5} \bm\Omega _{1,2}
   \bm\Omega _{3,4}}{(s_{1,2}+1) (s_{3,4}+1)}\!+\cdots-\!\frac{\nabla_4[\bm\Omega _{3,4} \MM_{1,2}[\xi
   _1]]}{(s_{1,2}+1) (s_{3,4}+1)}
\ea 
Here, the last entry, \texttt{Range[5]}, in {\texttt {In[10]}}    is used to denote the set of all punctures. It is noteworthy that the function {\texttt{BreakingOmegaCycles}} automatically assigns the two cycles with bookkeeping variables $\xi_1$ and $\xi_2$, eliminating the need to specify them in {\texttt {In[10]}}.

Utilizing the command {\texttt{OrderedCoefficient}}, we can derive the formula to break $V_2(1,2)V_2(3,4)$,
\ba 
\non
{\color {blue} {\texttt {In[11]:=}} }&  \,
{\texttt{OrderedCoefficient[}}  {\texttt {Out[10]}}  , {\texttt  {\{}} \eta_2,\eta_4, \xi_1,\xi_2{\texttt {\}] }}   
\\
\non
{\color {blue} {\texttt {Out[11]=}}}& \,
\frac{
   {\hat {\rm G}}_2^2-{\hat {\rm G}}_2 s_{2,3}
   f^{(1)}_{1,2} f^{(1)}_{2,3}
 +  \cdots
   -\nabla_4\left[f^{(1)}_{3,4} (2 s_{1,2}
   f^{(2)}_{1,2}+{\hat {\rm G}}_2)\right]
   }{\left(s_{1,2}+1\right)
   \left(s_{3,4}+1\right)}
\ea 

Simply running the command
\ba
\non
&{\texttt{BreakingOmegaCycles[\{1,2,3\},\{4,5,6\},Range[6]]}}, 
\\
\non
 &{\texttt{BreakingOmegaCycles}}    {\texttt{[\{1,2\},\{3,4\},\{5,6\}, Range[6]]}},
 \cdots
 \non
 \ea
allows one to recover all the double and triple cycle examples presented in the companion paper \cite{Rodriguez:2023qir}. The command also works seamlessly for four cycles, requiring only a few seconds even for an example with $n=10$,
  \ba
\non
 &{\texttt{BreakingOmegaCycles}}    {\texttt{[\{1,2\},\{3,4\},\{5,6\},\{7,8,9\},Range[10]]}} \qquad \,.
 \non
 \ea
 We have verified that the output of this command aligns perfectly with the formula \eqref{fourcycleinge}. This showcases the efficiency and accuracy of the command in handling various scenarios, even for relatively large values of $n$.
 
For a product of meromorphic Kronecker-Eisenstein $F$-cycles, the appropriate command to use is  then {\texttt{BreakingFCycles}} instead. For example,
\ba 
\non
{\color {blue} {\texttt {In[12]:=}} }&  \,{\texttt{BreakingFCycles[\{1, 2\}, \{3, 4\},\{5, 6\}, Range[7]]  }}   
\\
\non
{\color {blue} {\texttt {Out[12]=}}}& \,
-\frac{{\bm F}_{1,2} {\tilde x}_{2,5} {\tilde\MM}_{3,4}[\xi
   _2] {\tilde\MM}_{5,6}[\xi _3]}{(s_{1,2}+1) (s_{3,4}+1)
   (s_{5,6}+1)} +\frac{{\bm F}_{1,2} {\bm F}_{3,4} {\bm F}_{5,6} \ell \!\cdot\! k_2
   \ell \!\cdot\! k_4 \ell \!\cdot\! k_6}{(s_{1,2}+1) (s_{3,4}+1)
   (s_{5,6}+1)}
   \\
   \non &\qquad\qquad\qquad\qquad\qquad\qquad\qquad\qquad+ \cdots - \frac{\ell \!\cdot\! k_2 {\tilde\nabla}_4[{\bm F}_{1,2} {\bm F}_{3,4} {\bm F}_{5,6}
   {\tilde x}_{1,6}]}{(s_{1,2}+1) (s_{3,4}+1)
 (s_{5,6}+1)}
\ea
Similarly, to obtain identities for $g^{(w)}_{i,j}$, the command {\texttt{OrderedCoefficient2}} is employed.

By leveraging these commands, one can effortlessly generate numerous examples that demonstrate the utility of our formulae \eqref{ansa} and \eqref{ansaF}, along with their applications to Kronecker-Eisenstein series coefficients. Further explanations and examples are available in the accompanying notebook.

\section{Tadpoles, multibranch and connected multiloop graphs}
\label{multisec}

In the preceding sections, we tackled the scenario involving a product of isolated cycles, culminating in a closed-form expression. However, it is conceivable that more intricate configurations, beyond isolated cycles, may emerge in the generating functions of string integrands.

To unravel the complexities of a monomial in $\Omega(z_{ij},\beta_k, \tau)$ or its non-trivial coefficient $f^{w>0}_{ij}$, a graphical representation proves invaluable. In this representation, each instance of $\Omega(z_{ij},\beta_k, \tau)$ or $f^{w>0}_{ij}$ is depicted as a substantial edge linking nodes $z_i$ and $z_j$. For the cases involving $\Omega(z_{ij},\beta_k, \tau)$, we can further embellish the edges with the bookkeeping variables $\beta_k$. Although we opt to temporarily set aside this additional notation for the sake of clarity, the sketch still manages to capture the essence of the properties of monomials.

\begin{figure}[h]
\centering
  \subfloat[isolated cycle]{\begin{tikzpicture}[line width = 0.5mm,scale=1.2]
    \draw (0,0) circle (0.5);
    \filldraw (120:0.5) circle (1pt) node[below right=-1.5pt]{$z_1$} (-120:0.5) circle (1pt) node[above right=-1.5pt]{$z_2$} (0.5,0) circle (1pt) node[left=0pt]{$z_3$}
     ;
    \node at (-120:1) [left=0pt]{$\phantom{7}$};
    \node at (120:1) [left=0pt]{$\phantom{7}$};
    \node at (0,1) [left=0pt]{$\phantom{8}$};
        \node at (1.0,0) [left=0pt]{$\phantom{8}$};
    {$\phantom{8}$};
        \node at (-1.0,0) [left=0pt]{$\phantom{8}$};
    \end{tikzpicture}
    }
    \qquad
    \subfloat[tadpole]{\begin{tikzpicture}[line width = 0.5mm,scale=1.2]
    \draw   (0,0) circle (0.5);
    \draw (0.5,0) -- (1,0);
    \filldraw (120:0.5) circle (1pt) node[below right=-1.5pt]{$z_1$} (-120:0.5) circle (1pt) node[above right=-1.5pt]{$z_2$} (0.5,0) circle (1pt) node[left=0pt]{$z_3$} (1,0) circle (1pt) node[below =0pt]{$z_4$};
    \path (1,0) -- (0.5,0);
    \draw  (1.5,0) -- (1,0);
    \filldraw (1.5,0) circle (1pt) node[below=0pt]{$z_5$} 
    ;
    \node at (-120:1) [left=0pt]{$\phantom{7}$};
    \node at (120:1) [left=0pt]{$\phantom{7}$};
    \node at (0,1) [left=0pt]{$\phantom{8}$};
    \end{tikzpicture}
    }
     \qquad
    \subfloat[multibranch graph]{\begin{tikzpicture} [line width = 0.5mm,scale=1.2]
    \draw (0,0) circle (0.5);
    \filldraw (150:0.5) circle (1pt) node[below right=-1.5pt]{$z_3$} -- (150:1) circle (1pt) node[above=0pt]{$z_5$} (30:0.5) circle (1pt) node[below left=-1.5pt]{$z_1$} -- (30:1) circle (1pt) node[above=0pt]{$z_4$} (0,-0.5) circle (1pt) node[above=0pt]{$z_2$} 
    ;
    \path  (150:1) -- (150:0.5);
    \path (30:1) -- (30:0.5);
    \node at (1.3,0) {};
    \node at (-1.3,0) {};
    \node at (0,1) [left=0pt]{$\phantom{z_8}$};
    \node at (-120:1) [left=0pt]{$\phantom{z_7}$};
    \end{tikzpicture}
    }
    \qquad
    \subfloat[A product of multibranch and tree graphs  ]{\begin{tikzpicture}[line width = 0.5mm,scale=1.2]
    \draw  (0,0) circle (0.5);
    \draw (120:0.5) -- ++(120:0.5) (-120:0.5) -- ++(-120:0.5);
    \filldraw (120:0.5) circle (1pt) node[below right=-1.5pt]{$z_1$} (0:0.5) circle (1pt) node[left=0pt]{$z_3$} (-120:0.5) circle (1pt) node[above right=-1.5pt]{$z_2$} (120:1) circle (1pt) node[left=0pt]{$z_6$} (-120:1) circle (1pt) node[left=0pt]{$z_7$};
    \draw  (1.25,0) circle (0.5);
    \filldraw (1.25,-0.5) circle (1pt) node[above=0pt]{$z_5$} (1.25,0.5) circle (1pt) node[below=0pt]{$z_4$} -- ++(0,0.5) circle (1pt) node[left=0pt]{$z_8$};
    \filldraw (2,-0.5) circle (1pt) node[below=0pt]{$z_{10}$} -- (2,0.5) circle (1pt) node[above=0pt]{$z_9$};
    \path  (2,0.5) -- (2,-0.5);
    \path  (120:1) -- (120:0.5);
    \path  (-120:1) -- (-120:0.5);
    \path (1.25,1) -- (1.25,0.5);
   \begin{scope}[xshift=3cm]
           \draw (0,0) circle (0.5);
    \filldraw (120:0.5) circle (1pt) node[above left=-1.5pt]{$z_{11}$} (-120:0.5) circle (1pt) node[below left=-1.5pt]{$z_{12}$} (0.5,0) circle (1pt) node[right=0pt]{$z_{13}$}
     ;
    \node at (-120:1) [left=0pt]{$\phantom{7}$};
    \node at (120:1) [left=0pt]{$\phantom{7}$};
    \node at (0,1) [left=0pt]{$\phantom{8}$};
        \node at (1.0,0) [left=0pt]{$\phantom{8}$};
    {$\phantom{8}$};
        \node at (-1.0,0) [left=0pt]{$\phantom{8}$};
   \end{scope}
    \end{tikzpicture}
    }
    \qquad
      \subfloat[connected multiloop graph]{
\begin{tikzpicture}[line width = 0.5mm,scale=1.2,yshift=-1cm]
    \draw (0,0) circle (0.5);
    \filldraw 
    (0:0.5) circle (1pt) node[right =-.5pt]{$z_2$} 
    (+180:0.5) circle (1pt) node[left =-.5pt]{$z_1$};
    \draw (-180:0.5)--(0:0.5); 
    \node at (-180:1) [left=0pt]{$\phantom{7}$};
    \node at (0,1) [left=0pt]{$\phantom{8}$};
     \node at (0,-.9) [left=0pt]{$\phantom{8}$};
        \node at (1.6,0) [left=0pt]{$\phantom{8}$};
    {$\phantom{8}$};
        \node at (-1.5,0) [left=0pt]{$\phantom{8}$};
    \end{tikzpicture}
    }
        \caption{  
    Sketches of monomials of $\Omega(z_{ij},\beta_k, \tau)$  or its non-trivial coefficient $f^{w>0}_{ij}$.
Each $\Omega(z_{ij},\beta_k, \tau)$  or its non-trivial coefficient $f^{(w>0)}_{ij}$ is depicted by a thick line connecting node $z_i$ and $z_j$. Although the $\beta_k$ variables play a role in $\Omega(z_{ij},\beta_k, \tau)$, we choose to simplify the sketches by temporarily excluding them.  \label{sketch}}
       
     \end{figure}
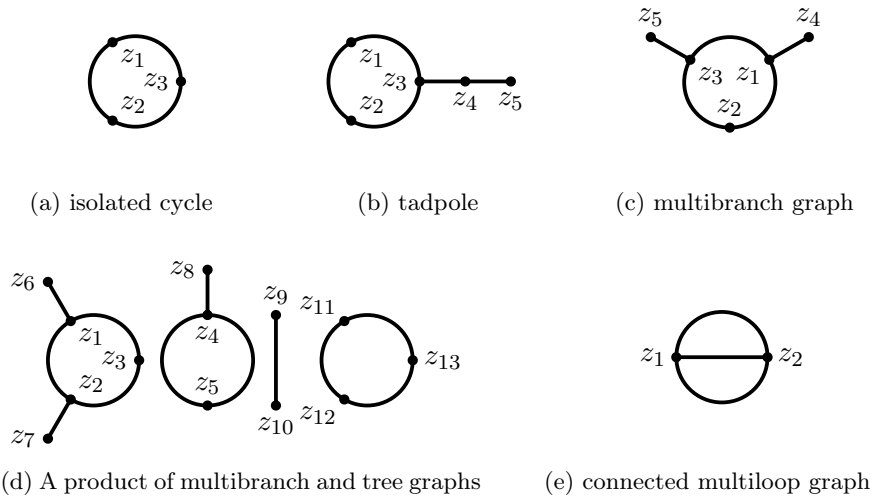

As illustrated in \cref{sketch}, attaching a single line to an isolated cycle results in a tadpole while attaching additional lines or trees produces a multibranch graph. Notably, isolated cycles and tadpoles can be considered as specific cases of multibranch graphs. Consequently, a universal generating function for massless string integrands would comprise a linear combination of products of multibranch  and tree graphs.

Developing a comprehensive formula to accommodate these varied cases is a formidable challenge. Rather, our focus shifts to elucidating the fundamental strategy for translating any product of multibranch structures into a basis form. This approach draws inspiration from tree-level treatments, such as in \cite{He:2018pol}, and encompasses two pivotal steps: first, we demonstrate the procedure for translating any tadpole into a basis form; subsequently, we extend this methodology to handle multibranch structures by reducing them as  tadpoles.  Further, in \cref{connected multiloop graphsec}, we initiate the discussion on the handling of connected multiloop graphs.

\subsection{Reducing tadpoles to chains by bruteforce}

When dealing with the product of cycles in previous sections, we encountered tadpoles in the intermediate state. In those specific cases, it was possible to transform all intermediate tadpoles into isolated $f$-$\Omega$ cycles using the fusion operation \eqref{eq:fusionr}. However, this elegant method may not be universally applicable. In this subsection, we provide a brute-force approach to decompose any tadpoles into chains.

Consider the tadpole $\CC_{(123)}(\xi) \Omega_{34}(\beta_1)\Omega_{45}(\beta_2)$ depicted in \cref{sketch} (b). Applying the single-cycle formula \eqref{stibp} with $z_a = z_3$ as the special point to break $\CC_{(123)}(\xi)$, we obtain,
\ba
\label{equation71}
(1{+}s_{123}) \CC_{(123)}(\xi) \Omega_{34}(\beta_1)\Omega_{45}(\beta_2)
=  \Big( \MM_{123}(\xi)- 
 {\bm \Omega}_{312}  \sum_{i=4}^{n}  x_{2,i} + &{\bm \Omega}_{321}  \sum_{i=4}^{n}  x_{1,i} 
\\
- 
\nabla_2 {\bm \Omega}_{312}   +\nabla_1 {\bm \Omega}_{321}  
\Big) &  
\Omega_{34}(\beta_1)\Omega_{45}(\beta_2) \,,
\non
\ea
where the total Koba-Nielsen derivative terms w.r.t. $z_1$ and $z_2$ can be omitted.
The combinatoric behaviour of \eqref{equation71} can be shown by
\ba 
\label{equation71graph}
&
 \begin{tikzpicture}[line width = 0.5mm,scale=1.2]
    \draw (0,0) circle (0.5);
      \draw [dashed, line width = 0.2mm] (2.5,0) circle (0.5);
          \node at (1.2,-.7) [right=0pt,font=\normalsize]{${\hat R}\!=\!\!\{6,\!\cdots\!,n,\!0\}$};
    \draw (0.5,0) -- (1,0);
    \filldraw (120:0.5) circle (1pt) node[below right=-1.5pt]{$z_1$} (-120:0.5) circle (1pt) node[above right=-1.5pt]{$z_2$} (0.5,0) circle (1pt) node[left=0pt]{$z_3$} (1,0) circle (1pt) node[below =-1.5pt]{$z_4$};
    \path (1,0) -- (0.5,0);
    \draw (1.5,0) -- (1,0);
    \filldraw (1.5,0) circle (1pt) node[below=-1.5pt]{$z_5$} ;
    \node at (3,0) [right=0pt,font=\normalsize]{$\displaystyle\;\; \to $};
    \end{tikzpicture}
\\ 
& \nonumber
\begin{tikzpicture}[line width = 0.5mm,scale=1.2]
        \node at (0,.5) [font=\normalsize]{$~$};
    \begin{scope}[ yshift=0.6cm]
          \draw [dashed, line width = 0.2mm] (2.2,0) circle (0.5);
          \node at (2.01,-.75) [right=0pt,font=\normalsize]{${\hat R}$};
           \node at (2.8,0) [right=0pt,font=\normalsize]{$,$} ;
    \draw (0,0) circle (0.6);
    \draw (0.6,0) -- (1.1,0);
    \filldraw (90:0.6) circle (1pt) node[below =-1.5pt]{$z_3$} (180:0.6) circle (1pt) node[ right=-1.5pt]{$z_1$} (-90:0.6) circle (1pt) node[above=-1.5pt]{$z_2$} (0:0.6) circle (1pt) node[left=-1.5pt]{$z_4$}  (1.1,0) circle (1pt) node[below =-1.5pt]{$z_5$};
    \path (1.1,0) -- (0.6,0);
    \end{scope}
    \begin{scope}[xshift=4.25cm,yshift=.6cm]
              \draw [dashed, line width = 0.2mm] (1.8,0) circle (0.5);
          \node at (1.61,-.75) [right=0pt,font=\normalsize]{${\hat R}$};
                     \node at (2.4,0) [right=0pt,font=\normalsize]{$,$} ;
    \draw (0,0) circle (0.7);
    \filldraw (0:0.7) circle (1pt) node[left =-1.5pt]{$z_5$} (72:0.7) circle (1pt) node[ below=-1.5pt]{$z_4$} (144:0.7) circle (1pt) node[right=-.5pt]{$z_3$} (216:0.7) circle (1pt) node[right=-.5pt]{$z_1$}  (-72:0.7) circle (1pt) node[ above=-1.5pt]{$z_2$};
    \end{scope}
    \begin{scope}[xshift=7cm,yshift=0.6cm]
               \node at (4.1,0) [right=0pt,font=\normalsize]{$,$} ;
                  \draw [dashed, line width = 0.2mm] (1,0) circle (0.5);
          \node at (0.81,-.75) [right=0pt,font=\normalsize]{${\hat R}$};
    \draw (1.5,0)node {\tikz{\draw [fill=black] (1,0) circle (1pt)}} --++(0.5,0)node {\tikz{\draw [fill=black] (1,0) circle (1pt)}} node [below] {$z_2$}  --++(0.5,0)node {\tikz{\draw [fill=black] (1,0) circle (1pt)}} node [below] {$z_1$}  --++(0.5,0)node {\tikz{\draw [fill=black] (1,0) circle (1pt)}} node [below] {$z_3$}  --++(0.5,0)node {\tikz{\draw [fill=black] (1,0) circle (1pt)}} node [below] {$z_4$}  --++(0.5,0)node {\tikz{\draw [fill=black] (1,0) circle (1pt)}} node [below] {$z_5$};
    \path  (0.5,0)--(0,0);
    \path  (1,0)--(0.5,0);     
        \path (1.5,0)--(1,0);
    \path  (2,0)--(1.5,0);
    \path  (2.5,0)--(2,0);
    \end{scope}
    \begin{scope}[ yshift=-1cm]
                  \node at (2.8,0) [right=0pt,font=\normalsize]{$,$} ;
              \draw [dashed, line width = 0.2mm] (2.2,0) circle (0.5);
          \node at (2.01,-.75) [right=0pt,font=\normalsize]{${\hat R}$};
    \draw (0,0) circle (0.6);
    \draw (0.6,0) -- (1.1,0);
    \filldraw (90:0.6) circle (1pt) node[below =-1.5pt]{$z_3$} (180:0.6) circle (1pt) node[ right=-1.5pt]{$z_2$} (-90:0.6) circle (1pt) node[above=-1.5pt]{$z_1$} (0:0.6) circle (1pt) node[left=-1.5pt]{$z_4$}  (1.1,0) circle (1pt) node[below =-1.5pt]{$z_5$};
    \path (1.1,0) -- (0.6,0);
    \end{scope}
    \begin{scope}[xshift=4.25cm,yshift=-1cm]
                  \node at (2.4,0) [right=0pt,font=\normalsize]{$,$} ;
              \draw [dashed, line width = 0.2mm] (1.8,0) circle (0.5);
          \node at (1.61,-.75) [right=0pt,font=\normalsize]{${\hat R}$};
    \draw (0,0) circle (0.7);
    \filldraw (0:0.7) circle (1pt) node[left =-1.5pt]{$z_5$} (72:0.7) circle (1pt) node[ below=-1.5pt]{$z_4$} (144:0.7) circle (1pt) node[right=-.5pt]{$z_3$} (216:0.7) circle (1pt) node[right=-.5pt]{$z_2$}  (-72:0.7) circle (1pt) node[ above=-1.5pt]{$z_1$};
    \end{scope}
     \begin{scope}[xshift=7cm,yshift=-1cm]
                   \node at (4.1,0) [right=0pt,font=\normalsize]{$,$} ;
                  \draw [dashed, line width = 0.2mm] (1,0) circle (0.5);
          \node at (0.81,-.75) [right=0pt,font=\normalsize]{${\hat R}$};
             \draw (1.5,0)node {\tikz{\draw [fill=black] (1,0) circle (1pt)}} --++(0.5,0)node {\tikz{\draw [fill=black] (1,0) circle (1pt)}} node [below] {$z_1$}--++(0.5,0)node {\tikz{\draw [fill=black] (1,0) circle (1pt)}} node [below] {$z_2$}  --++(0.5,0)node {\tikz{\draw [fill=black] (1,0) circle (1pt)}} node [below] {$z_3$}  --++(0.5,0)node {\tikz{\draw [fill=black] (1,0) circle (1pt)}} node [below] {$z_4$}  --++(0.5,0)node {\tikz{\draw [fill=black] (1,0) circle (1pt)}} node [below] {$z_5$};
     \path (0.5,0)--(0,0);
         \path (1,0)--(0.5,0);         \path  (1.5,0)--(1,0);
                     \path(2,0)--(1.5,0);
                 \path (2.5,0)--(2,0);
    \end{scope}
    \end{tikzpicture}   
\ea 
where a thick line connecting node $z_i$ and $z_j$ could be $\Omega(z_{ij},\beta_k, \tau)$  or $f^{(1)}_{ij}$.
The dashed cycles  represent the remaining particle set ${\hat R}$ following the conventions in \cref{fig:2}.
 However, in this context, they merely specify the summation range and total Koba-Nielsen derivative terms on the right-hand side of \eqref{equation71}.

The right-hand side of \eqref{equation71} introduces new $f$-$\Omega$ tadpoles, exemplified by $\Big( {\bm \Omega}_{312} f^{(1)}_{24} $ $\Omega_{34}(\beta_1) \Big)$ $ \Omega_{45}(\beta_2)$, and isolated $f$-$\Omega$ cycles, such as ${\bm \Omega}_{312} f^{(1)}_{25} \Omega_{34}(\beta_1)\Omega_{45}(\beta_2)$. Significantly, the tails of the new $f$-$\Omega$ tadpoles are consistently shortened, as illustrated by the transition from $\Omega_{34}(\beta_1)\Omega_{45}(\beta_2)$ to $\Omega_{45}(\beta_2)$ or $1$. This phenomenon is explicitly depicted in \eqref{equation71graph}. Through the iterative application of the F-IBP relation \eqref{stibp}, we can systematically transform $f$-$\Omega$ tadpole graphs into a summation of labeled trees.

For example, for the newly induced $f$-$\Omega$ tadpole
\ba
&(1+s_{1234})\Big( {\bm \Omega}_{312} x_{2,4}   \Omega_{34}(\beta_1) \Big) \Omega_{45}(\beta_2) 
\\
=\,\,&(1+s_{1234}) s_{24} \Big(  \Omega_{12}(\eta_2) \Omega_{24}(\zeta) \Omega_{43}(-\beta_1)  \Omega_{31}(-\eta_3)  \Big)  \Omega_{45}(\beta_2)   \big|_{\zeta^0} 
\nl
=\,\,&
(1+s_{1234}) s_{24} \Big(  \CC_{(1243)}(\xi) \Big|_{\substack{ \eta_2\to \eta_2-\zeta
\\
\eta_3\to \zeta-\xi-\eta_4
\\
\eta_4\to -\beta_1-\zeta
\\
\xi\to -\eta_3
 }}
  \Big)  \Omega_{45}(\beta_2)   \big|_{\zeta^0}\,,
\non 
\ea
we have
\ba
\label{equation73}
(1+s_{1234}&)\Big( {\bm \Omega}_{312} x_{2,4}   \Omega_{34}(\beta_1) \Big) \Omega_{45}(\beta_2) 
=
\Big[\Big(  \MM_{1243}(\xi) -   {\bm \Omega}_ {1243} \sum_{i=5}^n x_{3,i }
\\ &
-   {\bm \Omega}_ {1342} \sum_{i=5}^n x_{2,i }
+ \big(  {\bm \Omega}_ {1234}+{\bm \Omega}_ {1324} \big)\sum_{i=5}^n x_{4,i }     \Big)  \Big|_{\substack{ \eta_2\to \eta_2-\zeta
\\
\eta_3\to \zeta-\xi-\eta_4
\\
\eta_4\to -\beta_1-\zeta
\\
\xi\to -\eta_3
 }}
 (1+s_{1234}) s_{24} 
\Omega_{45}(\beta_2) \Big]  \Big|_{\zeta^0}\,.
\non 
\ea
New length-5 isolated $f$-$\Omega$ cycles such as $ {\bm \Omega}_ {1243} x_{3,5 } \Omega_{45}(\beta_2)$ are induced on the right-hand side of \eqref{equation73}, but we already have the tools to decompose them.

Regardless of the complexity of the tadpole, one simply needs to repeat the above operation a finite number of times until the tadpole is reduced to the basis.

\subsection{Reducing multibranch graphs as tadpoles}

In the previous subsection, we demonstrated the utility of the single-cycle formula \eqref{stibp}, specifically its feature of lacking one Koba-Nielsen derivative for a puncture, in breaking a tadpole. This approach initially seems inapplicable to multibranch graphs, such as $\CC_{(123)} (\xi)\Omega_{14}(\beta_1)\Omega_{35}(\beta_2)$, depicted in \cref{sketch} (c). 

Nonetheless, we can circumvent this issue with a straightforward strategy. Observe that removing a graph segment containing a portion of the cycle transforms the remainder into a labeled tree. This tree can then be expressed as a sum of labeled lines sharing a common starting point using Fay identities \eqref{fayid} (or \eqref{fayprac} in practical applications). 
Take, for instance, the chain $\Omega_{14}(\beta_1)\Omega_{12}(\eta_{23}+\xi)\Omega_{23}(\eta_{3}+\xi) \Omega_{35}(\beta_2)$ in \cref{sketch} (d), which we can interpret as a labeled tree rooted at point 1. Applying Fay identities and denoting $\beta_{12}=\beta_1+\beta_2$, we obtain,
\ba
\label{multibranchtotadpole}
&\Big( \Omega_{14}(\beta_1)\Omega_{12}(\eta_{23}+\xi)\Omega_{23}(\eta_{3}+\xi)
 \Omega_{35}(\beta_2) 
 \Big) \Omega_{31}(\xi)
 \\
 =\,\,&
 \Big( {\bm \Omega}_{14235} + {\bm \Omega}_{12435} + {\bm \Omega}_{12345} + {\bm \Omega}_{12354}  \Big) \Big|_{\substack{ \eta_2\to \eta_2
 \\
 \eta_3\to \eta_3+\xi-\beta_2
 \\
 \eta_4\to \beta_1
 \\
 \eta_5\to \beta_2
 } }\Omega_{31}(\xi)
 \nl
 =\,\,&
 \Big( 
 \Omega_{14}(\eta_{23}+\xi+\beta_1)\Omega_{42}(\eta_{23}+\xi)\Omega_{23}(\eta_{3}+\xi)
 \Omega_{35}(\beta_2) 
 \nl\,\,&
 +
 \Omega_{12}(\eta_{23}+\xi+\beta_1)\Omega_{24}(\eta_{3}+\xi+\beta_1)\Omega_{43}(\eta_{3}+\xi)
 \Omega_{35}(\beta_2) 
 \nl\,\,&
 +
 \Omega_{12}(\eta_{23}+\xi+\beta_1)\Omega_{23}(\eta_{3}+\xi+\beta_1)\Omega_{34}(\beta_{12} )
 \Omega_{45}(\beta_2) 
 \nl\,\,&
 +
 \Omega_{12}(\eta_{23}+\xi+\beta_1)\Omega_{23}(\eta_{3}+\xi+\beta_1)\Omega_{35}(\beta_{12})
 \Omega_{54}(\beta_1) 
 \Big) \Omega_{31}(\xi)\,,
 \non
\ea
whose combinatoric behavior can be shown by
\begin{align}
\raisebox{-1.2cm}{
   \begin{tikzpicture}[line width = 0.5mm,scale=1.2]
  \draw (-120:1) -- (-120:0.5);
  \draw (-120:0.5) arc (-120:120:0.5);
  \draw (120:0.5) ++(-0.3,0) arc (120:240:0.5);
  \filldraw (120:0.5) circle (1pt) node[below=0pt]{$z_1$} (-120:0.5) circle (1pt) node[above=0pt]{$z_3$} (0.5,0) circle (1pt) node[left=0pt]{$z_2$} (120:1) circle (1pt) node[left=0pt]{$z_4$} (-120:1) circle (1pt) node[left=0pt]{$z_5$};
  \filldraw (120:0.5) ++(-0.3,0) circle (1pt) node[below=0pt]{$z_1$} (-120:0.5) ++(-0.3,0) circle (1pt) node[above=0pt]{$z_3$};
  \draw (120:0.5) -- (120:1);
  \end{tikzpicture}
 }
  \to
  \raisebox{-.45cm}{
  \begin{tikzpicture}[line width = 0.5mm]
  \begin{scope}[yshift=0cm]
  \draw (0,0) -- (2,0);
  \draw (0,0) .. controls (0.5,0.5) and (1,0.5) .. (1.5,0);
  \foreach \x in {0.125,0.375,0.625,0.875} {
  \path (2,0) -- (0,0);
  }
  \filldraw (0,0) circle (1pt) node[below=0pt]{$z_1$} (0.5,0) circle (1pt) node[below=0pt]{$z_4$} (1,0) circle (1pt) node[below=0pt]{$z_2$} (1.5,0) circle (1pt) node[below=0pt]{$z_3$} (2,0) circle (1pt) node[below=0pt]{$z_5$};
  \end{scope}
  \begin{scope}[xshift=2.6cm,yshift=0cm]
  \draw (0,0) -- (2,0);
  \draw (0,0) .. controls (0.5,0.5) and (1,0.5) .. (1.5,0);
  \foreach \x in {0.125,0.375,0.625,0.875} {
  \path (2,0) -- (0,0);
  }
  \filldraw (0,0) circle (1pt) node[below=0pt]{$z_1$} (0.5,0) circle (1pt) node[below=0pt]{$z_2$} (1,0) circle (1pt) node[below=0pt]{$z_4$} (1.5,0) circle (1pt) node[below=0pt]{$z_3$} (2,0) circle (1pt) node[below=0pt]{$z_5$};
  \end{scope}
  \begin{scope}[xshift=5.2cm]
  \draw (0,0) -- (2,0);
  \draw (0,0) .. controls (0.25,0.5) and (0.75,0.5) .. (1,0);
    \foreach \x in {0.125,0.375,0.625,0.875} {
  \path (2,0) -- (0,0);
  }
  \filldraw (0,0) circle (1pt) node[below=0pt]{$z_1$} (0.5,0) circle (1pt) node[below=0pt]{$z_2$} (1,0) circle (1pt) node[below=0pt]{$z_3$} (1.5,0) circle (1pt) node[below=0pt]{$z_4$} (2,0) circle (1pt) node[below=0pt]{$z_5$};
  \end{scope}
  \begin{scope}[xshift=7.8cm,yshift=0cm]
  \draw (0,0) -- (2,0);
  \draw (0,0) .. controls (0.25,0.5) and (0.75,0.5) .. (1,0);
    \foreach \x in {0.125,0.375,0.625,0.875} {
  \path (2,0) -- (0,0);
  }
  \filldraw (0,0) circle (1pt) node[below=0pt]{$z_1$} (0.5,0) circle (1pt) node[below=0pt]{$z_2$} (1,0) circle (1pt) node[below=0pt]{$z_3$} (1.5,0) circle (1pt) node[below=0pt]{$z_5$} (2,0) circle (1pt) node[below=0pt]{$z_4$};
  \end{scope}
    \node at (7.5,0) [font=\normalsize]{$,$};
  \node at (2.3,0) [font=\normalsize]{$,$};
  \node at (4.9,0) [font=\normalsize]{$,$};
  \end{tikzpicture}
  }
  ~.
\end{align}

In this manner, we have outlined a systematic process for transforming any multibranch graph into a combination of tadpoles, which can subsequently be reduced to labeled trees using F-IBP.

\subsection{General treatment for a  product of multibranch  and tree graphs}
The most complex graphs, lacking interconnected cycles, resemble the structure depicted in \cref{sketch} (d). These are composed of isolated tadpole and multibranch graphs, possibly interspersed with various labeled trees. Utilizing the methodologies previously discussed, we can systematically convert tadpole and multibranch graphs into labeled trees attached to other connected components, consequently reducing the overall cycle count by one. This recursive approach ultimately eliminates all cycles, yielding results solely composed of labeled trees or their products, which can be directly simplified using Fay identities.

To better elucidate this concept, consider the example below,
\ba
\CC_{(12)}(\xi_1) \Omega_{13}(\beta_1)\Omega_{24}(\beta_2)\, \CC_{(56)}(\xi_2) \Omega_{57}(\beta_3)\,  \Omega_{68}(\beta_4)  \,.
\ea
Initially, we transform the multibranch $\CC_{(12)}(\xi_1) \Omega_{13}(\beta_1)\Omega_{24}(\beta_2)$ into tadpoles,
\ba\label{threetadpoles}
\CC_{(12)}&(\xi_1) \Omega_{13}(\beta_1)\Omega_{24}(\beta_2)
=
\Big( \Omega_{13}(\eta_2+\xi_1+\beta_1)\Omega_{32}(\eta_2+\xi_1) \Omega_{24}(\beta_2) 
\\
&+ \Omega_{12}(\eta_2+\xi_1+\beta_1)\Omega_{23}(\beta_{12}) \Omega_{34}(\beta_2) 
+ \Omega_{12}(\eta_2+\xi_1+\beta_1)\Omega_{24}(\beta_{12}) \Omega_{43}(\beta_1) 
\Big) \Omega_{21}(\xi_1)\,.
\non
\ea
Focusing on the first tadpole  on the  right-hand side of   \eqref{threetadpoles}, we apply F-IBP
\ba
&\Big( \Omega_{13}(\eta_2+\xi_1+\beta_1)\Omega_{32}(\eta_2+\xi_1)   \Omega_{21}(\xi_1) \Big)  \Omega_{24}(\beta_2) \,  \CC_{(56)}(\xi_2) \Omega_{57}(\beta_3)\Omega_{68}(\beta_4)  
\\
\overset{\rm IBP}=\,\,&
\Big(
\MM_{132}(\eta_2+\xi_1)\big|_{\eta_3\to \beta_1} -   \Omega_{21}(-\eta_2)\Omega_{13}(\beta_1) \sum_{i=4}^n x_{3,i}
+   \Omega_{23}(-\eta_2)\Omega_{31}(-\eta_2-\beta_1) \sum_{i=4}^n x_{1,i}
\Big)
\nl\,\,& 
\times  \Omega_{24}(\beta_2) \,  \CC_{(56)}(\xi_2) \Omega_{57}(\beta_3) \Omega_{68}(\beta_4)   \,.
\non
\ea 
The tadpole $\Big( \Omega_{13}(\eta_2+\xi_1+\beta_1)\Omega_{32}(\eta_2+\xi_1)   \Omega_{21}(\xi_1) \Big)  \Omega_{24}(\beta_2)$ is now reduced to chains and isolated cycles. The latter can be further decomposed until they also become chains  which may attach to the second multibranch $ \CC_{(56)}(\xi_2) \Omega_{57}(\beta_3) \Omega_{68}(\beta_4) $.  Regardless of the complexity, we have effectively reduced the cycle count by one and need only to further reduce the second multibranch $  \CC_{(56)}(\xi_2) \Omega_{57}(\beta_3) \Omega_{68}(\beta_4) $. Similar reduction steps apply to the other tadpoles in  \eqref{threetadpoles}. Though the final expression may appear intricate, we have demonstrated its reducibility to simpler forms.

\subsubsection{Comments on possible refinement \label{comments}}

Up to now in this section, we have demonstrated a systematic method to decompose any product of isolated cycles, tadpoles, or multibranch graphs into basis elements. Although this approach is effective, more efficient methods may exist, particularly in practical applications. A prime example is the product of pure isolated cycles discussed in the preceding sections.

In the context of bosonic or heterotic string integrands, we encounter another crucial element alongside elliptic functions $V_w(1,2,\cdots,m)$, defined as,
\begin{equation}
{\mathscr E}_i :=\sum_{\substack{j=1\\j\neq i}}^{n} \e_i\cdot k_j f^{(1)}_{ij}\,,
\end{equation}
where $\e_i$ denotes the gluon polarization \footnote{For a graviton, we need two copies of this.}. Products of ${\mathscr E}_i$ can potentially yield multibranch structures upon expansion, implying that our existing knowledge on handling $\CC_W$'s might be insufficient. However, employing a strategy analogous to that used at tree level \cite{He:2019drm}, we can treat ${\mathscr E}_i$ as a length-1 ``cycle''. By extending the definitions of genus-one fusions \eqref{eq:fusionr} and labeled forests \eqref{treeExpansion} to incorporate ${\mathscr E}_i$, we can develop a recursive formula akin to \eqref{ansa} for integrands involving ${\mathscr E}_i$. For instance, similar to \eqref{twocycle222}, for a single ${\mathscr E}_n$ paired with a single cycle, we derive,
\ba
\label{equation710}
 ( 1+s_{12\cdots,n-1})\CC_{(12\cdots,n-1)}&(\xi_1){\mathscr E}_n 
= 
\MM_{12\cdots m}(\xi_1) {\mathscr E}_n
  \\ 
 & -  
 \sum_{\substack{
a,b  
 =1 \\ a\neq b}
}^{n-1} \sum_{\substack{\rho \in A \shuffle B^{\rm T}\\ 
(a,A,b,B)=(1,2,\cdots,n-1) } }
 (-1)^{|B|} 
\,\e_n\!\cdot\! k_a \,f^{(1)}_{n a} \, \, (x_{b,n}+\nabla_b) \, {\bm \Omega}_{a,\rho,b} ,
\non 
\ea
The final term on the right side of \eqref{equation710} introduces isolated length-$n$ $f$-$\Omega$ cycles, which can subsequently be broken down using \eqref{stibp}. Despite this, a comprehensive derivation of a general formula to address an arbitrary number of ${\mathscr E}_i$ is beyond this paper’s scope, marking the end of our discussion on ${\mathscr E}_i$’s here.

\subsubsection{A product of meromorphic multibranch and tree graphs }

The methodologies outlined in this section are equally applicable within the chiral splitting framework. In this context, we define a meromorphic tadpole as a monomial where all instances of $\Omega$ in a tadpole are substituted with $F$, exemplified by $\CCF_{(12)}(\xi)F_{23}(\beta)$. Similarly, a meromorphic multibranch can be defined, such as $\CCF_{(12)}(\xi)F_{23}(\beta_1)F_{14}(\beta_2)$.

Given that $F_{i,j}(\beta_k)$ and $\Omega_{i,j}(\beta_k)$ obey identical Fay identities, as expressed in \eqref{fayid}, we can reduce any meromorphic multibranch to meromorphic tadpoles in the same manner as we do for $\Omega$'s. The equivalence established in \eqref{multibranchtotadpole} remains valid when substituting all occurrences of $\Omega$ with $F$.

Addressing meromorphic tadpoles, it is crucial to be mindful of the attaching point when applying the formula \eqref{stibpf}. Upon breaking the cycle in an original meromorphic tadpole, the tails in the resultant $g$-$F$ tadpoles are shortened. For example,
\ba
\label{equation711}
(1+s_{12}) \CCF_{(12)}(\xi) F_{13}(\beta_1) F_{34}(\beta_2)
= & \Big( \MMF_{12}(\xi)+ {F}_{12}(\eta_2) \, \ell\!\cdot\! k_2\,- {F}_{12}(\eta_2) \sum_{i=3}^{n} {\tilde x}_{2,i} \Big) 
\\ & \times F_{13}(\beta_1) F_{34}(\beta_2) - \tilde\nabla_2 \big( {F}_{12}(\eta_2) F_{13}(\beta_1) F_{34}(\beta_2) \big)\,,
\non
\ea
where the tail $F_{13}(\beta_1) F_{34}(\beta_2)$ in the meromorphic tadpole on the left-hand side of \eqref{equation711} becomes the shorter tail $F_{34}(\beta_2)$ in the $g$-$F$ tadpole $ {F}_{12}(\eta_2) {\tilde x}_{2,3} F_{13}(\beta_1) F_{34}(\beta_2)$ on the right-hand side. This recursive approach allows us to reduce any meromorphic tadpole to its basic form.

For products consisting of isolated meromorphic cycles, meromorphic tadpoles, or meromorphic multibranch graphs, we can sequentially break down the cycles using the aforementioned techniques.

\subsection{Towards connected multiloop graphs \label{connected multiloop graphsec}}

In generating functions for genus-one string integrands with massive external legs, a typical new structure beyond multibranch graphs is the connected multiloop graphs \cite{}, where two or more cycles interconnect, exemplified by $\prod_{i=1}^r \Omega_{12}( \chi_{i} ) $ for $r\geq 3$.  
In this paper, we embark on the relevant study by addressing the challenge of reducing $\prod_{i=1}^r \Omega_{12}( \chi_{i} ) $ to a basis, commencing with the two-loop case of $r=3$ as depicted in \cref{sketch} (e). This process presents a form of generalization from the $r=2$ case, discussed in section 2 of the companion paper \cite{Rodriguez:2023qir}. 
A comprehensive treatment of any connected multiloop graphs remains a subject for future exploration.

\subsubsection{Two connected loops}

As already used in \cite{Rodriguez:2023qir},
by carefully taking the limit $z_1\to z$ and $z_2\to -z$ in Fay identities (\ref{fayid}),  one can derive  the following identities 
\begin{align}
\Omega(z,\eta_1,\tau) \Omega({-}z,\eta_2,\tau) &= \Omega(z,\eta_1{-}\eta_2,\tau)
 \big( \hat g^{(1)}(\eta_2,\tau) - \hat g^{(1)}(\eta_1,\tau) \big) + \partial_z \Omega(z,\eta_1{-}\eta_2) \,.\label{variant}
\end{align}
From this, we deduce,
\ba
 \label{multiloopori}
\Omega_{12}(\chi_1) \Omega_{12}(\chi_2)\Omega_{12}(\chi_3)
=
\Omega_{12}(\chi_{12}) \Omega_{12}(\chi_3) \Big(\hat{g}^{(1)}(\chi_{12}, & \tau)  -\hat{g}^{(1)}(\chi_{1}, \tau)\Big)
\\ \non
&
\qquad -\Omega_{12}(\chi_3)\partial_{2} \Omega_{12}( \chi_{12} ) \,,
\ea 
where $\chi_{12}=\chi_{1}+\chi_{2}$. The term $\Omega \partial  \Omega$ prompts us to employ IBP relations, leading to two simultaneous equations
\ba
 \mathcal{I}_{n} \partial_{2}  \Big(\Omega_{12}( \chi_{12} )  \Omega_{12}(\chi_3)    \Big) & =  \partial_{2}\Big(
  \Omega_{12}( \chi_{12} )   \Omega_{12}(\chi_3)    \mathcal{I}_{n}  \Big) 
- \Omega_{12}( \chi_{12} )   
\Omega_{12}(\chi_3)
\Big(  \partial_{2} \mathcal{I}_{n} \Big)
\nl&
=
\Big(-s_{12} f_{12}^{(1)} +\sum_{i=3}^{n}  s_{2i} f_{2i}^{(1)} + \partial_{2} \Big) \Big(
 \Omega_{12}( \chi_{12} ) 
\Omega_{12}(\chi_3)   \mathcal{I}_{n} \Big)\,.
\label{ibp2ptmultiloop}
\ea
Together with 
\begin{align}
f^{(1)}_{12} \Omega_{12}(\chi_{12})&=\partial_{2}\Omega_{12}(\chi_{12})+\big(\hat{g}^{(1)}(\chi_{12})+\partial_{\chi_{1} }\big)\Omega_{12}(\chi_{12})\,,
\nl 
f^{(1)}_{12} \Omega_{12}(\chi_3)&=\partial_{2}\Omega_{12}(\chi_3)+\big(\hat{g}^{(1)}(\chi_3)+\partial_{\chi_3}\big)\Omega_{12}(\chi_3)\,,
\end{align}
they lead to the following two simultaneous equations,
\ba
&
    \Omega_{12}(\chi_3)  \partial_{2}  \Omega_{12}( \chi_{12} ) +\Omega_{12}( \chi_{12} )   \partial_{2} \Omega_{12}(\chi_3) -\Big(\sum_{i=3}^{n}  s_{2i} f_{2i}^{(1)} + \nabla_{2} \Big) \Big(
\Omega_{12}( \chi_{12} )  \Omega_{12}(\chi_3)    \Big)
\nl =\,\,& 
-s_{12}  \Omega_{12}( \chi_{3} ) \partial_{2}\Omega_{12}(\chi_{12})
-s_{12}  \Omega_{12}( \chi_{3} ) 
\big(\hat{g}^{(1)}(\chi_{12})+\partial_{\chi_{1}}\big)\Omega_{12}(\chi_{12})
\nl =\,\,& 
-s_{12}  \Omega_{12}( \chi_{12} ) \partial_{2}\Omega_{12}(\chi_3) -s_{12}  \Omega_{12}( \chi_{12} ) \big(\hat{g}^{(1)}(\chi_3)+\partial_{\chi_3}\big)\Omega_{12}(\chi_3)
\,.
\label{ibp2pttmultiloop}
\ea
Solving these two equations \eqref{ibp2pttmultiloop} for the two $\Omega   \partial \Omega$ terms,  we get
\ba\label{solmultiloop}
\Omega_{12}( \chi_{3} )   \partial_{2} \Omega_{12}(\chi_{12}) =\,\,& \frac{ {\cal A}-(1+s_{12}){\cal B}_{12}+{\cal B}_{3}}{2+s_{12}}\,,
\nl 
\Omega_{12}( \chi_{12} )   \partial_{2} \Omega_{12}(\chi_3) =\,\,& \frac{ {\cal A}-(1+s_{12}){\cal B}_3+{\cal B}_{12}}{2+s_{12}}\,,
\ea
with 
\ba
{\cal A}=\,\,&\Big(\sum_{i=3}^{n}  s_{2i} f_{2i}^{(1)} + \nabla_{2} \Big) \Big(
\Omega_{12}(\chi_3) \Omega_{12}( \chi_{12} )     \Big)  \,,
\nl
{\cal B}_{12}=\,\,&
 \Omega_{12}( \chi_{3} ) 
\big(\hat{g}^{(1)}(\chi_{12})+\partial_{\chi_{1}}\big)\Omega_{12}(\chi_{12})\,,
\nl 
{\cal B}_3=\,\,&   \Omega_{12}( \chi_{12} ) \big(\hat{g}^{(1)}(\chi_3)+\partial_{\chi_3}\big)\Omega_{12}(\chi_3)\,,
\ea

By incorporating the solution from \eqref{solmultiloop} into \eqref{multiloopori}, the following concise representation is achieved
\ba
\label{equation718}
\Omega_{12}(\chi_1) \Omega_{12}(\chi_2)\Omega_{12}(\chi_3)
 =& \frac{1}{2+s_{12}} \Big(
(3+2s_{12}) \hat{g}^{(1)}(\chi_{12})\!-\!(2+s_{12})\hat{g}^{(1)}(\chi_{1})\!+\!(1+s_{12})\partial_{\chi_{1}}\non
\\ 
\,\,&
-\hat{g}^{(1)}(\chi_{3})\!-\!\partial_{\chi_{3}}
-\sum_{i=3}^{n}  s_{2i} f_{2i}^{(1)} - \nabla_{2} 
\Big) \Omega_{12}( \chi_{12} ) \Omega_{12}( \chi_{3} ) \,,
\ea
successfully eliminating all structures of connected multiloop graphs and only leaving isolated cycles or $f$-$\Omega$ tadpoles, which are already understood and can be further simplified.

Note that   \eqref{equation718} is an exact formula with the Koba-Nielsen derivative of $z_1$ not participating on the right-hand side. Consequently, if there are chains directly attached to this graph of connected multiloop graphs solely through $z_1$ or ${\bar z}_1$ in the generating functions of comprehensive string integrands (e.g., $\Omega_{12}(\chi_1) \Omega_{12}(\chi_2)\Omega_{12}(\chi_3) \Omega_{14}(\beta)$), the Koba-Nielsen derivative term $\partial_2(\cdots)$ on the right-hand side of the formula transforms into a total derivative, which can then be disregarded, simplifying the expression further.

\subsubsection{General case}

In general, for any $r\geq 2$, one can derive
\ba
\label{multiloopgene}
\prod_{i=1}^r \Omega_{12}( \chi_{i} ) 
=\,\,& \frac{1}{r-1+s_{12}} \Big[
(2r-3+2s_{12}) \hat{g}^{(1)}(\chi_{12})-(r-1+s_{12})\hat{g}^{(1)}(\chi_{1})+(r-2+s_{12})\partial_{\chi_{1}}
\nl\,\,&-\sum_{i=3}^r \left( \hat{g}^{(1)}(\chi_{i})-\partial_{\chi_{i}} \right) -\sum_{i=3}^{n}  s_{2i} f_{2i}^{(1)} - \nabla_{2} 
\Big]\Omega_{12}( \chi_{12} ) \prod_{i=3}^r \Omega_{12}( \chi_{i} )  \,,
\ea
successfully reducing the number of cycles by one. Through recursive application, such connected multiloop graphs  can be simplified down to a basis, showcasing the versatility of this approach for a broad range of cases.
This comprehensive demonstration underscores the robustness of the method, affirming its capacity to simplify any polynomial of $\Omega_{i,j}(\beta_k)$ down to its basis components.

Following a parallel structure to \eqref{multiloopgene}, we also establish,
\ba
\prod_{i=1}^r \! F_{12}( \chi_{i} ) & 
= \frac{1}{r-1+s_{12}} \Big[
(2r-3+2s_{12}) {g}^{(1)}(\chi_{12})\!-\!(r\!-\!1\!+\!s_{12}){g}^{(1)}(\chi_{1})\!+\!(r\!-\!2+s_{12})\partial_{\chi_{1}}
\nl\,\,&-\sum_{i=3}^r \left( {g}^{(1)}(\chi_{i})-\partial_{\chi_{i}} \right) -\sum_{i=3}^{n}  s_{2i} g_{2i}^{(1)} + \, \ell\!\cdot\! k_2\, - \tilde\nabla_{2}
\Big] F_{12}( \chi_{12} ) \prod_{i=3}^r F_{12}( \chi_{i} )  \,,
\ea
validating the applicability of this methodology within the chiral splitting formalism as well.

\section{Discussion\label{secdis}}

In our companion paper \cite{Rodriguez:2023qir}, we significantly refined Fay-identities and integration-by-parts (F-IBP) methodologies applied to one-loop string integrals with Koba-Nielsen factors, focusing on Kronecker-Eisenstein series and their associated coefficients $f^{(w)}(z_i{-}z_j,\tau)$ and $g^{(w)}(z_i{-}z_j,\tau)$. We managed to express these elements in terms of conjectural chain topology bases for generating functions of one-loop string integrals \cite{Mafra:2019ddf,Mafra:2019xms, Gerken:2019cxz}. Building upon these advancements, the present study broadens the scope of the recursive strategies introduced in \cite{Rodriguez:2023qir}, encompassing a greater variety of Kronecker-Eisenstein cycles and unraveling the elegant
combinatorial structure of their F-IBP reductions. Utilizing single-cycle formulae derived previously, we successfully decomposed products of any number of Kronecker-Eisenstein cycles into a chain basis, albeit introducing certain total Koba-Nielsen derivative terms in the process. To facilitate application and accessibility, we have embedded our main results within a {\tt Mathematica} framework. Our study does not just stop at cyclic products; it delves into more general configurations of the Kronecker-Eisenstein series and coefficients that naturally appear in the moduli-space integrand of genus-one string amplitudes. These additional contributions are represented through tadpoles, multibranch structures, and connected multiloop graphs.

This paper succinctly formulates a method to break down products of isolated cycles of the Kronecker-Eisenstein series. However, for the most general massless string integrands without coupling terms from left- and right-moving sectors, we provide a comprehensive yet potentially intricate conceptual framework for application. This intricacy becomes more apparent when dealing with particular string integrands, such as those in heterotic strings with external graviton vertices, which we acknowledge may require additional efforts for streamlined basis decomposition, as detailed in \cref{comments}. We also showcase recursive reduction in connected multiloop graphs, typically appearing in the case of massive external string states. A complete exploration of combinatorial toolboxes for handling any polynomial of the Kronecker-Eisenstein series remains an open avenue for future work.

In terms of enhancing computational methods in string theories, this paper improves tools for $\ap$-expansions for genus-one integrals. Thanks to this work, clarifying the physical relevance of basis coefficients can be taken more easily. This leads to optimized computations, validation of the chain bases, and advancements in low-energy expansions of one-loop string amplitudes \cite{Mafra:2019ddf, Mafra:2019xms, Gerken:2020yii,Gerken:2020aju}. 

Moreover, our decomposition techniques pave the way for decomposing genus-one string amplitudes into gauge-invariant kinematic functions and exposing double-copy structures in general one-loop open-string amplitudes akin to those for maximal supersymmetry \cite{Mafra:2017ioj, Mafra:2018qqe}. In particular, this discovery prompts the possibility of uncovering analogous structures and loop-level double-copy relations in heterotic and bosonic theories, thereby extending the quantum-field-theory building blocks at tree level \cite{Mafra:2011nv, Zfunctions, Carrasco:2016ldy, Azevedo:2018dgo}.

Our work not only generates relations between string theory amplitudes via equivalent relations among string integrands, but also reveals much more relationships among partial loop integrands of field theories such as Yang-Mills, GR, Einstein-Yang-Mills theories etc. in the field theory limit, a domain where our results are particularly applicable \cite{He:2016mzd,He:2017spx}. By breaking all $f^{(w)}_{ij}$ loops at finite $\tau$ and $\a'$, introducing solely tachyon poles, and then proceeding to the field theory limit to remove these poles, we establish relations for partial loop integrands devoid of poles. This enables the extraction of polynomial BCJ numerators in that representation. 
Certain elliptic functions 
denoted by $V_w(1,2,\cdots, m)$ in the literature has also shown to be useful to 
produce loop integrands with quadratic propagators in the one-loop Cachazo-He-Yuan formula \cite{Feng:2022wee,Geyer:2015bja,Geyer:2015jch,Geyer:2017ela,He:2015yua,Cachazo:2015aol,Edison:2021ebi,Dong:2023stt,Xie:2024pro,Mafra:2018pll,Mafra:2018qqe} and its $\a'$ uplift, areas poised to benefit significantly from our techniques for handling $V_m(1,2,\cdots, m)$ and their products.

This work also paves the way for mathematical explorations, particularly in connection with elliptic multiple zeta values \cite{Enriquez:Emzv,Broedel:2014vla}, modular graph forms \cite{DHoker:2015wxz,DHoker:2016mwo}, and elliptic polylogarithms \cite{BrownLev, Broedel:2014vla, Broedel:2017kkb, Ramakrish, DHoker:2018mys, Broedel:2019tlz, DHoker:2020hlp}. It guides the verification of conjectural $n$-point integral bases and their generalizations \cite{Broedel:2019gba, Broedel:2020tmd, Kaderli:2022qeu},
 drawing parallels with Feynman integrals in particle physics. These connections suggest that twisted de Rham theory could provide a unified and robust framework to comprehend genus-one string integrals, their reductions, monodromy relations, and open-closed string relations \cite{Mastrolia:2018uzb, Frellesvig:2019kgj, Mizera:2019vvs, Frellesvig:2020qot, Caron-Huot:2021xqj, Caron-Huot:2021iev, Duhr:2023bku, Aomoto87, Mizera:2017cqs, Mizera:2017rqa, Mizera:2019gea, Tourkine:2016bak, Hohenegger:2017kqy, Tourkine:2019ukp, Casali:2019ihm, Casali:2020knc, Broedel:2018izr, Gerken:2020xfv, Stieberger:2021daa, Stieberger:2022lss,Mazloumi:2024wys}. Existing mathematical frameworks that leverage twisted cohomology setups have already demonstrated that meromorphic Kronecker-Eisenstein series form a basis under certain conditions, and a generalization of this could substantiate the conjectures presented herein \cite{ManoWatanabe2012,ghazouani2016moduli,goto2022intersection, Felder:1995iv, DHoker:1988pdl, DHoker:1989cxq,Bhardwaj:2023vvm}.

Last but not least, our study underscores the potential and efficacy of combinatorial toolboxes for tree-level string integrand basis decomposition at the genus-one level \cite{He:2018pol, He:2019drm}. It is of course interesting to study the potential extensions to genus-two scenarios \cite{DHoker:2023vax}. 
A natural follow-up step is to study the potential extensions to bases of Koba-Nielsen integrals for higher-genus string amplitudes. As a higher-multiplicity generalization of the derivatives of Green functions in the two-loop five-point \cite{DHoker:2020prr,DHoker:2020tcq,DHoker:2021kks} and three-loop four-point amplitudes \cite{Gomez:2013sla,Geyer:2021oox}, it would be interesting to construct generating functions of higher-genus string integrals from the integration kernels of \cite{DHoker:2023vax}. 

\section*{Acknowledgments}

We would like to thank Carlos Rodriguez and Oliver Schlotterer   for  constructive feedback on our manuscript and collaboration on related topics.  Our appreciation also
extends to Freddy Cachazo and Song He for useful discussions. 
The research of Y.Z.\ was  supported in part by a grant from the Gluskin Sheff/Onex Freeman Dyson Chair in Theoretical Physics and by Perimeter Institute. Research at Perimeter Institute is supported in part by the Government of Canada through the Department of Innovation, Science and Economic Development Canada and by the Province of Ontario through the Ministry of Colleges and Universities.
The research of Y.Z.\ was also  supported by the Knut and Alice Wallenberg Foundation under the grant KAW 2018.0116: From Scattering Amplitudes to Gravitational Waves.



\appendix


\section{Notations\label{appendixno}}

In continuation of the discussions in our companion paper \cite{Rodriguez:2023qir}, we utilize several terminologies and notations that are extensively defined therein. To aid readers, we recapitulate these expressions, particularly those introduced in \cref{section:two} and \cref{seccode}.

\subsection*{Integration domains and measures}

For the integration of open and closed string amplitudes, as delineated in \eqref{openamp} and \eqref{closedamp}, their respective domains and measures are detailed below,
\ba\label{openamp2}
& \int_{\rm op} d {\bm \mu}_n^{\rm op} ~\phi:=\sum_{\text {top }} C_{\text {top }} \int \limits_{D^{\tau}_{\text {top }}} %
\frac{d \tau }{(\operatorname{Im} \tau)^{\frac{D}{2}}}
 \int \limits_{D^{z}_{\text {top }}} 
d z_{2}   \ldots d z_{n} ~\phi\,,
\\
&\label{closedamp2}
 \int_{\rm cl} d {\bm \mu}_n^{\rm cl} ~\phi := \int \limits_{\mathfrak{F}} 
 \frac{d^{2} \tau }{(2 {\rm Im} \,\tau)^{\frac{D}{2}}}
\int \limits_{\mathfrak{T}_\tau^{n-1}}
d^{2} z_{2}   \ldots d^{2} z_{n} \,
~ \phi \,.
\ea
Here, for open strings,
the summation extends over two topologies—cylinder and Moebius-strip. The \( C_{\text{top}} \) factors, known as color or Chan-Paton factors, along with the integration domains \( D^{\tau}_{\text{top}} \) and \( D^{z}_{\text{top}} \) for \( \tau \) and \( z_i \) respectively, are discussed in detail in \cite{green1988superstring}. The exponent of \( \Im \tau \) is determined by the spacetime dimension  \( D \).

For closed strings,
the integration of the modular parameter \( \tau \) is performed over the fundamental domain \( \mathfrak{F} \) of the modular group ${\rm SL}_2(\mathbb Z)$. The integration over punctures \( z_2,\ldots,z_n \) spans the toroidal worldsheet \( \mathfrak{T}_\tau \), which is characterized by a standard parallelogram in the complex \( z_i \)-plane, having corners at \( 0,1,\tau{+}1, \) and \( \tau \). 

Even though $z=0$ was set using the translation invariance on both open and closed string
worldsheets at genus one, we consider $z_1$ to be generic throughout our study.

\subsection*{Koba-Nielsen factors}
The Koba-Nielsen factors for open and closed string amplitudes \eqref{openamp}, \eqref{closedamp}, as well as for chiral splitting \eqref{openampF}, are detailed as follows
\ba
&\mathcal{I}_{n}^{\text {op}}  := \exp \bigg({-}\sum_{i<j}^{n} s_{i j}\left[\log \left|\theta_{1}(z_{i j}, \tau)\right|-\frac{\pi}{\operatorname{Im} \tau}\left(\operatorname{Im} z_{i j}\right)^{2}\right]\bigg)\, ,
\\
&
\mathcal{I}_{n}^{\rm cl}  := \exp \bigg({-}\sum_{i<j}^{n} s_{i j}\left[\log \left|\theta_{1}(z_{i j}, \tau)\right|^{2}-\frac{2 \pi}{{\rm Im} \,\tau} \left(\operatorname{Im} z_{i j}\right)^{2}\right]\bigg)\,,
\label{defkn}
\\
&
\mathcal{J}_{n}(\ell) := \exp \bigg({-} \sum_{1 \leq i<j}^{n} s_{i j} \log \theta_{1} (z_{i j}, \tau )+ \sum_{j=1}^{n}   z_{j} \left(\ell \!\cdot\! k_{j}\right)+ \frac{\tau}{4 \pi i} \ell^{2}\bigg)\,.
\label{chiKN}
\ea
Despite treating \( s_{ij} \) independently, the translation invariance of \( \mathcal{J}_{n}(\ell) \) necessitates momentum conservation along the loop momentum's direction, embodied in the condition \(\sum_{j=1}^{n}    \left(\ell \cdot k_{j}\right) =0\).

\subsection*{${\pmb M}_{12\cdots m}(\xi)$ and $\tilde{\pmb M}_{12\cdots m}(\xi)$}

The doubly-periodic function 
\(\MM_{12\cdots m}(\xi)\) is a crucial component in the single-cycle formula \eqref{rhorhog}, and it is defined by the following elaborate expression,
 \begin{align}
&\!\!\!\!\MM_{12\cdots m}(\xi) :=
 \sum_{b =2}^{m} 
\!\!\!\!\!\!
\sum_{\substack{\rho \in \{2,3,\cdots, b-1\}
\\
\quad \shuffle \{m,m-1,\cdots, b+1 \}
}} 
\!\!\!\!\!\!\!\!\!\!\!\!\!\!\!\!
 (-1)^{m-b} \Bigg(
\sum_{i=1}^m \! s_{i b} \,\partial_{\eta_b}{-}
  \sum_{i=2}^m \! s_{i b}\, \partial_{\eta_i}
 {+} (1{+}s_{12\cdots m}) v_1(\eta_{b}, \,\eta_{b+1,\cdots,m}{+}\xi )   
 \nl
 &\quad \quad \quad \quad \quad  
  - \hat g^{(1)}(\eta_b) 
 -\sum_{i=2}^{b-1} { S}_{i,\rho} 
v_1(\eta_{b },\,\eta_{i,i+1,\cdots, b-1 })
-\sum_{i=b+1}^m { S}_{i,\rho} 
v_1(\eta_{b },\,\eta_{b+1,b+2,\cdots, i })  
 \Bigg){\bm \Omega}_{1,\rho,b}  
\notag \\
&\quad + \!\!\!\!\!\!\!\!\! \!\!\!\sum_{1\leq p<u<v<w<q\leq m+1} 
 \!\!\! \!\!\! \!\!\!\!\!\!\!\!\!
(-1)^{m+u+v+w}\, \Big(
v_1(\eta_{u+1,\cdots,w-1},-\eta_{u,\cdots,w-1})+v_1(\eta_{u,\cdots,w}-\eta_{u+1,\cdots,w}) 
 \Big) 
\non \\
&\quad \quad \quad \quad \quad   \times
\Big(\sum_{i=q}^m s_{vi}+\sum_{i=1}^p s_{vi} \Big) \! \! \! \!
 \sum_{\substack{\rho \in \{2,3,\cdots,p\}\shuffle \{m,m-1,\cdots, q\}\\
\gamma \in \{p+1,p+2,\cdots,u-1\}\shuffle \{v-1,v-2,\cdots, u+1\}\\
\pi \in \{v+1,v+2,\cdots,w-1\}\shuffle \{q-1,q-2,\cdots, w+1\}
 }} 
 \sum_{\sigma\in \{\gamma,u\}\shuffle \{ \pi, w\}} 
 \!\!\!\!\!\!\!
 \quad 
 {\bm \Omega}_{1,\rho,v,\sigma}
 \,,
 \label{rhorhoMM} 
 \end{align}
where $
S_{j, \rho}:=
s_{1j }+
\sum_{ i \in \rho} s_{ i j} \ {\rm if}
~ j\notin \rho\,, $
 otherwise $
s_{ 1 j}+ \! \sum_{ i \in \rho \atop i \text { precedes } j \text { in } \rho}
\! \! \! \! \! s_{ i j} \ \ \ \ 
$ 
and 
\ba 
& v_1(\eta, \xi ):= \hat g^{(1)}(\eta)+\hat g^{(1)}(\xi)-\hat g^{(1)}(\eta{+}\xi)=g^{(1)}(\eta)+g^{(1)}(\xi)- g^{(1)}(\eta{+}\xi)\,,
\\ 
& \qquad {\rm with}\quad 
\hat g^{(1)}(\eta,\tau) := g^{(1)}(\eta,\tau) + \frac{ \pi \eta}{\Im \tau} 
= \frac{1}{\eta} - \eta \hat {\rm G}_2(\tau)
-\sum_{n=4}^{\infty} \eta^{n-1} {\rm G}_n(\tau)\,.
\ea 
Here, the holomorphic Eisenstein series are derived from the Kronecker-Eisenstein series evaluated at the origin, represented as
\begin{equation}
{\rm G}_w(\tau) := \sum_{(m,n) \neq (0,0)} \frac{1}{(m\tau + n)^{w}} = - f^{(w)}(0,\tau)\, ,\qquad w\geq 4  \,,
\label{1.6}
\end{equation}
with modular weight \( (w,0) \). This is represented through absolutely convergent double sums over integers \( m,n \) for \( w\geq 4 \).
Although the analogous limit \( z\rightarrow 0 \) of \( f^{(2)}(z,\tau) \) is not well-defined, we come across a non-holomorphic yet modular variant of the weight-two Eisenstein series 
\begin{equation}
\label{G2hat}
\hat {\rm G}_2(\tau) := \lim_{s\rightarrow 0}\sum_{(m,n) \neq (0,0)} \frac{1}{(m\tau + n)^{2} \,|m\tau + n|^s} \, .
\end{equation}
Subsequently, the meromorphic version \( {\rm G}_2(\tau) \) of \( \hat {\rm G}_2(\tau) \) is given by \( {\rm G}_2(\tau) := \hat {\rm G}_2(\tau) + \frac{\pi}{\Im \tau} \).

In the context of the meromorphic functions \( \MMF_{12\cdots m} (\xi) \), they can be systematically derived from the doubly-periodic functions \( \MM_{12\cdots m} (\xi) \) through a simple substitution, 
\ba
\label{MMFeq}
\MMF_{12\cdots m} (\xi) = \MM_{12\cdots m} (\xi)\big|_{{\bm \Omega}_{1 \alpha(2)\ldots \alpha(m)} \to {\bm F}_{1 \alpha(2)\ldots \alpha(m)},\quad\hat g^{(1)}(\eta) \to g^{(1)}(\eta)}\,.
\ea

\subsection*{Elliptic functions and their breaking}

Elliptic functions \(V_w(1,2,\ldots,m)\) for a general \(w\) are constructed from the products of the Kronecker-Eisenstein series as shown below,
\begin{align}
&F(z_{12},\eta,\tau) F(z_{23},\eta,\tau) \ldots F(z_{m,1},\eta,\tau) \notag \\
&= \Omega(z_{12},\eta,\tau) \Omega(z_{23},\eta,\tau) \ldots \Omega(z_{m,1},\eta,\tau)
\notag \\
&=: \eta^{-m} \sum_{w=0}^\infty \eta^w V_w(1,2,\ldots,m|\tau) \ , \label{1.1a}
\end{align}
which yields 
\begin{align}
V_w(1,2,\ldots,m) &= \! \! \! \! \! \sum_{k_1{+}k_2{+}\ldots{+}k_m=w}\! \! \! \! \!  
f^{(k_1)}_{12} f^{(k_2)}_{23}\ldots 
f^{(k_{m-1})}_{m-1,m} f^{(k_m)}_{m1} 
\label{altVw} \\
&= \! \! \! \! \!  \sum_{k_1{+}k_2{+}\ldots{+}k_m=w} \! \! \! \! \!  
g^{(k_1)}_{12} g^{(k_2)}_{23}\ldots 
g^{(k_{m-1})}_{m-1,m} g^{(k_m)}_{m1}\,,
\notag
\end{align}
with cyclic identification $z_{m+1}=z_1$. 
Here, \(V_m(1,2,\ldots,m)\) prominently features an \(f\)-cycle \(f^{(1)}_{12}f^{(1)}_{23} \cdots f^{(1)}_{m-1,m} f^{(1)}_{m1}\).

 As elucidated in the companion paper \cite{Rodriguez:2023qir}, a direct application of \eqref{rhorhog} to string integrands involves decomposing the elliptic functions $V_m(1,2,\ldots,m)$ into a basis. First, according to \eqref{1.1a},
 one can easily establish a connection between $V_m(1,2,\ldots,m)$ and the cycle $\CC_{(12\cdots m)}(\xi)$ by extracting its coefficients of bookkeeping variables $\eta_i$ and $\xi$ in a specific order,
\begin{align}
V_m(1,2,\cdots ,m) =\CC_{(12\cdots m)}(\xi)\big|\big|_{\eta_2^0,\eta_3^0,\cdots,\eta_m^0} := \big(\cdots \big(\big( \CC_{(12\cdots m)}(\xi)\big|_{\eta_2^0} \big)\big|_{\eta_3^0} \big)\cdots \big|_{\eta_m^0} \big)  
\label{coeff}\,.
\end{align}
Then we just need to perform the same operation on the right-hand side of \eqref{rhorhog} to break $V_m(1,2,\ldots,m)$,
\begin{align}
\label{vopen}
V_m(1,2,\cdots, m) 
=\,\,& \frac{\MM_{12\cdots m}(\xi)\big|\big|_{\eta_2^0,\eta_3^0,\cdots,\eta_{m}^0,\xi^0}}{ 1+s_{12\cdots m}}-
 \frac{1}{ 1+s_{12\cdots m}}\sum_{
b =2}^{m}  (-1)^{m-b} \\
 &\quad 
 \times \! \! \! \! \sum_{\rho \in \{2,3,\cdots, b-1\}\shuffle \{m,m-1,\cdots, b+1 \}} \! \! \! \bigg(
\sum_{i=m+1}^{n} s_{bi} f_{bi}^{(1)} 
 +\nabla_b 
\bigg){\bm \Omega}_{1,\rho,b}\big|\big|_{\eta_2^0,\eta_3^0,\cdots,\eta_{m}^0}  
 \,. \notag
 \end{align}

Similar operations work for a product of elliptic functions.

\bibliographystyle{JHEP}
\bibliography{citesIBP}{}

\end{document}